\documentclass[aps,prc,twocolumn,showpacs,preprintnumbers,amsmath,amssymb]{revtex4}

\usepackage{color} %-- use \textcolor{color}{text} to write the text in color
\usepackage{ dsfont }
\usepackage{graphicx}

\usepackage{fancyhdr}
\usepackage{setspace}
\usepackage{amssymb}
\usepackage{relsize}
\usepackage{ tipa }
\usepackage{varwidth}
\usepackage{rotating}

\usepackage{pdflscape}

\usepackage{hvfloat}

\def\pu1 {$^{241}$Pu$^*$\mbox{ }}
\def\bra       {\langle}
 \def\ket       {\rangle}
\def\pua {$^{239}$Pu\mbox{ }}
\def\pub {$^{240}$Pu\mbox{ }}

\newcommand{\etal}{{\it et al.}}

\begin{document}
%consistently to neutron-induced fission cross sections 

\title{Revisiting fission-probability data using $\mathcal{R}$-matrix Monte-Carlo simulations : \\ application to Pu fissile isotopes over the 4 to 8 MeV excitation energy range }

\author{O.~Bouland}
\affiliation{CEA, DEN, DER, SPRC, Physics Studies Laboratory, Cadarache, F-13108 Saint-paul-lez-Durance, France}
\email{olivier.bouland@cea.fr}
\author{B.~Jurado}
\affiliation{CENBG, CNRS/IN2P3-Universit\'e de Bordeaux, Chemin du Solarium, B.P. 120, F-33175 Gradignan, France}

\begin{abstract}
This article describes an original approach to analyze simultaneously cross sections and surrogate data measurements using efficient Monte Carlo extended $\mathcal{R}$-matrix theory algorithm based on unique set of nuclear structure parameters. The alternative analytical path based on the manifold Hauser-Feshbach equation was intensively used in this work to gauge the errors carried by the surrogate-reaction method commonly taken to predict neutron-induced cross sections from observed partial decay probabilities. Present paper emphasizes in particular a dedicated way to treat direct reaction entrance and prior decay excited nucleus outgoing channels widths correlations. Present smart theoretical foundation brought the opportunity to apply successfully our method to both fission-probability data and directly measured neutron cross sections according to Pu fissile isotopes; namely the $^{237, 238, 240, 242~and~244}$Pu$^*$ nuclei. This new capability opens genuine perspectives in matter of 'evaluation process' from foreseen fission- and $\gamma$-decay probabilities simultaneously measured as derived data will become available. 
\end{abstract}

%\textcolor{blue}{COPY SUBMITTED TO PHYS. REV. C}
\date{\today}

%\keywords{}

\maketitle
\section{\label{s:Introduction}Introduction}

A wealth of experimental neutron-induced fission cross-section data for actinides and higher transuranic nuclides has been collected up over decades and is still being added. However the idea to supplement this database with particle-transfer-induced reactions has been raised a long time ago~\cite{cra:70b}. In point of fact, the surrogate technique was promoted from mid-sixties by Britt \etal~\cite{bri:65} at Los Alamos Scientific Laboratory. The original goal was to identify the positions of major low-lying collective bands near fission saddle by measuring fission-fragment angular correlations. Within the next decade, Back \etal~\cite{bac:74} investigated direct nuclear reactions for studying sub-threshold fission barrier vibrational structures with relatively low fission probabilities. Over the years, a variety of surrogate vectors have been used as stripping $(d,p)$ and pickup $(p,d)$ reactions, ($^3He,p$), ($^3He,d$) and ($^3He,t$) charge-exchange reactions or even two-neutron transfer reactions as $(t,p)$ and $(p,t)$ reactions. Analytical modeling of these observed direct-reaction fission probabilities were performed under several simplifications contained in the so-called 'surrogate-reaction method' (SRM). Early promising neutron-induced fission cross section comparisons~\cite{cra:70b} obtained either from SRM extrapolation or Neutron Physics Spectroscopy (NPS) measurements led to agreement within 10\% to 20\% at neutron energy above the nucleus pairing energy but exhibited larger deviations at lower energies. Major limitations in surrogate data  extrapolation were promptly noticed~\cite{cra:70b,bri:79} with the difficulty to estimate a) the compound nucleus formation cross section by neutron absorption, b) the possible influence of angular momentum differences between neutron capture and direct entrance reactions and c) the validity of the Weisskopf-Ewing (WE) hypothesis on reaction decay probability spin-parity independence~\cite{wei:40}.        

Last decade, surrogate reactions received newest interest in terms either of simulation~\cite{you:03a,you:03b} or experimental investigation (on the spur of study~\cite{pet:04}; see also a list of measurements  in Ref.~\cite{esc:12}) to infer neutron-induced partial cross sections. From the very beginning~\cite{cra:70b}, the SRM has been thought as to be very helpful for target material with unsuitable lifetimes (less than several days) or with high radio-toxicity. Present design of advanced reactor systems (e.g; accelerator driven systems and generation IV nuclear power reactors) strengthens our desideratum for a conclusive outcome on surrogate data feedback. Two clear benefits are expected: higher quality nuclear data uncertainty assessments on regular reactor fuel nuclides and actual experimental alternative to achieve suitable evaluated 'neutron-induced' cross section data for higher transuranic nuclei (as the Am, Np and Cm elements) that were described as 'exotic' actinides by Britt and Wilhelmy~\cite{bri:79} in the seventies. 
%\textcolor{red}{HERE}
Using mathematical derivatives of experimental probabilities, the authors of study~\cite{rom:12} have recommended to not use the WE approximation to retrieve neutron-induced capture or fission cross sections from surrogate experiments. Even though, the authors suggested to use those derivatives to extract complementary information to NPS such that fission barrier heights. %Quoting Romain: These confirm that the Weisskopf-Ewing approximation should not be used on a surrogate experiment in order to extract neutron capture or neutron induced fission cross sections.
By echo to their assertion, we can state that carrying severe approximations was understandable in the seventies because of computer limitations, lack of precise information on nuclear level densities across the deformation and the difficulties for achieving confident optical model calculations over large range of nuclides. Nowadays the bulk of those approximations can be overridden even if huge difficulties persist in the estimation of the various direct, pre-equilibrium and compound nucleus process fractions.
%See also Escher 2012 page 374 et surtout page 387
From the experimental side, we still have to face the tremendous  challenge of deriving unambiguously the Prior Decay Excited Nucleus (PDEN) individual probabilities, $P^{J^\pi} (E_x)$; with $J$ and ${\pi}$ being respectively the total angular momentum and parity of the state of the nucleus formed at a specific excitation energy $E_x$ across the surrogate reaction. This must entail building the experimental conditions such that a variety of energies and angles for the observed outgoing particle becomes available.
%See Escher  p359 
%return each $P^{J^\pi} (E_x)$ spin-parity decay component and not the weighted sum \big[$\sum_{J^\pi} \mathcal{F}_{sp}^{CN}(E_{x},J,{\pi})  P^{J^\pi} (E_{x})$\big]
%\big[$\sum_{J^\pi}\mathcal{P}^{J^\pi}(E_x)$\big] ; with $\mathcal{F}_{sp}^{CN}(E_{x},J,{\pi})$ the fraction of compound nucleus~\footnote{{To prevent confusion between target and residual nuclei,  the '$\star$' notation is usually favored in this article for the residual nucleus such that $^AElement^{ *}$.}} (CN) formed by surrogate reaction in a state of total angular momentum $J$ and parity $\pi$ at the corresponding excitation energy $E_x$.  
%\mathcal{F}_{sp}^{CN}(E_{sp},J,{\pi}) \times  P_{c'}^{J^\pi} (E_{c'})

In a previous paper~\cite{bou:13}, we have enlightened the actual possibility to carry one-dimensional fission barrier extended $R$-matrix simulations accurate enough to make predictions of low-energy neutron-induced fission cross sections for the isotopes of given family for which no neutron spectroscopy measurements exist. 
%This has been accomplished thanks to Monte Carlo samplings of class I and class II state parameters of the actinide double humped fission barrier and to model input parameters in part obtained from macroscopic-microscopic nuclear structure calculations~\cite{Mol:16}. Ultimately, we were able to confirm the fissility character (suggested by Fig.~4 of Ref.~\cite{you:03b} or even much earlier by Cramer and Britt from (t,pf) reaction extrapolations~\cite{cra:70b}) of the very short-lived $^{243}$Pu ($\tau_{1/2}$=4.95h) target nucleus although this isotope is still erroneously assigned to the fertile isotope family in the standard ENDF/B-VII.1 and JEFF-3.2 evaluated libraries. 
However in that study, we would not have been able to achieve such reasonable prediction without experimental fission probability data constrains. %induced from relevant $(d,p)$, $(t,p)$, ($^3He,t$) or even ($^3He,d$) direct reactions. 
This 'surrogate data' aspect was not documented in the earlier publication and is the first aim of present article. Subsequently we will realize that present input supplies another perspective for forthcoming  surrogate data extraction  by segregating the various terms of the Hauser-Feshbach equation and quantifying them in magnitude or possible error compensation for fission and capture neutron-induced cross section extrapolation. Present analyses cover the 4 to 8 MeV excitation energy range meaning the fluctuating domain below and above neutron emission threshold ($S_n$) that is the most impacted by the SRM  hypotheses. We imagine that below $S_n$ where only $\gamma$- and fission decays compete, there are few arguments to expect better agreement between neutron and surrogate physics spectroscopies in matter of fission cross section extrapolation than in terms of capture feedback. At the end of the day,  we would like to demonstrate that hereby representation of particle-transfer induced reaction data can indeed bring valuable complementary information in terms of cross section evaluation for neutron reactor applications.  

Across this paper, we do not cope with observed $\gamma$-decay probabilities because of the historical absence of such experimental data but new perspectives in that era have opened up very recently~\cite{duc:16}. Some partial feedback on this question is already unveiled over  preliminary analysis~\cite{bou:16b} of recent surrogate experiments~\cite{duc:16} collecting simultaneously $\gamma$- and fission-decay probabilities according to the $^{238}$U$(^3He,^4He)^{237}$U$^{*}$ reaction. The final data analysis is the topic of a separate publication~\cite{mar:19}.                   

This article is organized as follows. An overview of surrogate history and associated modeling strategy is first advertised. This is continued by the description of our original approach for surrogate reaction data that was made available using our in-house $\mathcal{AVXSF\mbox{-}LNG}$ ({\it AVerage CROSS Section}  {\it F}ission - {\it L}ynn and {\it N}ext {\it G}eneration)  program. This makes possible the investigation of the behavior of the manifold Hauser-Feshbach equation components whenever surrogate reactions are involved; namely spin-parity population fractions, in-out channel width fluctuation correction factors and reaction decay probabilities which calculation is based in this work on efficient Monte Carlo extended $\mathcal{R}$-matrix theory algorithm. Inaccuracies brought by the SRM in regards to the above  baseline in case of both fertile and fissile heavy target nuclides are emphasized. By contrast to a previous paper~\cite{bou:13} focussing on cross section observables, surrogate-like probability simulations might request modeling of possibly observed $\beta$-vibrational resonances. Dedicated handling of those structures in the fission strength function is then documented. Last section consists of the application of present approach to both fission-probability data and directly measured neutron fission cross sections according to Pu fissile isotopes; namely the $^{237, 238, 240, 242~and~244}$Pu$^*$ nuclei. Once consistency is obtained between fission probabilities and neutron fission cross section data simulations using an unique set of nuclear structure parameters, fission barrier heights can be assigned with much higher degree of confidence and even more so for fissile isotopes which fission threshold is not accessible by NPS techniques. This article concludes on performances of present approach off the historical SRM path and emphasizes new perspectives brought in matter of 'evaluation process' from foreseen fission- and $\gamma$-decay probabilities simultaneously measured as derived data will become available.

\section{\label{s:surrogateTh}Surrogate reactions as substitute to neutron physics spectroscopy}

As raised in the introduction, surrogate measurements are a substitute technique  to determine reaction cross sections for nuclei  that are difficult to measure directly by NPS or to predict with some degree of confidence from systematics or theory. The surrogate technique comes as an alternative to form the nucleus, $A^*$~\footnote{{To prevent confusion between the NPS target nucleus and the final nucleus prior to decay,  the '$\star$' notation is used in this article for the latter such that $^AElement^{ *}$.}} usually formed thru NPS as $n+(A\mbox{-}1)\rightarrow A^*$, that we want to measure decay properties . Alternatively another projectile-target combination, more accessible experimentally, can be chosen such that $projectile+(surrogate\mbox{ }target)\rightarrow A^* + ejectile$. %; reaction from which the decay of $C^*$ is observed in coincidence with the $ejectile$ that assigns unambiguously the reaction involved. 
By measuring the number of coincidences between the observable characterizing the exit channel ($c'$) pursued and the $ejectile$ %(i.e; the outgoing direct-reaction particle) 
that signs the nucleus to be analyzed, normalized to the total number~\footnote{Actually corrected for the experimental detector efficiency.} of surrogate events recorded, the experimental probability $\mathcal{P}^{A^*}_{surr,c'}(E_x)$ is derived. 
%$\mathcal{P}^{A^*}_{surr,c'}(E_x)$ is derived. 
%Combination (see later in the text) of $\mathcal{P}^{A^*}_{surr,c'}(E_x)$, corresponding to ${A^*}$ formation at excitation energy $E_x$ by given surrogate reaction  and subsequent decay in channel $c'$, with calculated neutron-induced ${CN\equiv A^*}$ formation cross section is then routinely carried out using the WE frame to retrieve the desired neutron reaction cross section.
%and compared, whenever possible, with neutron-incident cross sections that were  directly measured. 
%(labeled '$sp$' since the single-particle entrance channel signature is preserved here)
\subsection{\label{ss:Historical}Historical surrogate modeling}

The starting point and appropriate formalism for describing compound nucleus (CN) reactions is Hauser-Feshbach (HF) statistical theory~\cite{hau:52}, meaning the pure Hauser-Feshbach equation~\cite{bou:13} % (as stated by Eq.~1 of Ref.~\cite{bou:13}) 
that carries 'Niels~Bohr independence of formation and decay of given CN' approximation~\cite{boh:36}. More realistic picture of the interaction involves $W_{c,c'}$, the customary in-out-going channel width fluctuation correction factor~\footnote{$W_{c,c'}$ applied to neutron cross sections will be described as 'WFCF' along present paper.} (WFCF). The average partial cross section $\sigma_{cc'}$ formulation for an entrance channel $c$ and exit channel $c'$  applied to neutron-induced reactions at given neutron energy $E_n$ is then addressed as  
% Equation~1 of Ref.~\cite{bou:13} of the average partial cross section $\sigma_{cc'}$ for an entrance channel $c$ at given incident neutron energy $E_n$ and exit channel $c'$ (i.e; the pure Hauser-Feshbach equation that carries the ÔBohr hypothesisÕ of independence of the formation and decay of the CN in first order) associated with the so-called in-out-going channel width fluctuation correction factor (WFCF), $W_{cc'}$, applied to neutron-induced reactions becomes
\begin{eqnarray}
{\sigma_{n,c'}}(E_n)&=& \mathlarger{\sum_{J^\pi}}  \Biggl[ \sigma_{n}^{CN}(E_n,J,{\pi})  \nonumber \\
\times  \sum _{s'={|I'-i'|}}^ {{I'+i'}}&& \sum _{l'={|J-s'|}}^ {J+s'}  {\frac{T_{c'}^{J^{\pi_{(l's')}}} (E_{c'})}  {\sum_{c''} {T_{c''}^{J^{\pi_{(l"s")}}} (E_{c''})}}}\times W_{n,c'}^{J^\pi} \Biggr]\mbox{ , }\mbox{ }\mbox{ }\mbox{ }\mbox{ }
\label{eq:hfwccp}
\end{eqnarray}
where $\sigma_{n}^{CN}(E_n,J,{\pi})$ is the neutron-induced partial compound nucleus formation cross section related to   given ($J,\pi$) couple; the expression of which is, 
\begin{eqnarray}
\sigma_{n}^{CN}(E_n,J,{\pi}) = \pi \lambdabar^2 g_{J} \sum _{s={|I-\frac{1}{2}|}}^ {{I+\frac{1}{2}}} \sum _{l={|J-s|}}^ {J+s}  T_n^{J^{\pi_{(ls)}} }(E_n)\mbox{,}
\label{eq:NC}
\end{eqnarray}
with $g_{J}$, the statistical spin factor or weight according to total angular momentum $J$ as $g_{J}=(2J+1)/(2(2I+1))$ and  $T_n^{J^{\pi_{(ls)}}}$, the neutron entrance transmission coefficients.

SRM postulates that the WFCF can be neglected (equivalent to say $W_{n,c'}\approx 1$) although, by matter of fact, we know that this correction is substantial~\cite{hil:03} for opening channels up to  $1$~MeV above neutron emission threshold energy ($S_n$) as far as actinide neutron cross sections are concerned. Specialized to fission decay, the amount of correction depends on both the number of fission channels involved and the magnitude of their average widths. Larger the sub-barrier  effect is (case of fertile heavy isotopes), larger is the amount of fluctuations: $W_{n,f}\approx 35\%$ at 1 keV neutron-incident energy for fertile  isotopes to be juxtaposed with the 20$\%$ observed for fissile isotopes~\cite{bou:14}. Obviously $W_{n,f}$ tends to unity as the number of open channels increases. In the present study, fluctuation calculations have been carried up to a maximum excitation energy of $2.1$~MeV above the neutron binding energy for conservative statistical treatment. The question of the actual WFCF behavior above and below neutron emission threshold for both fission and capture channels in surrogate context is among the items advertised in the present article.

The absence of WFCF is indeed the basic hypothesis behind the SRM. Equation~\ref{eq:hfwccp} switches back to pure Hauser-Feshbach formulation that can be written in a concise manner as, 
\begin{eqnarray}
{\sigma_{n,c'}}(E_n)=\mathlarger{\sum_{J^\pi}}  \Biggl[\sigma_{n}^{CN}(E_n,J,{\pi}) \times  \mathcal{B}_{c'}^{J^\pi} (E_{c'})\Biggr]\mbox{ ,}
\label{eq:hfstd1}
\end{eqnarray}
where $\mathcal{B}_{c'}^{J^\pi} (E_{c'})$, the CN partial decay probability into channel $c'$, is also commonly referred as branching ratio (BR) to channel $c'$ in surrogate literature. Equation~\ref{eq:hfstd1} can be factorized as
\begin{eqnarray}
{\sigma_{n,c'}}(E_n)=\sigma_{n}^{CN}(E_n)\mathlarger{\sum_{J^\pi}}  \Biggl[\frac{\sigma_{n}^{CN}(E_n,J,{\pi})}{\sigma_{n}^{CN}(E_n)} \times  \mathcal{B}_{c'}^{J^\pi} (E_{c'})\Biggr]\mbox{, }\mbox{ }
\label{eq:hfstd2}
\end{eqnarray}
to make provision for  $[\sigma_{n}^{CN}(E_n,J,{\pi})/\sigma_{n}^{CN}(E_n)]$, the fraction of CN formed in a $(J,\pi)$ state that would be described as $\mathcal{F}_{n}^{CN}(E_{n},J,{\pi})$ in surrogate literature. This suggests to unfold the experimental (coincidence) probability  accordingly such that
\begin{eqnarray}
\mathcal{P}_{surr,c'}^{A^*} (E_x)=  \sum_{J^\pi}\Biggl[  \mathcal{F}_{surr}^{A^*}(E_x,J,{\pi}) \times  \mathcal{B}_{c'}^{J^\pi} (E_{c'})\Biggr]\mbox{. }\mbox{ }
\label{eq:pex}
\end{eqnarray}
Straightforward connection between neutron-induced cross section and the surrogate probability, mostly measured as a function of a single variable~\footnote{If we except touchy probability measurements performed at several ejectile detection angles that can return additional information in terms of total angular momentum and parity quantum numbers.} meaning the excitation energy, is routinely made and defines the SR Method. Weisskopf and Ewing~\cite{wei:40} suggested that BR quantities could be independent of spin and parity consideration; meaning substituting $\mathcal{B}_{c'}^{J^\pi}\mbox{ by }\mathcal{B}_{c'}$. In the framework of independence between formation and decay processes %=WCFC=1
supplemented by the latter hypothesis, branching ratios are pulled out of the spin-parity sum in both Eqs.~\ref{eq:hfstd1} and~\ref{eq:pex}. Combination of resulting simplified equations leads to the equivalence
\begin{eqnarray}
{\sigma_{n,c'}^{SRM}}(E_n) \equiv \Biggl[\sum_{J^\pi} \sigma_{n}^{CN}(E_n,J,{\pi}) \Biggr]\times  \mathcal{P}_{surr,c'}^{A^*} (E_{c'})\mbox{ ,}\label{eq:hfstdsp}\\
\mbox{since  }\sum_{J^\pi} \mathcal{F}^{A^*}_{surr}(E_{x},J,{\pi})=\mathds{1}\label{eq:unit}.
\end{eqnarray}
%\\

%~\footnote{Literature commonly describes as 'WE limit of Hauser-Feshbach theory', the threshold energy from which the WE frame should be valid.}

At first sight, this {\it surrogate strategy} supplies suitable estimate of the intended neutron-induced cross section without any need (Eq.~\ref{eq:pex}) to be able to 1) measure individual decay probabilities  and 2) predict the $\mathcal{F}^{A^*}_{surr}(E_x,J,{\pi})$ excited state population fractions. By using  Eq.~\ref{eq:hfstdsp}, assumption is made that the CN neutron-induced formation cross section, $\sum_{J^\pi} \sigma_{n}^{CN}(E_n,J,{\pi})$ can be ideally modeled using suitable optical model potential.\\

We realize promptly that
\begin{itemize}
\item the absence of WFCF at least hides issues related to undeniable partial outgoing channel widths correlations over fluctuating energy range. On the other side, using classic definition of WFCF can distort surrogate data to neutron cross section conversion since we expect conceptual WFCF differences between surrogate and NPS measurements, 
\item the straightforward use of the unitarity property (Eq.~~\ref{eq:unit}) washes out $(J,\pi)$ population fraction disparities expected   between neutron-induced and surrogate reactions. It is equivalent to state that final results are not sensitive to the actual entrance or exit spin-parity probability distributions. This goes against last decade insights showing that transfer reactions populate excited state spins twice higher on average than those produced by neutron-induced reactions~\cite{bout:12} %Fig5.5 boutoux page 180
and  that $\gamma$- and fission-BR can be influenced by the angular momentum and parity of the decaying nucleus as enlightened in Ref.~\cite{esc:12}. 
%The latter statement is justified in the absence of fission by the competition with neutron evaporation carrying low relative orbital momenta and since transfer reactions are known to populate CN state average spins higher than the ones carried by neutron-induced reactions and the $\gamma$ decay width little influenced by the $(J,\pi)$ couple of the CN state,          
\end{itemize}
 %\textcolor{blue}{
 Above arguments suggest that any successful use of the Surrogate Reaction Method (Eq.~\ref{eq:hfstdsp}) relies more on a case-to-case situation than on a systematic rule although we expect to fulfill the conditions at higher excitation energy (reaching the so-called WE limit). \\

\subsection{Questionable SRM predictions\\ as a matter of fact} 
%WE prescription 
The lack of high confidence level on the use of the historical surrogate conversion technique counteracts unfortunately the conveniency of the method. We could have guessed that such straightforward conversion was carrying by itself the seeds against the use of the technique. Finally it would be very likely that the SRM will lead to the level of confidence commonly aimed on nowadays major actinide neutron-induced cross section evaluations (a few $\%$ uncertainty). On the opposite, surrogate experiments are undoubtedly of great help in the case of poorly known nuclides which half-lives range from minutes to hours. This is well exemplified by the case of the ($^{243}Pu+n$) system for which capture and fission cross section uncertainties integrated over a LWR reactor type neutron spectrum were respectively estimated to 275~\% and 118~\% using the EAF2007/UN data library~\cite{cab:10}. Lowering huge uncertainties is clearly doable by complementing the cross section evaluation process with surrogate data. But beyond that statement, transfer reaction data analysis of fissile target isotope supplies invaluable estimation on barrier heights that lie below neutron separation energy. 
%This is what we claimed without large documentation support in the first paper~\cite{bou:13} describing our $R$-Matrix Monte Carlo analysis of the neutron-induced resonance cross sections for a range of Pu isotopes but consistently performed with transfer reaction data using our $\mathcal{LNG}$ ({\it AVerage CROSS Section}  {\it F}ission)  programme~\cite{lyn:74a}. 
Figure~16 of paper~\cite{bou:13}, that plots the fission cross section of the ($^{243}Pu+n$) system, well testifies the extreme impact of substantial change in $^{244}Pu^*$ barrier heights by bringing that nucleus from the fertile to the fissile family. Present $R$-Matrix Monte Carlo simulation does not carry most of the approximations involved routinely by the surrogate strategy, neither the simplifications adopted in the calculation of the two-peaked fission barrier resonant penetration by Younes and Britt~\cite{you:03a,you:03b}. We expect therefore that present step-by-step surrogate demarch will bring convincing arguments in favor of indisputable inclusion of surrogate data in standard evaluation task for a step further towards low uncertainty evaluated neutron-induced reaction cross sections. Present work intends to solve most of the dilemma raised by recent surrogate analyses~\cite{rom:12,duc:16} that suggest we could work with confidence from experimental fission probabilities but not from $\gamma$-decay probabilities.

\section{\label{s:Dedicated}Dedicated $\mathcal{LNG}$ approach to surrogate reaction data analyses}

%Last section, we have traced back the theoretical path extensively used over five decades for surrogate reaction analyses from which, we can address corresponding master equation applied to fission decay. This reads
From above theoretical background, the surrogate-like probability dedicated to fission decay can be addressed :
\begin{eqnarray}
\mathcal{P}_{surr,f}^{A^*} (E_{x})=\sum_{J^\pi} \mathcal{F}_{surr}^{A^*} (E_{x},J,{\pi}) \times   \mathcal{B}_{f}^{J^\pi} (E_{x})\times W_{surr,f}^{J^\pi}\mbox{ ,}
\label{eq:hfstdspA}
\end{eqnarray}
where $W_{surr,f}$ is the Surrogate-dedicated WFCF factor that corrects for partial channel width fluctuations correlations across flux conservation in surrogate context. For better display, the excitation energy $(E_{x})$ dependence of the SWFCF has been dropped from Eq.~\ref{eq:hfstdspA}; we will embrace this notation throughout this paper. We must stress that the amount of correlation  between formation and decay processes is related to the nature of reaction mechanisms leading to the PDEN. The question of the memory preservation plays a major role in surrogate measurements because of the necessary high energy of the incident charged particle to overcome the coulomb barrier (with $24$~MeV $^3He$ beam in recent $^{238}U(^3He,^4He)^{237}U^*$ measurement~\cite{duc:16}); the interaction time being not long enough to wash out any prior history making of fragile support the use of the CN hypothesis. \\   

We realize that accurate simulation of experimental fission probabilities requires the best knowledge of the three quantities involved in right hand-side of Eq.~\ref{eq:hfstdspA}. The description of the $\mathcal{LNG}$ approach to deal with those quantities is the aim of this chapter.     

\subsection{\label{ss:Compound}Excited nucleus spin-parity population as function of entrance vector for $\mathcal{LNG}$ calculations} 

\begin{figure}[t]
\center{\vspace{1.cm}
\resizebox{0.95\columnwidth}{!}{
\includegraphics[height=5cm,angle=0]{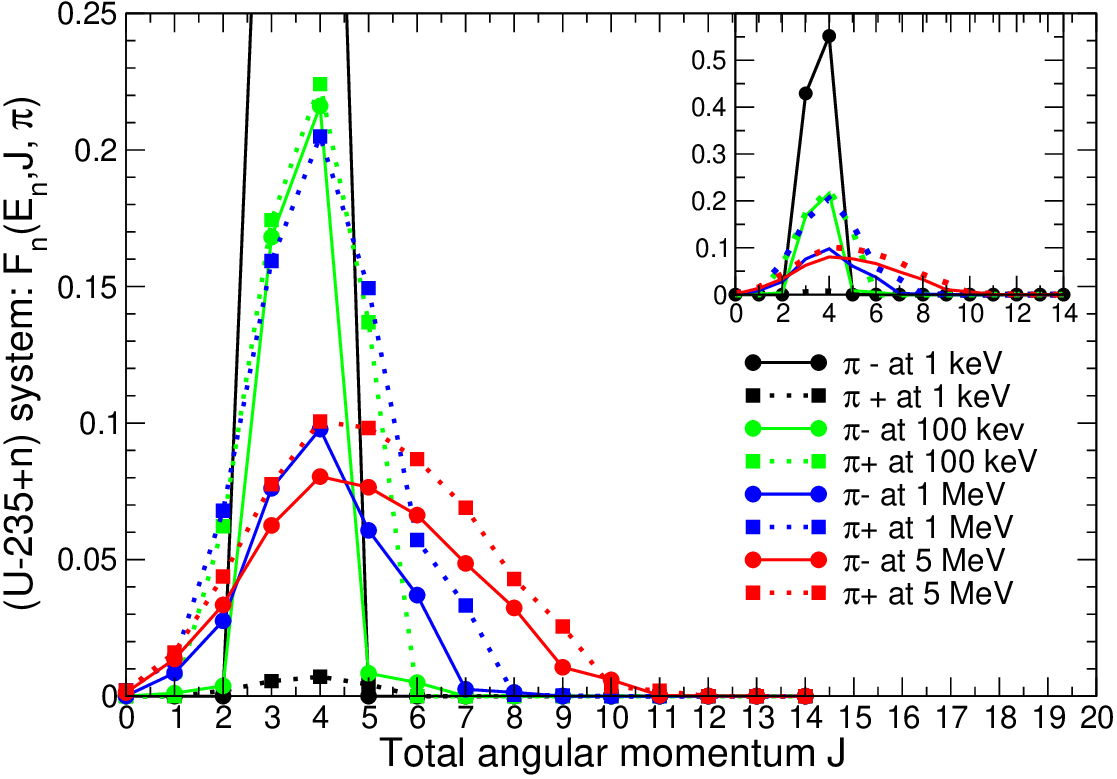}}}
\caption{(Color online) Distributions of total angular momenta associated to the neutron-induced reaction ($n+^{235}U$) resulting from $\mathcal{LNG}$ calculations for selected neutron energies. The foreground plot focuses on  fast neutron incident energies whereas the inset graphic shows the full picture according to reduced scale. Solid lines, connecting dots and addressing negative $\pi$ (respectively dashed lines with squares for positive $\pi$), are drawn to guide the eye.}
\label{fig:U236*JpiDist}
\end{figure}

By reference to Eq.~\ref{eq:hfstdspA}, we begin our discussion by focusing on first ingredient, $\mathcal{F}_{surr}^{A^*}(E_x,J,{\pi})$, which level of knowledge remains unsatisfactory in terms of surrogate entrance vector behavior. For neutron-induced reactions, the CN formation cross section is easily calculated from an appropriate optical-model potential or using the $\mathcal{LNG}$ approach (Eq.~\ref{eq:NC}). As described in previous paper~\cite{bou:13}, the latter involves the computing of elastic neutron channel transmission coefficients using the general form established by Moldauer~\cite{mol:61},
\begin{eqnarray}
T_n^{J^{\pi_{(ls)}}}=1-exp\left(-2\pi S_l  \right),\label{tnn}
\end{eqnarray}
where $S_l$ is the energy dependent neutron strength function for given relative orbital momentum $l$. Literature on heavy nuclides supplies accurate values of $S_l$ only for $s$- and $p$-wave elastic channels extracted from resolved resonance region analyses and average cross section fits below $300$ keV. For present plutonium isotope family calculations, we have simply assumed $S_l=1.044\times10^{-4}$ for even-$l$ waves and $S_l=1.48\times10^{-4}$ for odd-$l$ waves according to the rule of thumb that implies similar strength function values for even l-waves  (resp. for odd l-waves). The even and odd values assumed in this work are mostly within the uncertainties addressed in associated literature ($\pm0.1\times10^{-4}$ and $\pm0.4\times10^{-4}$ at best respectively for $s$- and $p$-waves). 
%On this ground, we obtain easily the partial CN formation cross section, $\sigma_{n}^{CN}(E_n,J,{\pi})$, that normalized to the total CN formation cross section, $\sum_{J^\pi} \sigma_{n}^{CN}(E_n,J,{\pi})$, supplies the CN spin-parity population distribution according to incoming   neutrons. This reads,
%\begin{eqnarray}
%\mathcal{F}_{c}^{CN}(E_c,J,{\pi}) = \frac{\sigma_{c}^{CN}(E_c,J,{\pi})}{\sum_{J^\pi} \sigma_{c}^{CN}(E_c,J,{\pi})}\mbox{ with }c%\equiv n \mbox{ . }
%\label{eq:Fc}
%\end{eqnarray}
We observe that  definition of $\mathcal{F}_{n}^{CN}(E_n,J,{\pi})$ is analogue to mothball both energy and level parity dependence of the neutron entrance transmission coefficients in the statistical spin factor (Eq.~\ref{eq:NC}) such that $g_{J}\rightarrow$g$_{J,\pi}(E_n)$. For benchmarking $\mathcal{LNG}$ entrance route based on the neutron transmission coefficients by Moldauer against the results obtained more directly from appropriate optical model potential (as mentioned by Escher \etal.~\cite{esc:12}), we plot on Fig.~\ref{fig:U236*JpiDist} the distributions of total angular momenta corresponding to the neutron-induced reaction $\bigr(n+^{235}U\bigl)$ calculated with the $\mathcal{LNG}$ code for selected neutron energies. The foreground plot, corresponding to  neutron incident energies $\geq100~keV$, is quite close to Fig.20 of Ref.~\cite{esc:12}. As expected, we observe that  $J^\pi$ CN states other than $3^-$ and $4^-$ ($s$-waves) are populated only significantly at high neutron energies.\\

%In matter of SRM, the neutron-induced cross section is reconstructed on the ground of Eq.~\ref{eq:hfstdsp} which, involves $\mathcal{P}_{surr,c'}^{A^*}(E_{x})$,  the measured  total decay probability across channel $c'$ and $\sum_{J^\pi} \sigma_{c}^{CN}(E_c,J,{\pi})$, the 
%sum of the various entrance spin-parity CN formation cross sections. However we know that a necessary condition for the validity of the WE frame relies on an idealized matching between neutron-induced and surrogate spin-parity entrance distributions; reading $\mathcal{F}_{surr}^{CN}(E_{x},J,{\pi}) \equiv \mathcal{F}_{n}^{CN}(E_n,J,{\pi})$. 

\begin{figure}[t]
\center{\vspace{0.cm}
\resizebox{0.95\columnwidth}{!}{
\includegraphics[height=5cm,angle=0]{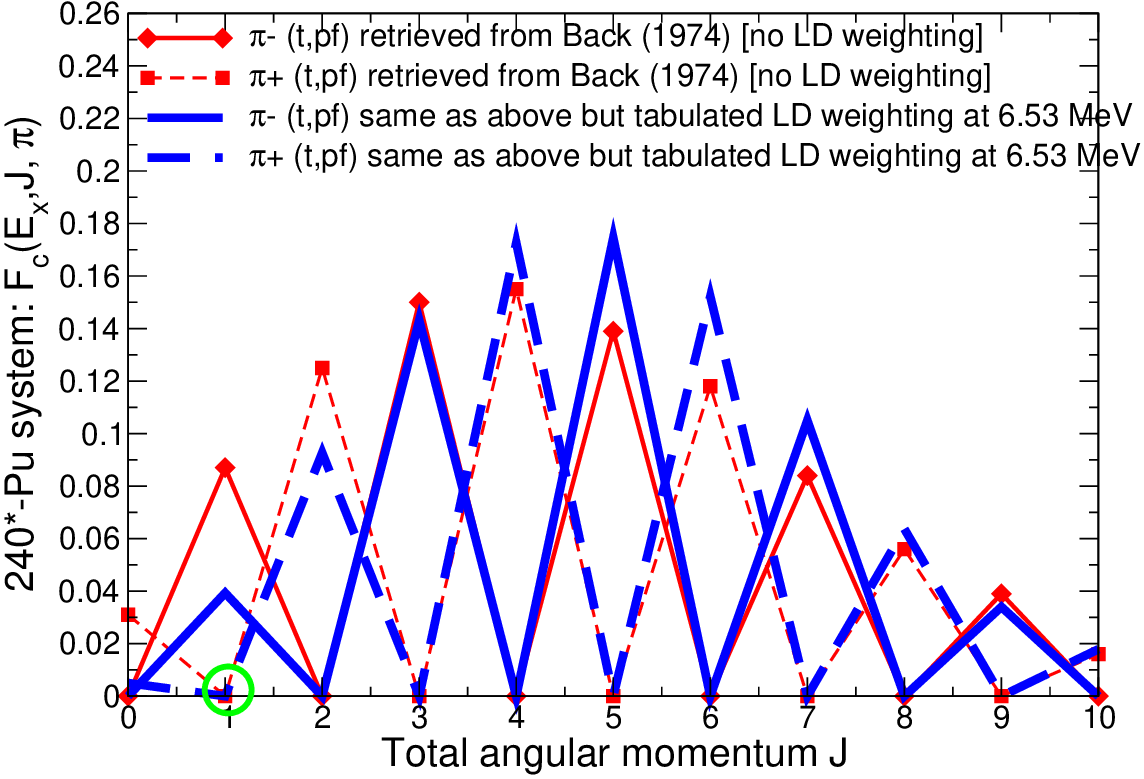}}}
\caption{(Color online) Peculiar behavior of total angular momentum population distribution according to the $^{240}Pu^*$ as formed by direct $(t,p)$ reactions. The above data are retrieved using Eq.~\ref{eq:jpiCN} from early DWBA theoretical calculations as quoted by Back \etal~~\cite{bac:74}. Lines connecting dots and addressing negative $\pi$ (respectively dashed lines with squares for positive $\pi$) are drawn to guide the eye. Thin lines correspond to using Eq.~\ref{eq:jpiCN} with uniform Level Density (LD) whereas thick ones address energy-dependent population distribution according to LD calculations described in Ref.~\cite{bou:13}.}
\label{fig:Pu240*JpiDist_tp}
\end{figure}

\begin{figure}[t]
\center{\vspace{0.5cm}
\resizebox{0.95\columnwidth}{!}{
\includegraphics[height=5cm,angle=0]{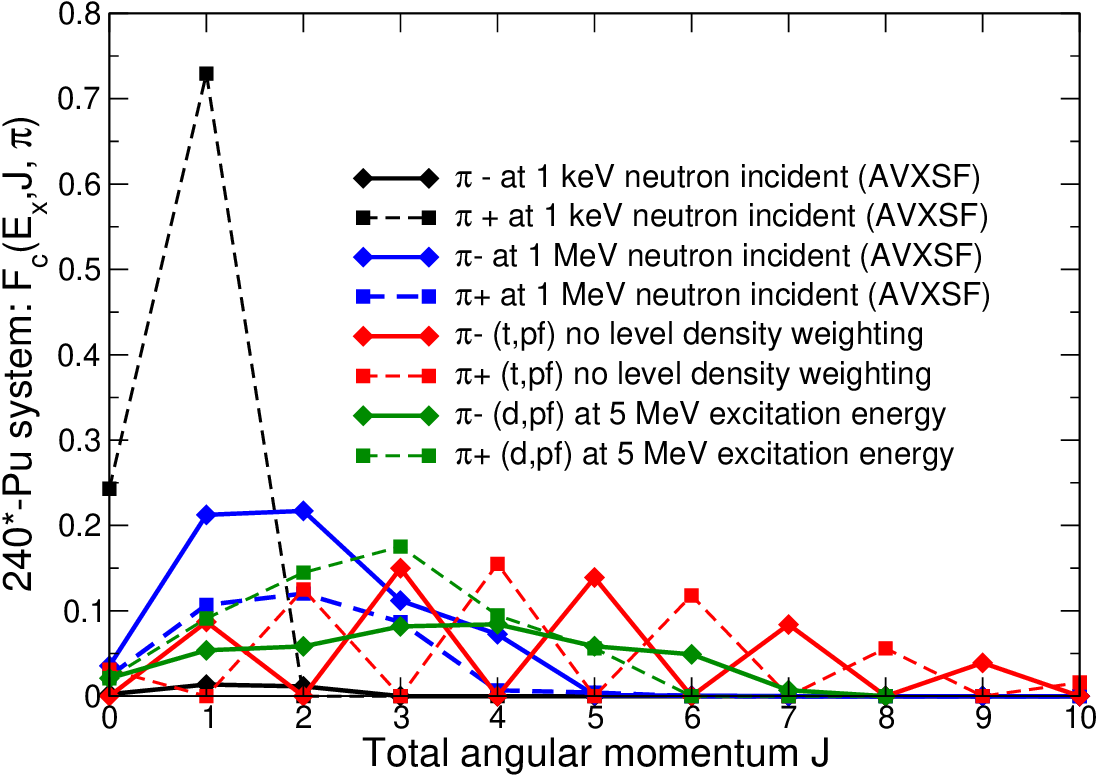}}}
\caption{(Color online) Distributions of total angular momentum population corresponding to $^{240}$Pu$^*$ formed either by neutron-induced reaction ($n+^{239}$Pu) at both $1$~keV and $1$~MeV or pictured from DWBA calculations early quoted by Back \etal~~\cite{bac:74} ($(t,p)$ reaction) or Andersen \etal~\cite{and:70} ($(d,p)$ reaction).  Solid lines connecting dots and addressing negative $\pi$ (respectively dashed lines with squares for positive $\pi$) are drawn to guide the eye.}
\label{fig:Pu240*JpiDist_all}
\end{figure}

Last decade literature pointed out that a necessary condition for the validity of the SRM relies on an idealized matching between neutron-induced and surrogate spin-parity entrance distributions; reading $\mathcal{F}_{surr}^{A^*}(E_{x},J,{\pi}) \equiv \mathcal{F}_{n}^{CN}(E_n,J,{\pi})$. Indeed this equivalence has been questioned since the pioneer times~\cite{cra:70b}. The answer was  inferred from the early finite range interaction distorted-wave Born approximation (DWBA)  $(t,p)$, $(d,p)$ and ($^3He,d$) cross section calculations reported by Back \etal~(cf. Fig.7 of Ref.~\cite{bac:74}) although those calculations carry large uncertainties and should be refined using coupled-channel optical model formalism for instance. {\it Regarding present paper, we simply stick} with early direct calculations made by  Andersen \etal~\cite{and:70} for the particular case of the $^{239}$Pu$(d,pf)$ one-particle stripping reaction and with results quoted by Back \etal~\cite{bac:74} for the remaining $(d,p)$, ($^3He,d$) and  $(t,p)$ surrogate reactions. We have used for calculating Eq.~\ref{eq:hfstdspA} the $J^\pi$  entrance fractions as a function of excitation energy supplied by Andersen \etal~\cite{and:70} in graphic form~\footnote{With the distance between neighboring curves representing the spin-parity probability of the upper curve.} and related to PDEN  states in $^{240}$Pu$^*$.
% the one-neutron state transferred into a shell-model orbit of the target nucleus (i.e; assumed to form a CN. 
On the opposite, Ref.~\cite{bac:74} returns preferentially the distribution of orbital angular momentum according to the transfer of a particle (separated out of the light incident projectile) to given single-particle shell of the target nucleus. The conversion of the latter to spin-parity distribution has been made in this work following the $j-j$ spin-orbit vectorial coupling scheme. In case of $(t,p)$ reactions on even-even target isotopes owning zero intrinsic spin and positive parity, obtention of the weighting factors, $\mathcal{F}_{surr}^{A^*}(E_{x},J,{\pi})$, were assessed from stripping theory where we assume that the released proton is scattered from the short-range part of the proton-target nucleus interaction as if the di-neutron (constituting with the proton the incident triton) was not present. On this assumption, we can consider a two-body problem with a  di-neutron trapped into a shell with single-particle-type character of the target nucleus. %three-body problem as  p + di-neutron + target
The PDEN is set up from the bound residual nucleus meaning the target nucleus (noted $0$) complemented by the  di-neutron cluster (noted $12$). Applying $j-j$ coupling scheme to this system leads to 
\begin {equation}
    \overrightarrow J_{PDEN}= \overrightarrow I_{0} +  ( \overrightarrow i_{12} + \overrightarrow l_{12}) \mbox{ ,}\\
\label{eq:jj12}
\end {equation}
with $\overrightarrow i_{12}$  the intrinsic spin of the di-neutron cluster. The PDEN state parity, $\pi_{PDEN}$, for $l_{12}$ given relative orbital angular momentum of the di-neutron and the target nucleus is ruled accordingly
\begin {equation}
    \pi_{PDEN} = (-1)^{l_{12}}*\pi_{0}*\pi_{12}  \mbox{ ,}
\label{eq:pi12}
\end {equation}
with $\pi_{12}$ and $\pi_{0}$, the respective parities of the di-neutron cluster and even-even target nucleus (meaning $\pi_{0}=0^+$). Assuming also anti-symmetrical intrinsic spin character ($i_{12}=0$) and positive parity for the di-neutron cluster, the exact equivalence $J_{PDEN} \equiv l_{12}$ is verified. 
%between the CN total angular momentum and the relative orbital angular momentum of the di-neutron transferred into a shell with single-particle-type character of the target nucleus ($J_{residual} \equiv l_{12}$). 
Since the parity of the excited state is also driven by the even or odd character of $l_{12}$, even $J$ are built solely with positive parity and odd $J$ with negative parity. In case of symmetrical intrinsic spin ($i_{12}=1$), the conversion will lead to additional possibilities as   $\mid l_{12}-1\mid \leq J_{PDEN} \ \leq (l_{12}+1)$ but the lowest single-particle state, $^3P_0$ in spectroscopic notation, is expected to occupy a higher shell and is thus less favored. Figure~\ref{fig:Pu240*JpiDist_tp} illustrates that peculiar feature for triton-induced direct reactions on even-even ($0^+$) target nucleus in correspondence with {\it natural-parity}~\footnote{Meaning positive-parity states of even angular momenta and negative-parity states of odd angular momenta.} state configuration ($i_{12}^\pi=0^+$).     
  Expending $(t,p)$ reactions context to non-even-even target nucleus, the spin-parity state population distribution  expected in a PDEN from given angular momentum transfer cross section, $\sigma(l)$, is  
\begin {eqnarray}
 \mathcal{F}_{surr}^{A^*}(E_{x},J,{\pi}) &=& \rho(E_x,J,\pi) \times \nonumber \\
 \mathlarger{\sum _{j={|J-I|}}^ {{J+I}} \sum _{l ={|j-i|}}^ {j+i} } && \frac{P_l \times \delta (\pi_J, \pi_i \pi_{I} (-1)^l)}{\mathlarger{\sum _{j={|J-I|}}^ {{J+I}} \sum _{j={|j-i|}}^ {{j+i}} 1}}  \label{eq:jpiCN}\\
\mbox{ with }  P_l &=&   \frac{(2l+1)\sigma_l }{\mathlarger{\sum _{l} (2l+1)\sigma_l }}\mbox{ .} \nonumber
\end {eqnarray}
Figure~\ref{fig:Pu240*JpiDist_tp} obviously invalidates the hypothesis of comparable neutron and $(t,p)$ reaction spin-parity entrance distributions. Spin-parity distribution comparisons with less quirky signatures as involved in $(d,p)$ processes, deliver similar verdict as exemplified by the $^{239}$Pu$(d,p)^{240}$Pu$^*$ DWBA distribution calculated by  Andersen \etal~\cite{and:70} (shown hereby on Fig.~\ref{fig:Pu240*JpiDist_all}). Authors' positive parity distribution, closed to a truncated gaussian distribution centered about $J=3\hbar$ with dispersion of $1.5\hbar$, requires high-energy neutron-induced reactions. 

\subsection{\label{ss:Wspf}In-out-going channel width fluctuation correction factor modeling}

We know from Moldauer~\cite{mol:61} that the in-out-going channel width fluctuation correction factor, $W_{c,c'}$, plays in low energy neutron-induced reactions a major role in averaging over partial width distributions to assess average cross sections. This was well quantified in a model comparison by Hilaire \etal~\cite{hil:03}. In matter of surrogate reactions, the question of the correlation between entrance and exit channel widths is obviously sensitive because it depends on the direct entrance reaction type. For instance in a $(t,pf)$ reaction as considered later in this paper, two neutrons in paired orbits are stripped into the field of a target nucleus that has two neutrons less than the residual nucleus.  Its formation cross section is then proportional to the reduced width~\footnote{$\gamma^2$ is known as the reduced width whereas $\gamma$ is the so-called reduced width amplitude which is the value of the internal eigenfunction $X_\lambda$ at the entrance to channel $c$. Observed width, $\Gamma_c$, and reduced width are connected~\cite{lyn:68} thru centrifugal/Coulomb penetrabilities meaning $ \Gamma_{c}^{1/2} \equiv \gamma_{c}\sqrt{2P_c}.$}, $\gamma^2$, for a 'di-neutron' channel for which the reaction energy threshold close to twice the neutron separation energy. We guess that the correlation between the entrance 'di-neutron' reduced width and the outgoing partial reaction channel widths of the PDEN, restricted at low excitation energy to 1) single neutron emission  (above $S_n$), 2) $\gamma$ emission and 3) fission, will be much likely of third order magnitude and can be easily ignored. However other direct reaction 'vectors' that have been considered in this work as ($^3He,d$) and ($^3He,t$)  charge-exchange reactions where a proton is pulled into the field of a target nucleus (e.g; $^{240}Pu$), must be considered with more attention. The question of the correlation between the entrance and exit proton channel widths, $\Gamma_p$, is however simplified by the fact that above proton emission energy ($S_p=4.48~MeV$ in $^{241}Am$), the charged-particle penetrability becomes appreciable only when the exit proton (if any) energy tends to the Coulomb barrier value meaning $B \approx \bigl[1.44~Z_{p}Z_{240}\bigr]/\bigl[1.60[A_p^{1/3}+A_{240}^{1/3}]\bigr] \approx$ 11.7 MeV and so, its proton emission width. In matter of stripping $(d,p)$ reactions, the question of the correlation between formation and decay widths is more tricky. Later reaction can be illustrated as a deuteron sweeping past the target nucleus with its proton repelled by the Coulomb field and the neutron coming into close enough proximity to the target to be pulled into one of the (bound) single-particle orbit in the nuclear field. If the excitation energy of the single-particle neutron is above neutron emission energy, the corresponding elastic neutron emission width is non-zero and there might be interference between both partial widths. Theory related to width fluctuation effects for that type of surrogate reactions has been developed by Kerman and Mc Voy~\cite{ker:79} but is not raised in present work because of the following argumentation:  for fissile isotopes, particle-transfer-induced fission data are studied to infer fission barrier height values lying below the neutron separation energy. In addition, right above $S_n$ the neutron width remains small compared to the total width; statement that is still reinforced since surrogates populate high total angular momentum levels. In consequence, the assumption that the formation width does not participate in the width fluctuation correction sounds quite reasonable also for $(d,p)$ reactions. {\it Present SWFCF calculations have been carried without considering any relationship  between formation channel 
%(in practice the direct-reaction ultimate nucleus population fractions) 
and related decay channel} such we can state
\begin{eqnarray}
{\left\langle\frac{\mathcal{F}_{surr}^{A^*} \Gamma_{c'}} {\Gamma_{tot}} \right\rangle} &\equiv&  \mathcal{F}_{surr}^{A^*}{\left\langle\frac{ \Gamma_{c'}} {  \Gamma_{tot}} \right\rangle} = \mathcal{F}_{surr}^{A^*} {\left\langle\frac{ \Gamma_{c'}} {\sum_{j}  \Gamma_{j} + \Gamma_{cst}} \right\rangle}\nonumber\\
&=&\mathcal{F}_{surr}^{A^*} {\frac{ \left\langle{\Gamma}_{c'}  \right\rangle} {\left\langle\sum_{j}  { {\Gamma_{j}}   + \Gamma_{cst}   } \right\rangle }}\times W_{surr,c'}\mbox{ ,}
\label{eq:RSRconcept}
\end{eqnarray}  
%\left\langle{\frac{\Gamma_{c}^{J^{\pi_{(ls)}} }\Gamma_{c'}^{J^{\pi_{(l's')}} } } {\sum_{c''} {\Gamma_{c''}^{J^{\pi_{(l''s'')}} } }    } }\right\rangle={\frac{ \left\langle{\Gamma}_{c}^{J^{\pi_{(ls)}} } \right\rangle \left\langle{\Gamma}_{c'}^{J^{\pi_{(l's')}} }  \right\rangle} {\sum_{c''} { \left\langle{{\Gamma}_{c''}^{J^{\pi_{(l''s'')}} }}  \right\rangle }    } }\times W_{cc'}\mbox{ .}
where $\Gamma_{cst}$ represents a global non-fluctuating lumped channel merging various constant (cst)  components meaning  radiative decay, delayed fission in second well and fission over outer barrier  continuum transition states. Very small channel width values corresponding to the highest orbital angular momenta are also mothballed in $\Gamma_{cst}$ in order to speed up  the computation  of fluctuation factors  as well as weakly fluctuating channel contributions; according to width distribution Degrees of Freedom (DoF) larger than $2.0$ units. On above statement, the entrance surrogate channel width would be identified to the dissimilar entrance single-particle channel width. $\sum_{j}\Gamma_{j}$ regroups the  most fluctuating decay channel widths (i.e; largest particle-emission and fission channels). Numerical evaluation of SWFCF in the framework of Hauser-Feshbach statistical theory is carried out analogously to the general single variable integral established by Dresner~\cite{Dre:57} assuming that the  partial width of given channel, $c$, can be represented by a $\chi^2$ distribution with $\nu_c$ degrees of freedom. This results to 
\begin{eqnarray}
&&{\left\langle\frac{ \Gamma_{c'}} {\sum_{j}  \Gamma_{j} + \Gamma_{cst}} \right\rangle}_{\lambda_{I}}=\frac{\nu_{c'}}{2} \prod_{j} \left( \frac{\nu_{j}\Gamma_{cst}}{ 2\bar\Gamma_j } \right)^{\frac{\nu_{j}}{2}}   \times \mbox{  } \mbox{  } \mbox{  }  \nonumber \\ 
&& \int_{0}^{\infty} dt  \Biggl[e^{-t}  \biggl[  t+  \frac{\nu_{c'}\Gamma_{cst}}{ 2\bar\Gamma_{c'} } \biggr]^{-1}   \prod_{j}  \left(t+ \frac{\nu_{j}\Gamma_{cst}}{ 2\bar\Gamma_j }\right) ^{-{\frac{\nu_{j}}{2}}}        \Biggr] \mbox{  } \mbox{  }  \mbox{  } \mbox{  } 
 \label{eq:tfmu}
\end{eqnarray}
while that quantity  expressed for the lumped channels reduces to
\begin{eqnarray}
{\left\langle\frac{ \Gamma_{cst}} {\sum_{j}  \Gamma_{j} + \Gamma_{cst}} \right\rangle}= \prod_{j} \left( \frac{\nu_{j}\Gamma_{cst}}{ 2\bar\Gamma_j } \right)^{\frac{\nu_{j}}{2}} \times \nonumber \\
  \int_{0}^{\infty} dt  \frac{e^{-t} }  {    {\prod_{j}\left( t+ \frac{\nu_{j}\Gamma_{cst}}{ 2\bar\Gamma_j }\right) }     ^{-{\frac{\nu_{j}}{2}}}} \mbox{ . } \mbox{  }
  \label{eq:tfc}
\end{eqnarray}

%Among  the hypotheses carried by the SRM (Eq.~\ref{eq:hfstdsp}) that we want to emphasize in this paper, are indeed the in-out-going channel width fluctuation correction factor shape and magnitude according to transfer probability measurement simulation as suggested by equations above. 

The SWFCF is among the items we want to emphasize in this paper. Beyond its specific shape and magnitude behavior (illustrated later in the text), we recall that statistical treatment specialized to fission decay probability involves an additional width fluctuation correction factor brought by class-II state properties, so-called $W_{II}$~\cite{bou:14}, whose impact is usually disregarded in standard average cross section evaluation codes. For present illustration of $W_{c,c'}$ and $W_{II}$ effects, modified formulation of the analytical Eq.~\ref{eq:hfstdspA} is envisioned although real situation commands avoiding decoupling fine structure width fluctuations from class-II state width fluctuations brought by double fission barrier treatment. More reliable calculations based on Monte Carlo simulations are favored in this work for valuable partial cross sections or surrogate-like probabilities modeling.\\      

Usual WFCF quantities needed in this work to calculate NPS average partial cross sections, were implemented on the ground of equations 17 and 16 of Ref.~\cite{lyn:70} that correspond respectively to treatment of the elastic ($c'=c$) and non-elastic partial channels ($c'\neq c$). By its formulation analogous to latter non-elastic partial channel width correlation treatment, Eq.~\ref{eq:tfmu} sets weak relationship between the entrance surrogate channel width and any exit reaction channel.

%We then realize that those two equations are subsets of Eq.~\ref{eq:tfmu} in which the question of the existence of relationship between the entrance width and any exit channel width has been estimated meaningless in first order. 

\subsection{\label{ss:MC}In-house fission decay probability calculation using the Monte Carlo route}
     
A major difficulty in fission decay probability calculations lies in poor model representation of the fission channel. Consistently to common state-of-art evaluation, this study uses the well-known Hill and Wheeler~\cite{hil:53} transmission coefficient formula that is based on two common approximations: a unique one-dimensional fission path and a representation of each single-humped fission barrier as inverted parabola. The approximation of a unique one-dimensional fission path appears well justified in case of the plutonium isotopes we study here, as predicted by static Finite Range Droplet Model (FRDM) calculations of M\"oller {\it et al.}~\cite{Mol:16}.  The asymmetric mode remains the main contribution to fission until past the outer saddle point from which the symmetric path becomes relevant. And even when these two modes co-exist, they remain distinct because of the existence of a significant separating ridge (at least 1 MeV above the upper valley). We can nevertheless argue that triaxiality is observed at the inner barrier (see Table XI of Ref.~\cite{cap:09}) and must be taken into account somehow. This is achieved  in our calculations by modulating, for instance, the circular frequency associated to the $\gamma$-axis primary phonon vibration excitation.  Whenever the axial symmetry is recovered (at the outer barrier for instance), the softness towards this axis is released by supplying a high phonon quantum value. Our original approach in matter of A.~Bohr transition states above fundamental fission humps has been  described in Ref.~\cite{bou:13}. It relies on ad hoc sequences of individual transition states above fundamental barriers at low excitation energies built consistently with our combinatorial Quasi-Particle-Vibrational-Rotational (QPVR)  calculations  that are performed to construct Level Densities (LD) on top of the individual transition state sequence. Over the upper energy range, detailed resonance structure is of much less importance. The fission cross-section mainly depends on both level densities of the compound nucleus at barrier deformations and the level density of target nucleus at normal deformation, which controls the competing inelastic neutron scattering reaction. Therefore, special attention was  paid to modeling level density functions and interpreting fits to them where these are required for matching neutron-induced cross-section data. Regarding the validity of the inverted parabola approximation, we have shown in our previous paper~\cite{bou:13} (Fig.~7) that this latter assumption clearly appears to be well justified only for the heaviest isotopes of plutonium (above mass 241). %Additional work~\cite{tam:15} is on-going for studying the influence of replacing one-dimensional double-humped parabolic fission barriers by realistic one-dimensional barrier shapes obtained from projection of potential energy surface fission paths; the latter being built directly from nuclear structure calculations such as FRDM~\cite{Mol:16}. 
We underline that present work is restrained to inverted parabola fission barrier calculations.\\ 

We can then argue that our fission decay probability calculations carry the most accurate and physical approach available routinely for low excitation energy (i.e; lower than second-chance fission) fission cross section calculations. % as argumented in previous paper~\cite{bou:13}. 
In this sense, this is a smart complement to the work made by Younes and Britt~\cite{you:03a,you:03b} on the inference of neutron-induced fission cross sections by fission-probability data regarding to the Pu isotope family. In particular our theory is based on an extension of $\mathcal{R}$-matrix theory to the fission deformation variable as outlined by Bjornholm and Lynn~\cite{bjo:80}. Since this theory has been exhaustively described in \cite{bou:13}, we will not get into in present paper.
%only communicate on its main bricks for the understanding of present article.    
\subsubsection{Typical analytical coupling formulae modeling double-humped potential}

\paragraph{\label{stat}Statistical Regime -}

For excitation energies above the fission barrier, the fission transmission coefficient is calculated assuming statistical regime. In the framework of a double-humped fission barrier whose two humps are uncorrelated and described by a (single-humped) Hill-Wheeler approach~\cite{hil:53}, standard probability treatment (Ref.~\cite{bjo:80}, page 752) supplies the average probability~\footnote{Probability or branching ratio regarding present surrogate topic.} of $d$irect fission for a compound nucleus formed in first well across the whole barrier (i.e; $\mathcal{B}_F\approx T_{f_d}/\bigl[T_{f_d} + T_{I}\bigr]$ with $T_{I}$ covering all decay types other than fission within first well (I)). In terms of fission transmission coefficient related to given transition channel $\mu$, this reads 
\begin{eqnarray}
T_{f}(\mu)\equiv T_{f_d}(\mu)=\frac{T_I T_A T_B(\mu)}{T_A T_{II} + T_I (T_A+T_B+T_{II})}\mbox{ ,}\\\nonumber
\label{eq:statRegimeFullExpression}
\end{eqnarray}that reduces to the well-known statistical transmission coefficient for fission if absorption ($T_{II,\gamma}$) and particle emission ($T_{II,x}$) are negligible in the second well (II). The hypothesis $T_{II}= \bigl[T_{II,\gamma}+T_{II,x}\bigr]\rightarrow 0$ is equivalent to consider here that the fraction of the wave is transmitted right across the secondary well without absorption. It leads to   
\begin{eqnarray}
T_{f_d}(\mu)\approx\frac{T_A T_B(\mu)}{T_A+T_B}\mbox{ ,}
\label{eq:statRegime}
\end{eqnarray}
where $T_{A}$ and $T_{B}$ are the  total fission transmission coefficients over barriers $A$ and $B$ calculated from the well-known Hill-Wheeler expression~\cite{hil:53}.

\paragraph{\label{stat}Sub-barrier excitation energies-}
At sub-barrier and near-barrier excitation energies, the detailed structure of class-II levels %(i.e; the nucleus states which wave functions lie predominantly in the second minimum of the potential fission barrier) 
has significant impact on  $T_f$. Since the bulk of the strength of $T_f$ is concentrated in a narrow energy interval about a class-II level, the actual average fission probability magnitude will be rather recovered by an average over the large energy intervals separating the class-II states. The consequence is a noticeable reduction of the average fission probability resulting from the use of Eq.~\ref{eq:statRegime}. This is equivalent to consider the fission transmission coefficient as a sum of the $direct$ term with an $indirect$ term; the latter manifesting the class-II nucleus structure. On the assumption of uniform class-I and class-II level spacings (so-called 'picket fence model'), Lynn and Back~\cite{lyn:74b} have worked out a formulation for $\mathcal{B}_{f,IS}$, the average fission probability  including intermediate structure ($IS$) that, in the limit of $complete$ $damping$ in second well (i.e; no direct fission, only indirect fission after inner barrier tunneling) reduces to
%\vspace{-0.05cm}
\begin{eqnarray}
\mathcal{B}_{f,IS}= \left[ 1+ \left(\frac{    T_{I} }  {T_{f}}\right)^2  +  \left(\frac{ 2T_{I} }  {T_{f}} \right) coth \left(\frac{    T_A+T_B }  {2}\right)  \right] ^{-1/2}\mbox{ .} \mbox{ }%\mbox{ }
\label{eq:LB} 
\end{eqnarray}
We underline that a variety of analytical formulae valid only under specific class-I/II coupling situations are developed in literature~\cite{bjo:80}. Whenever the fission model introduces the class-II states nucleus structure, %(i.e; when partial or full damping in the secondary well occurs)
a correlation factor between the class-II ($\lambda_{II}$) coupling, $\Gamma_{\lambda_{II(\downarrow)}}$, and fission, $\Gamma_{\lambda_{II(\uparrow)}}$, widths is expected. Using again the general form established by Moldauer~\cite{mol:61}, we can define $T_A=2\pi\bra\Gamma_{\lambda_{II(\downarrow)}}\ket_{_{II}}/{ D_{II}}$ and $T_B=2\pi\bra\Gamma_{\lambda_{II(\uparrow)}}\ket_{_{II}}/D_{II}$ with $\Gamma_{\lambda_{II,tot}}\approx \Gamma_{\lambda_{II(\downarrow)}}+\Gamma_{\lambda_{II(\uparrow)}}$ and $D_{II}$, the associated mean resonance spacing. % associated to the class-II states.     
By analogy to WFCF treatment, we address $W_{II}(\mu)$ as the 'class-II state WFCF' for given outer Bohr fission channel ($\mu$),\\  
\begin{equation}
{\left\langle \frac{{\Gamma_{\lambda_{II}\downarrow} \Gamma_{\lambda_{II}\uparrow}}(\mu) } {\Gamma_{\lambda_{II}}} \right\rangle}_{\lambda_{II}} =
 W_{II}(\mu) \frac{{ {\left\langle\Gamma_{\lambda_{II}\downarrow} \right\rangle}{\langle\Gamma_{\lambda_{II}\uparrow}(\mu) \rangle} }}{{{\langle \Gamma_{\lambda_{II}}\rangle}} }  \mbox{ .}
\label{eq:Rconcept}
\end{equation} 
%\vspace{-0.5cm}
In view of Eq.~\ref{eq:Rconcept}, the Eq.~\ref{eq:hfwccp} becomes more complicated in matter of fission such that
\begin{eqnarray}
 \sigma_{nf}(E_n) &=& \mathlarger{\sum_{J^\pi}} \Biggl[ \sigma_{n}^{CN}(E_n,J,{\pi}) \times \nonumber \\
&& \Big[ \mathsmaller{\sum_{\mu\in J^\pi}}  \mathcal{B}_f(E_{f},\mu)  W_{II}(\mu) \Bigr] W_{n,f}^{J^\pi} \Biggr]  \mbox{ , } \mbox{  }  
\label{eq:FXSdecoupled} 
\end{eqnarray}
or equivalently in terms of surrogate-like decay probability
\begin{eqnarray}
&&\mathcal{P}_{surr,f}^{A^*} (E_{x})=\sum_{J^\pi} \Biggl[  \mathcal{F}_{surr}^{A^*}(E_{x},J,{\pi}) \times \nonumber \\
 && \Big[ \mathsmaller{\sum_{\mu\in J^\pi}}  \mathcal{B}_f(E_{x},\mu)  W_{II}(\mu) \Bigr]\times W_{surr,f}^{J^\pi}\Biggr] \mbox{ ,} \mbox{ }\mbox{ }
\label{eq:Pdecoupled}
\end{eqnarray}
%where $\mathcal{B}_f(\mu)$ is calculated under statistical (Eq.~\ref{eq:statRegime}) or intermediate structure (Eq.~\ref{eq:LB}) regime. Because of their analogy, estimations of both class-II and class-I widths fluctuating factors (similarly to $W_{surr,c'}$ of Eq.~\ref{eq:RSRconcept}) can be made by numerical integration of the general one dimension integral stated by Dresner~\cite{Dre:57}. 
Recent calculations~\cite{bou:14} %regarding  neutron-induced average fission cross sections of $^{239}$Pu  and $^{240}$Pu  
have enlightened that global amount of correction due to the $W_{II}$ (Eq.~\ref{eq:Rconcept}) is of same order of magnitude for  fertile ($20\%$) and  fissile ($10\%$) target nuclides although the sub-barrier effect, as expected, is in the fissile case much smaller. %(23$\%$ compared to 80$\%$ at 1 keV neutron-incident energy). 
This returns a $10\%$ to $20\%$ estimate on the error brought by the absence of $W_{II}$ (mainly) below {\it300}~keV incident neutron energy  in calculated average fission cross section by codes that rely only on the standard width fluctuation corrected Hauser-Feshbach formulation (Eq.~\ref{eq:hfwccp}). 

However still more trouble is expected because of the IS pattern in average cross section calculation. Classically we assume that statistical fluctuations of the class-II partial fission widths, $\Gamma_{\lambda_{II}\uparrow}$,  exhibit an independent Porter-Thomas~\cite{por:56}  ($\nu=1$) distribution across $n$ fully open Bohr fission channels. If each average partial fission width according to given outer Bohr channel, is equal, then the distribution of the total fission widths is ruled by a  $\chi^2$ law with $\nu_f=n$ DoF.  In Hauser-Feshbach statistical theory with adequate $W_{n,f}$ factor, the associated DoF $\nu_{f}$ must be equal to the number of open channels at the outer barrier. However the IS, that lowers the transmission across the outer fission channels, manifests as an actual reduction of $\nu_{f}$. In our calculations, each  effective value of $\nu_{f}$ has been derived by maximum likelihood method from the value of the double barrier $\mathcal{R}$-matrix excited nucleus state fission width averaged over 1600 Monte Carlo trials. The results were presented~\cite{bou:14} as a function of the inner barrier transparency for a range of fully open outer barrier channels. The conclusion was that an ideal one fission channel according to single hump situation (i.e; no IS and $\nu_{f}\rightarrow 1$) is recovered only  when the inner barrier DoF, $\nu_{A}$, is sizable. In any other coupling situation, $\nu_{f}$ is strongly reduced by the IS ($0< \nu_{f}<1$) and, any subsequent $W_{n,f}$ calculation will have to be corrected accordingly. In practice this is equivalent to substitute $W_{n,f}(\nu_f)$ by $W_{n,f}(\nu_f^{eff})$ in Eq.~\ref{eq:FXSdecoupled} (resp. for Eq.~\ref{eq:Pdecoupled}) with $\nu_f^{eff}$, the effective DoF. This is another source of error that is usually compensated by other parameter variation during the fitting process on experimental data.\\ 

\subsubsection{Monte Carlo Calculations of the Intermediate Structure}
Real situation commands to avoid the decoupling hypotheses of Eq.~\ref{eq:FXSdecoupled} since class-I and class-II state width statistical properties are obviously correlated across the intermediate fission barrier.
%through the coupling elements $\langle X_{\lambda_I}^{(I)}|H_{c}|X_{\lambda_{II}}^{(II)} \rangle$ (Eq.~\ref{eq:mix}).   
Although, Eq.~\ref{eq:FXSdecoupled} supplies valuable estimate of the average neutron-induced fission cross section,  we also realize that exact solution for this equation relies on the possible derivation of an analytical expression  pertaining to the actual coupling strength situation. A powerful alternative to analytical formulae is our Monte Carlo-type (MC) method~\cite{bou:13} which presents the advantage of returning average  of either reaction cross sections or surrogate-like decay probabilities taking full account of the various  parameter statistical  fluctuations under relevant coupling conditions. Our MC approach simulates $R$-matrix resonance properties, relevant to each selected class-II state and neighboring class-I states (over at least a full class-II energy spacing), using a chain of pseudo-random numbers for a fine-tuned selection process based on both level width and spacing statistical distributions with suitable averages.   For backup, the simulated MC average total cross section (resp. the total decay probability) according to given spin-parity is compared with the entrance compound nucleus formation cross section (resp. the PDEN population fraction) as expressed in Eq.~\ref{eq:FXSdecoupled} (resp. Eq.~\ref{eq:Pdecoupled}), making allowance to slight magnitude renormalization whenever computing accelerations carried by the in-house MC procedure bring differences. 
% BLABLA Using the above Monte Carlo simulation, we want to quantify in next sections, the possible error expected on computed surrogate fission probabilities due in particular to the above approximations (usual decoupling hypothesis, disregard of $W_{II}$ and incorrect estimation of $\nu_{f}$) that are carried routinely in evaluation data tasks. 
%We underline that the accurate Monte Carlo procedure is carried each time as possible in this work and in most cases supersedes the analytical default route (Eq.~\ref{eq:FXSdecoupled}) and that, especially at deep sub-barrier excitation energies (i.e; for very weak coupling situation between class-I and class-II states) where Lynn and Back  formula (Eq.~\ref{eq:LB}) fails. 
%At final the MC simulation supplies directly the best estimate of neutron-induced cross sections or surrogate-like decay probabilities according open reactions %(neutron emission, $\gamma$ and fission decays). 
This MC procedure carries the compact formulation of the pre-cited equations meaning 

\begin{eqnarray}
{\sigma_{nf}}(E_n)=\mathlarger{\sum_{J^\pi}}  \Biggl[\sigma_{n}^{CN}(E_n,J,{\pi}) \times \mathcal{B}_{f,MC-xs}^{J^\pi} (E_{f})\Biggr]\mbox{ ,}
\label{eq:xsmc}
\end{eqnarray}
in terms of MC fission cross sections %(in correspondence with the analytical form of Eq.~\ref{eq:FXSdecoupled}) 
or %expressed in terms of 
surrogate-like MC fission probability,% (vs resp. Eq.~\ref{eq:Pdecoupled}),
\begin{eqnarray}
\mathcal{P}_{surr,f}^{A^*}(E_{x})&=&\sum_{J^\pi} \Biggl[  \mathcal{F}_{surr}^{A^*}(E_{x},J,{\pi}) \times \mathcal{B}_{f,MC-surr}^{J^\pi} (E_{x})\Biggr] \mbox{ } \mbox{ }\mbox{ .}\mbox{ } \mbox{ }
\label{eq:Pmc}
\end{eqnarray}

%where $W_{sp,f}$ is the WFCF factor correlating from one side, formation and fission decay widths of  fine structure CN resonances and from the other side, the latter widths with the total width appearing in the denominator of the pure Hauser-Feshbach equation~\ref{eq:hfwccp}. Correlations between formation and decay widths are created by the existence of direct and semi-direct mechanisms that restricts the validity of N.~Bohr hypothesis of pure CN process. On the other side, correlations between the various partial decay widths must be seen as a failure of statistical theory which impact is inversely proportional to the number of channels opened for a given reaction.\\   
%------

\subsection{\label{ss:MC} In-house $\gamma$ decay probability Monte Carlo calculation using $\mathcal{LNG}$}

We realize immediately that calculating Eqs.~\ref{eq:xsmc} or~\ref{eq:Pmc} involves simultaneous transmission coefficient calculations of competing $\gamma$-channels, neutron elastic channel(s) as well as open inelastic channels  to satisfy to total flux conservation. As quoted in section~\ref{ss:Compound}, neutron channels are modeled on the ground of Eq.~\ref{tnn} but the calculation  of $\gamma$-channel transmission coefficients borrows classic narrow resonance approximation limit according to  weak strength functions meaning 
\begin{eqnarray}
T_{\gamma}^{J^{\pi} }=   2\pi\frac{\bra\Gamma^{J^\pi}_{\gamma}\ket}{D_{J^\pi}}.
\end{eqnarray}
where $D_{J^\pi}$ is the mean average resonance spacing for given spin $J$ and parity $\pi$. % $S$-wave average resonance parameters at $S_n$ adopted in this work for the Pu isotope family and energy dependence parameters above $S_n$ have been compiled in Table III of~\cite{bou:13}. 
However the question of the energy dependence below $S_n$ for the fissile isotopes is also sensitive. The $\mathcal{LNG}$ default route %for calculating $T_{\gamma}^{J^{\pi}}$ uses the evaluated $s$-wave $\gamma$ average resonance parameter value associated with 
involves the equiprobable parity composite prescription of Gilbert and Cameron~\cite{gil:65} for $D_{J^\pi}$ which parameters have been adjusted so that to reproduce the experimental $s$-wave mean level density at $S_n$ (see Table III~\cite{bou:13}). In present objective of as good quality surrogate data modeling as possible, we can not afford the default semi-empirical approach and we rather rely our calculations on the QPVR LD procedure~\cite{bou:13}. % as also described in  previous paper (section III). 
For even-even PDEN, lowest excited states are built solely from pure collective excitations up to the breaking of the neutron or proton pairing energy whereas the spectrum of odd-N-even-Z fissioning nuclei involves single-quasineutron states that carry vibrational states. As excitation energy increases, meaning above the breaking of a neutron or proton pairing energy regarding an even-even nucleus, multi-quasi-particle states carrying multi-phonon vibrational states show up in the level spectrum. Finally rotational bands following classical rules, are built on those bandheads. %We will come back later on as far as fission-decay probability Monte Carlo calculations designed for surrogate modeling, are concerned. 
QPVR LD have also been used for better estimates of $\bra\Gamma^{J^\pi}_{\gamma}\ket$. Full model description and numerical applications are available in Refs.\cite{bou:13,bou:18}.

\section{\label{s: Validity}Scanning SRM framework}
\subsection{\label{ss:WFCFcomparison} WFCF versus SWFCF shape and magnitude}
%\setcounter{subsubsection}{0}
%Once having expressed various WFCF expressions meaning the familiar in-out-going channel width correction formula (Eq.~\ref{eq:Sconcept}), the extreme limit surrogate variant (Eq.~\ref{eq:RSRconcept}) and finally the dedicated fission channel extra correction (Eq.~\ref{eq:Rconcept}), 
Following equations settled above, we are able to quantify the global impact of $W_{n,f}$, $W_{surr,f}$ and $W_{II}$ for both categories of heavy nuclides: those which are fertile or fissile according neutron-induced  reactions. Reader  might also refer to the exhaustive study by Hilaire \etal~\cite{hil:03} about the various approximated WFCF formulae and expectations  when applied to heavy and light nuclides. We start this chapter by verifying if WFCF expected features are reproduced by present numerical calculations.   

\subsubsection{WFCF features according to heavy target nuclides}

\begin{figure}[t]
\center{\vspace{1.cm}
\resizebox{0.95\columnwidth}{!}{
\includegraphics[height=5cm,angle=0]{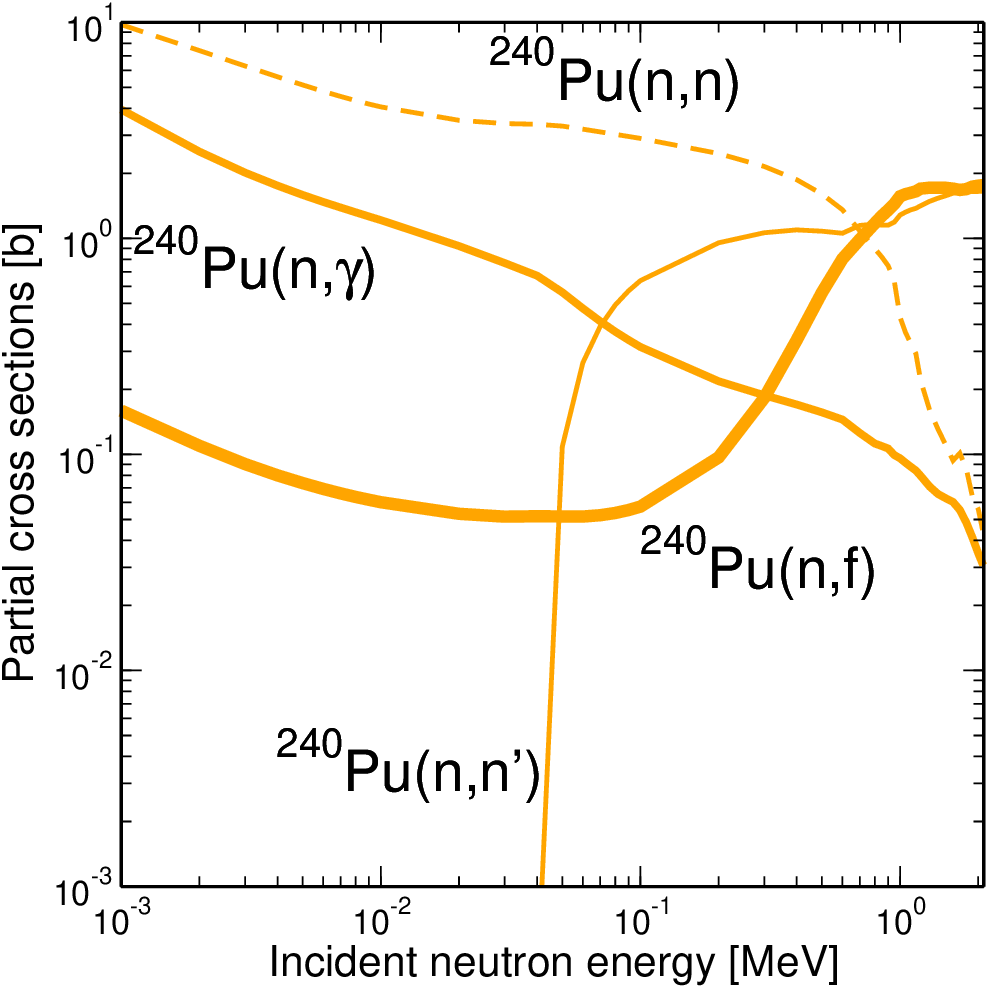}}}
\caption{(Color online) $^{240}$Pu neutron-induced partial cross sections computed analytically with $\mathcal{LNG}$. Thick, medium-thick and thin solid curves, and dashed line correspond  respectively to the $(n,f)$, $(n,\gamma)$, total inelastic $(n,n'_{tot})$ and elastic $(n,n)$ cross reactions.}%
\label{fig:xsgfi241*}
\end{figure}

\begin{figure}[t]
\center{\vspace{1.cm}
\resizebox{0.95\columnwidth}{!}{
\includegraphics[height=5cm,angle=0]{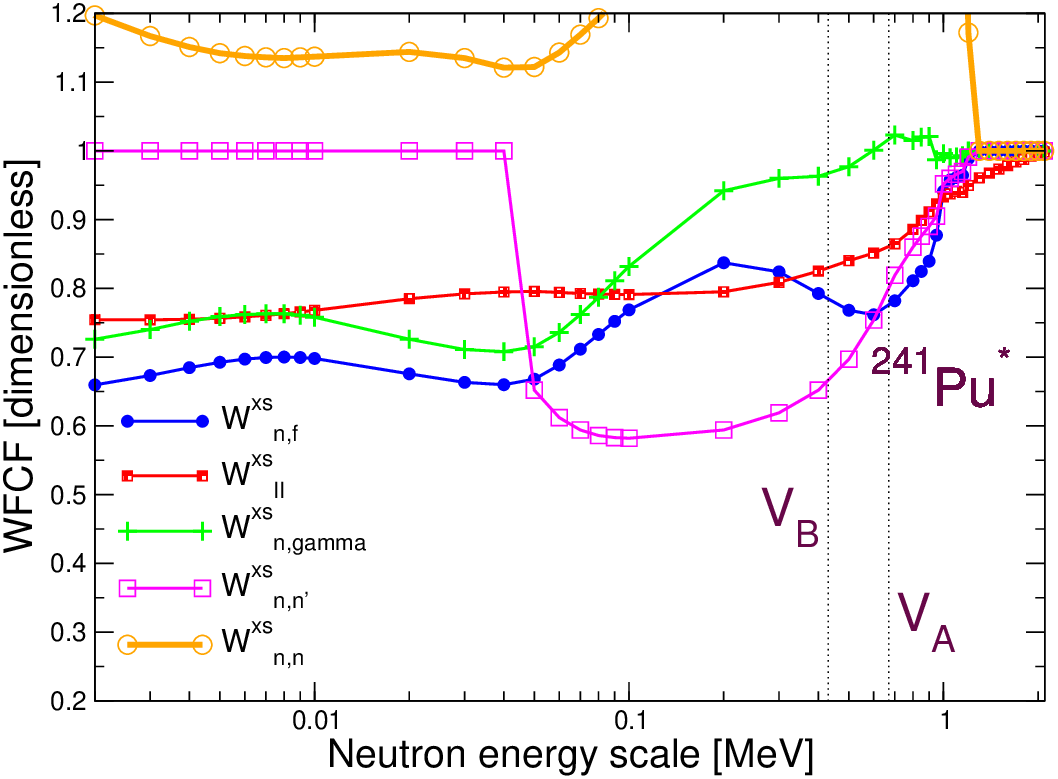}}}
\caption{(Color online) Comparison of (global) $W_{n,f}$, $W_{II}$, $W_{n,\gamma}$, $W_{n,n}$ and $W_{n,n'}$ curves according the (\pub  +n) system and average cross sections (superscript label '{\it xs}'). The WFCF coefficients displayed must be seen as global coefficients related to each partial reaction (elastic, inelastic, fission and capture) but integrated over all $J^{\pi}$ compound nucleus excited states. Note: present $W_{n,f}$ and $W_{II}$ curves differ from Ref.~\cite{bou:13} (figure 4) by their smoother character because Dresner numerical integrations are here performed on the whole fluctuating energy range. The default route involves jointly to Dresner integration, appropriate WFCF asymptotic formulae for saving computing time.} % The magnitude of the average sub-barrier tunnelling effect (green-double-dash-dot curve) is significantly larger than the estimated overall fluctuation factor ($W_{nf}(\nu_f^{eff}) \times W_{II}$ product; black-solid line) as expected for manifest cases of sub-threshold fission.}%
\label{fig:WFCFxsAllPu241*}
\end{figure}

\begin{figure}[t]
\center{\vspace{1.cm}
\resizebox{0.95\columnwidth}{!}{
\includegraphics[height=5cm,angle=0]{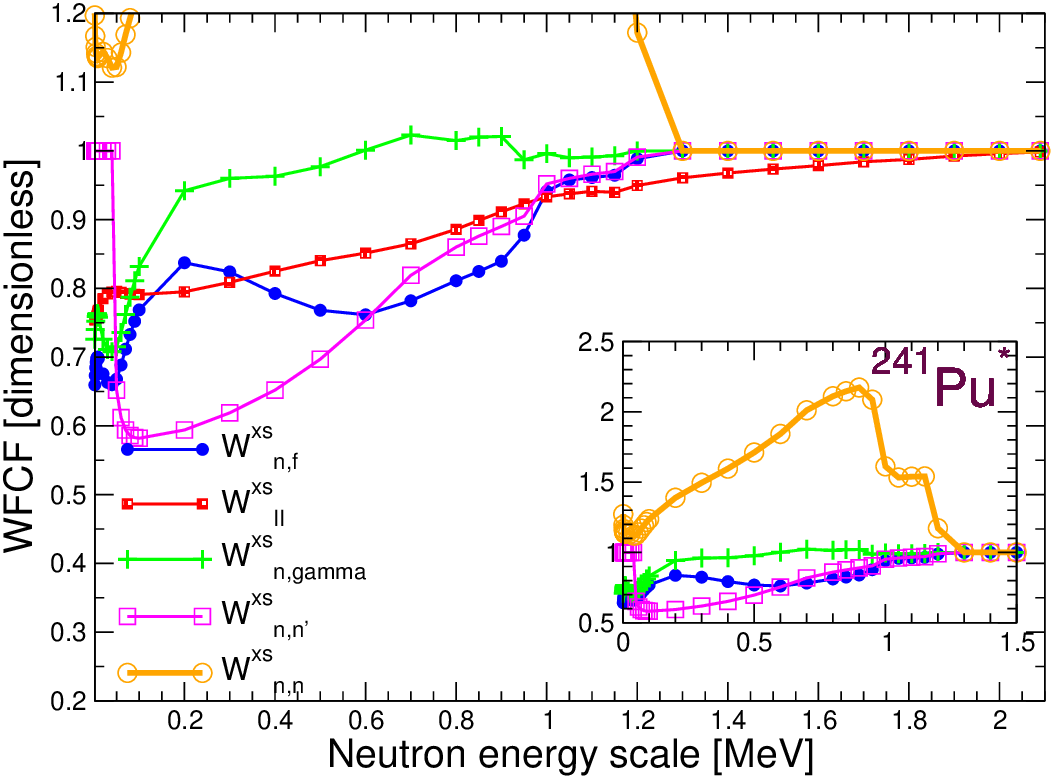}}}
\caption{(Color online) Same as Fig.~\ref{fig:WFCFxsAllPu241*} but using x-axis linear scale. In addition an inset image displays the whole WFCF picture and shows the expected strong elastic enhancement.}%
\label{fig:WFCFxsAllBPu241*}
\end{figure}

\begin{figure}[t]
\center{\vspace{1.cm}
\resizebox{0.95\columnwidth}{!}{
\includegraphics[height=5cm,angle=0]{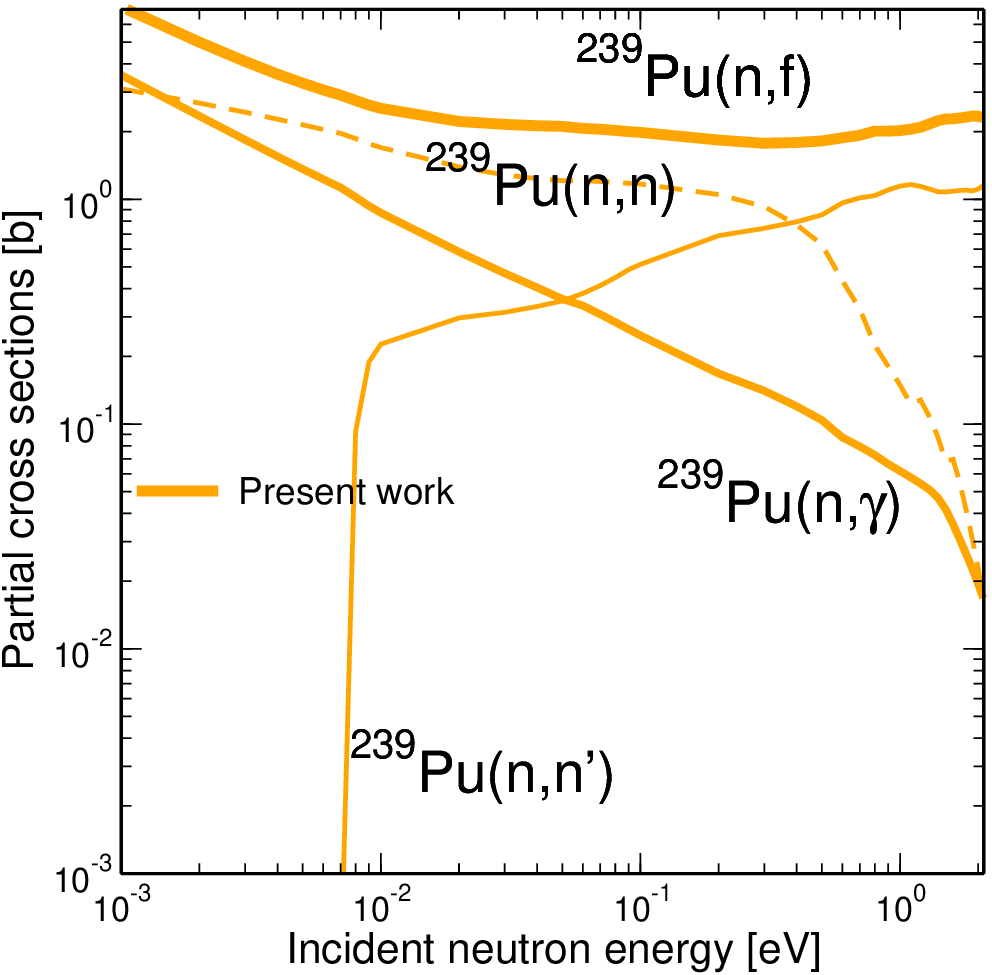}}}
\caption{(Color online) $^{239}$Pu neutron-induced partial cross sections computed analytically with $\mathcal{LNG}$. Legends are identical to Fig.~\ref{fig:WFCFxsAllPu241*}.} 
\label{fig:xsgfi240*}
\end{figure}
%Thick, medium-thick and thin solid curves and, dashed line correspond  respectively to the $(n,f)$, $(n,\gamma)$, total inelastic $(n,n'_{tot})$ and elastic $(n,n)$ cross reactions.}%

\begin{figure}[t]
\center{\vspace{1.cm}
\resizebox{0.95\columnwidth}{!}{
\includegraphics[height=5cm,angle=0]{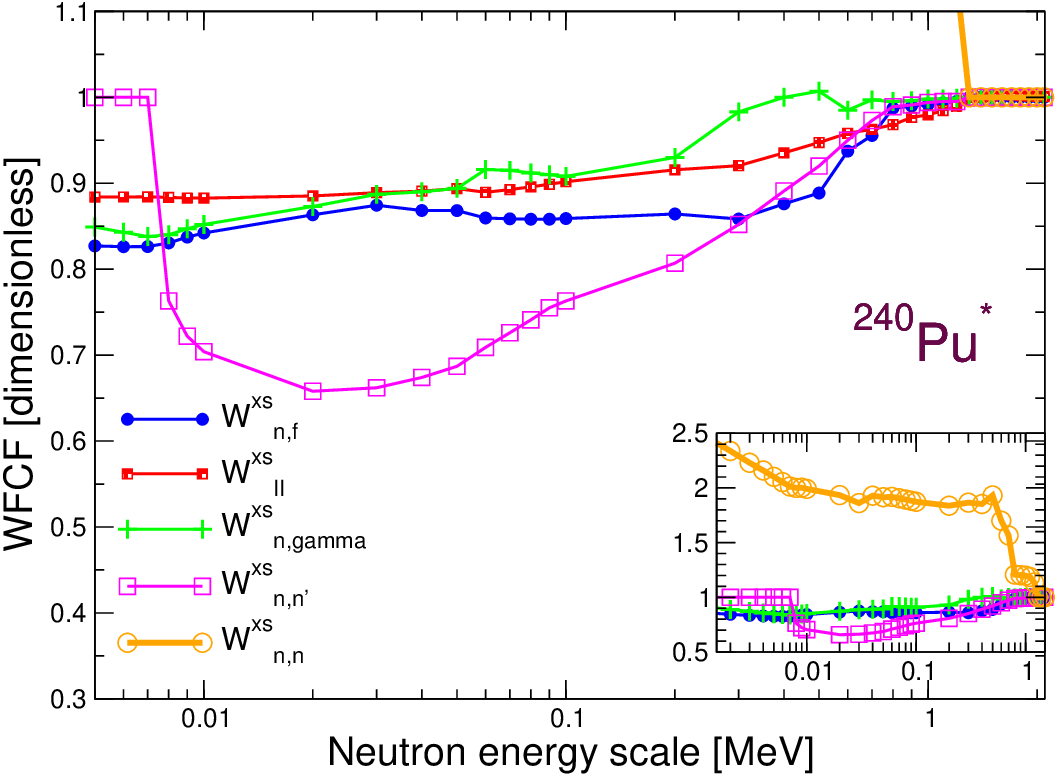}}}
\caption{(Color online) Comparison of  (global) $W_{n,f}$, $W_{II}$, $W_{n,\gamma}$, $W_{n,n}$ and $W_{n,n'}$ curves according the (\pua  +n) reaction. The inset image displays the whole WFCF picture and shows the expected strong elastic enhancement.}%
\label{fig:WFCFxsAllBPu240*}
\end{figure}
%\newpage

\begin{figure}[t]
\center{\vspace{1.cm}
\resizebox{0.95\columnwidth}{!}{
\includegraphics[height=5cm,angle=0]{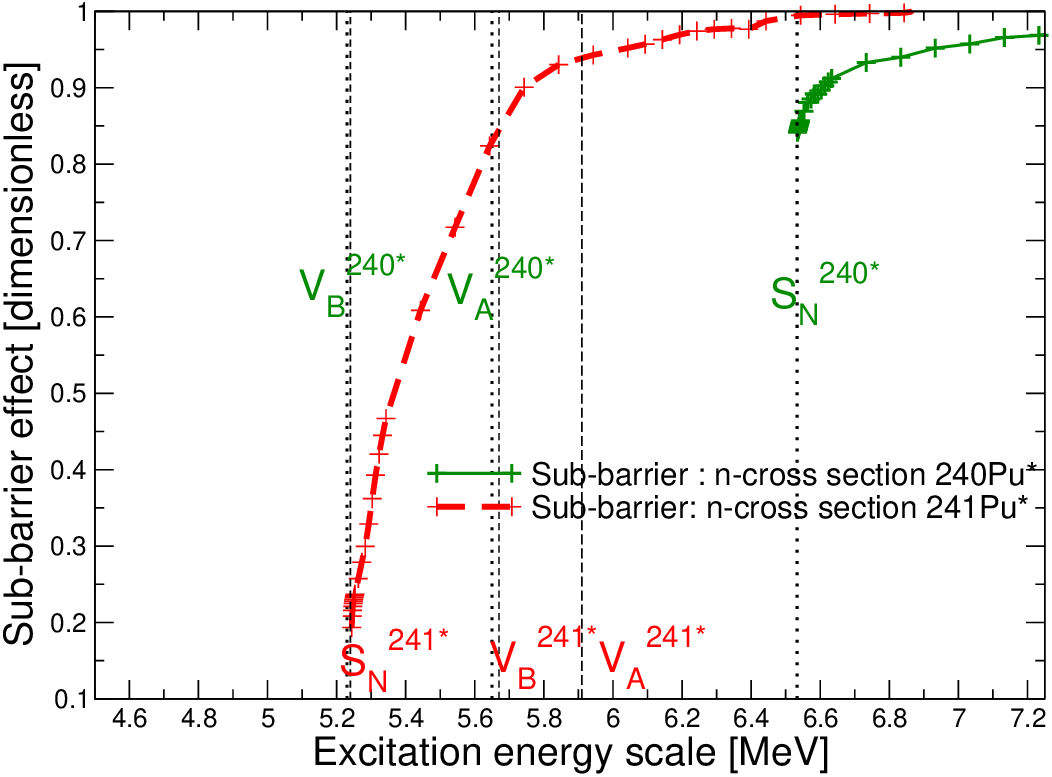}}}
\caption{(Color online) Sub-barrier tunneling estimates based on Lynn and Back~\cite{lyn:74b} formulation of $\mathcal{B}_{f,IS}$ (Eq.~\ref{eq:LB}). Right above $S_n$, the maximum of sub-barrier effect recorded for the fissile target nucleus (green solid curve) of about $15\%$, is in contrast to the $80\%$ strong impact according to the fertile case (red dashed curve).}
\label{fig:subPu240*VsPu241*Z}
\end{figure}

\begin{figure}[t]
\center{\vspace{1.cm}
\resizebox{0.95\columnwidth}{!}{
\includegraphics[height=5cm,angle=0]{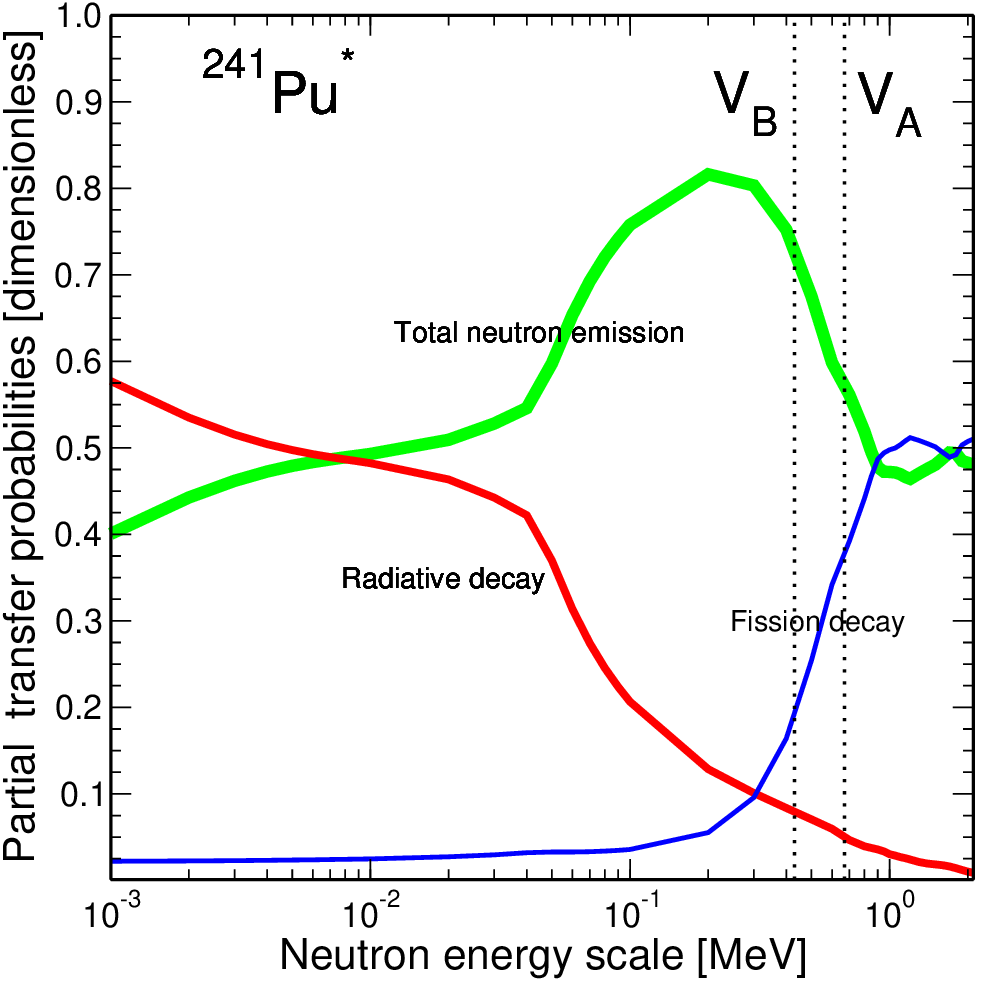}}}
\caption{(Color online) $^{241}$Pu$^*$ neutron-fed transfer partial decay probabilities computed analytically with $\mathcal{LNG}$. Thick, medium-thick and thin solid curves correspond  respectively to total neutron emission, $\gamma$ and fission by this nucleus.}% The dashed curve represents the 'inelastic' neutron emission channel contribution that opens above $40~keV$.}%
\label{fig:Pgfi241*}
\end{figure}

\begin{figure}[t]
\center{\vspace{1.cm}
\resizebox{0.95\columnwidth}{!}{
\includegraphics[height=5cm,angle=0]{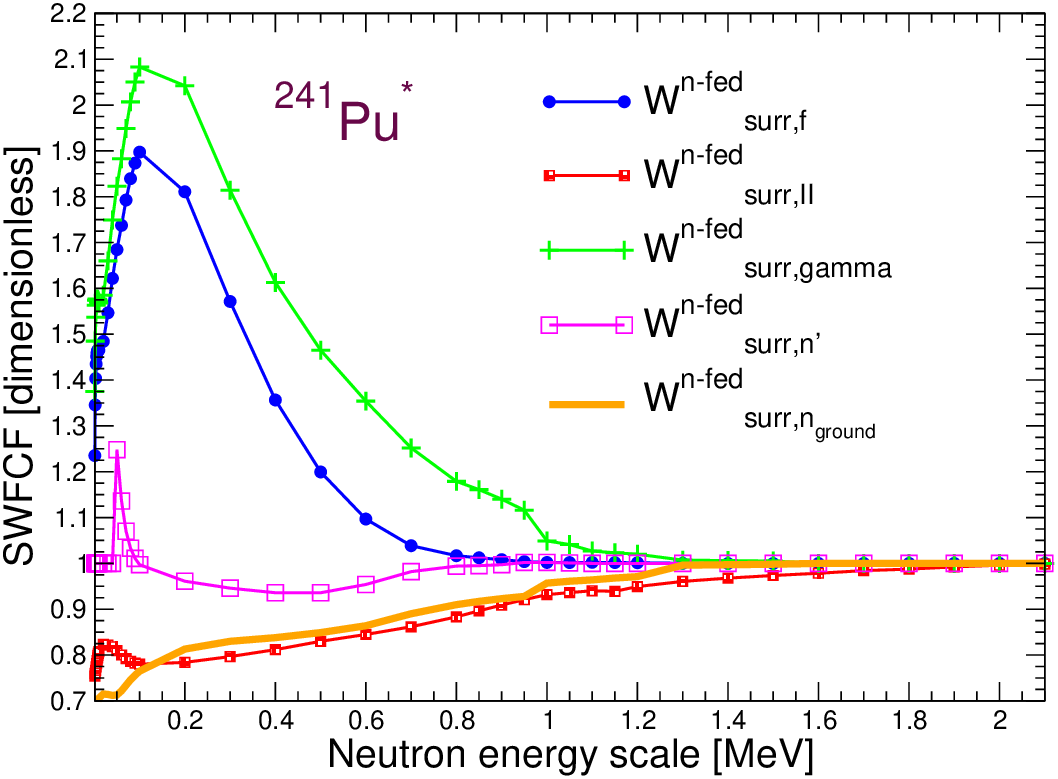}}}
\caption{(Color online) Comparison of $W_{surr,f}$, $W_{surr,II}$, $W_{surr,\gamma}$, $W_{surr,n_{ground}}$ and $W_{surr,n'}$ curves corresponding to associated surrogate-like partial decay probabilities in $^{241}Pu^*$. Those calculations are fed with an entrance population identical to the neutron-incident calculated population; ($n-fed$) superscript.} 
\label{fig:WFCFtransAllPu241*}
\end{figure}

\begin{figure}[t]
\center{\vspace{1.cm}
\resizebox{0.95\columnwidth}{!}{
\includegraphics[height=5cm,angle=0]{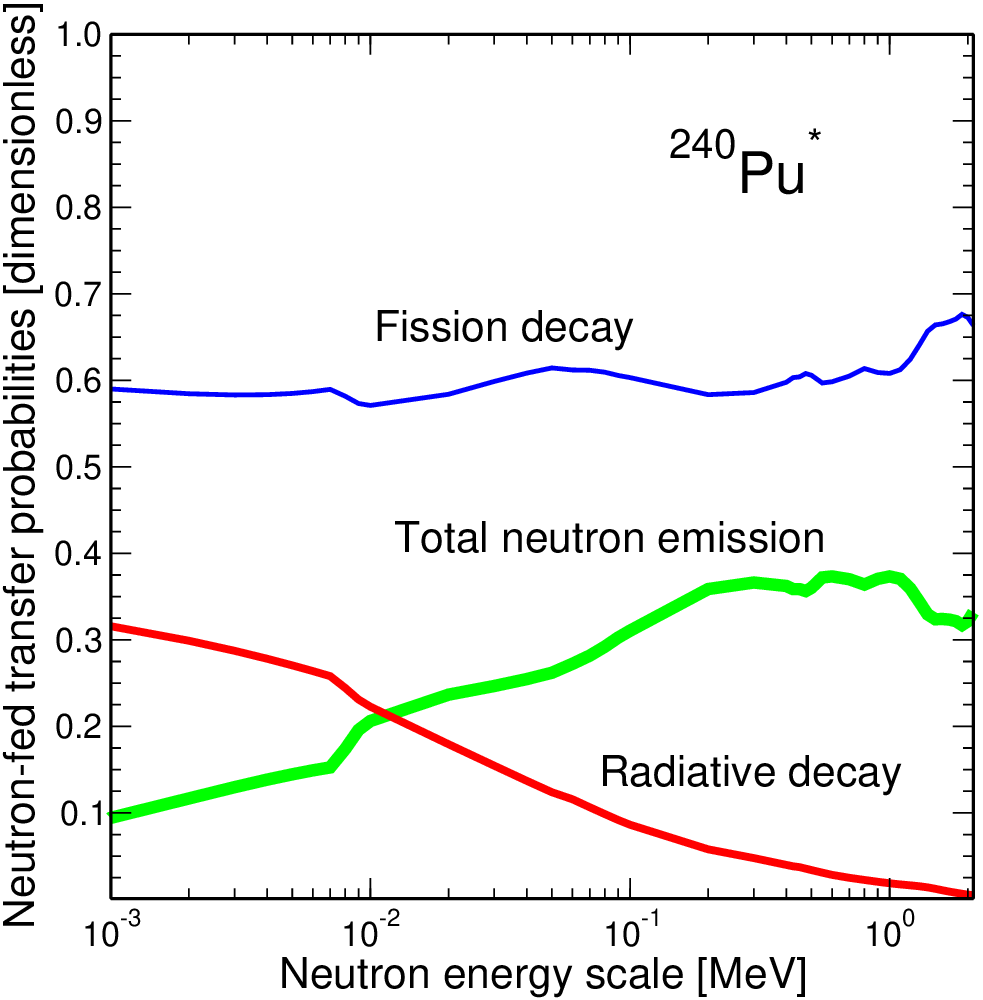}}}
\caption{(Color online) $^{240}$Pu$^*$ neutron-induced transfer partial decay probabilities computed analytically over neutron spectroscopy fluctuating energy range. Thick, medium-thick and thin solid curves correspond respectively to total neutron emission, $\gamma$ and fission decay probabilities.}%
\label{fig:Pgfi240*}
\end{figure}

\begin{figure}[t]
\center{\vspace{1.cm}
\resizebox{0.95\columnwidth}{!}{
\includegraphics[height=5cm,angle=0]{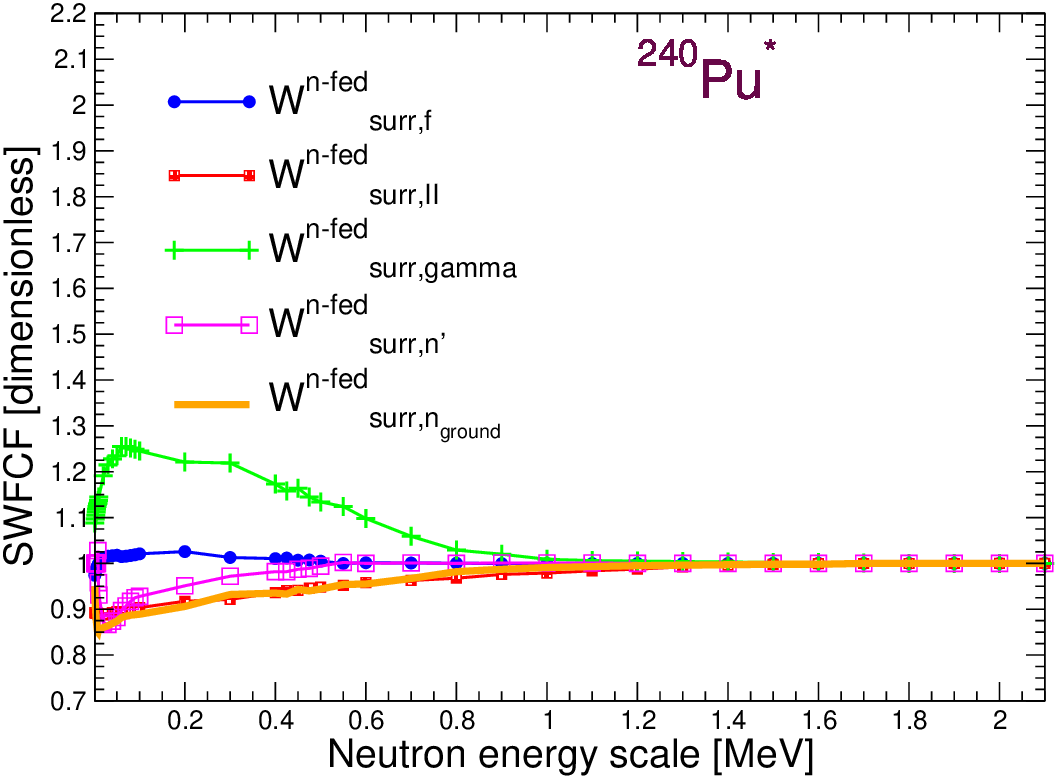}}}
\caption{(Color online) Comparison of $W_{surr,f}$, $W_{surr,II}$, $W_{surr,\gamma}$, $W_{surr,n}$ and $W_{surr,n'}$ patterns   corresponding to the neutron-fed surrogate-like  $^{240}Pu^*$ decay probabilities. We verify that $W_{II}$ factors computed by $\mathcal{LNG}$ using either surrogate (Eq.~\ref{eq:Pmc} - This Fig.)  or cross section (Eq.~\ref{eq:xsmc} - Fig.\ref{fig:WFCFxsAllBPu240*}) routes and with identical entrance population ($\mathcal{F}_{surr}^{^{240}Pu^*}\equiv\mathcal{F}_{n}^{^{240}Pu^*}$), give close results.}%
\label{fig:WFCFtransAllPu240*}
\end{figure}
\paragraph{Fertile nuclides}

We illustrate this category with $^{241}$Pu$^*$ formed by neutron capture. We should recover in terms of WFCF the well-known behavior of a medium-mass target nucleus with substantial capture cross section. Below fission threshold meaning a few hundred of keV according to inner and outer fundamental barrier heights (noted $V_A$ and $V_B$ on next graphics), only scattering and capture reactions are open. In addition if neutron energy lies below inelastic threshold ($E_n<50$~keV on Fig.~\ref{fig:xsgfi241*}), it all depends on a competition between  elastic and capture. Since both cross section magnitudes are of same order, elastic enhancement is already significant at low energy (more than $+10\%$; Fig.~\ref{fig:WFCFxsAllPu241*}, see $W^{xs}_{n,n}$) but becomes much larger (up to $+120\%$; Fig.~\ref{fig:WFCFxsAllBPu241*}, see inset) when crossing inelastic threshold energy. The observed drop in the inelastic cross section is quite large (about $-40\%$) and in agreement with the amount of flux redistributed to the elastic channel. The depreciation of the inelastic cross section is still amplified by the gradual disappearance of the competitive capture channel which is no more capable to bring back some neutron flux to the elastic (on the opposite to low energies where the decrease in the capture cross section reaches $-25\%$). The correction applied on the fission cross section due to WFCF is substantial ($W^{xs}_{n,f}\approx 0.80$) for this fertile target isotope and reaches  unity only above $1$~MeV where the total number of open  channels involved becomes very large (attested on Fig.~\ref{fig:xsgfi241*} by full opening of both fission and inelastic channels). The peculiar correction due to statistical fluctuations of class-II state widths is rather constant ($W^{xs}_{II}<0.80$)  over the whole fluctuation range until again the total number of playing fission channels becomes large; meaning from the energy where the inelastic cross section reaches its plateau and both $V_A$ and $V_B$ - see Fig.~\ref{fig:WFCFxsAllPu241*} - are actually overpassed.\\ 

\paragraph{Fissile nuclides}
Advising $^{240}$Pu$^*$ formed by neutron interaction sounds to be the logical extension to raise the issue of fissile nuclides. On the opposite to the fertile case, the amount of flux redistributed to the elastic channel is quite large right above $S_n$ (report to inset of Fig.~\ref{fig:WFCFxsAllBPu240*}) since low energy fission and capture cross section magnitudes  are comparable (Fig.~\ref{fig:xsgfi240*}). Above first inelastic threshold and by analogy to the $^{241}$Pu$^*$ case, the elastic enhancement strength is strongly supported by the flux borrowed from the inelastic channels. Regarding $W^{xs}_{II}$ (Fig.~\ref{fig:WFCFxsAllBPu240*}), the correction cannot be neglected ($W^{xs}_{n,f}\approx 0.90$ up to $S_n+300$~keV) although average sub-barrier effect that lowers the {\it statistical regime} fission probability ($\mathcal{B}_f$; Eq.~\ref{eq:FXSdecoupled}) by substituting $\mathcal{B}_f^{stat.\mbox{ }regime}$ with $\mathcal{B}_{f,IS}$, is much smaller for fissile than for fertile nuclides (Fig.~\ref{fig:subPu240*VsPu241*Z}).

\subsubsection{Foreseen WFCF features according to surrogates}

According to the  WFCF {\it extreme limit} that we postulated for surrogate measurements (Eq.~\ref{eq:RSRconcept}), we are now able to address the main features as function of PDEN excitation energy. Since we aim to argue on (global) WFCF features according to both neutron-induced and particle-induced transfer reactions, we have chosen to retrieve SWFCF factors from transfer probabilities (Eq.~\ref{eq:Pdecoupled}) that use, as input, population fractions  identical to the neutron-induced cross section calculations performed above (i.e.; $\mathcal{F}_{surr}^{A^*\equiv CN}(E_x,J,{\pi})$). This guaranties consistent framework for the next  comparisons. To understand what we are going to observe on particle-induced transfer reaction SWFCF, we must remember that we postulated no in-out-going channel widths correlation but channel widths correlations between the many exit channels remain preserved.\\
%$W^{n-fed}_{surr,c'}$
\paragraph{Fertile nuclides}

We are now back to the $^{241}$Pu$^*$ system, but with specialization to 'neutron-induced' transfer reactions (so-called 'neutron-fed' in next sections). We shall  reinforce our argumentation by showing the $^{241}$Pu$^*$ neutron-fed transfer reaction decay probabilities  based on Eq.~\ref{eq:Pdecoupled}. According to the fertile nuclide category,  we visualize on Fig.~\ref{fig:Pgfi241*} a tiny surrogate-like fission decay probability until fission threshold be overpassed. Right below 1 MeV neutron energy, the surrogate-like radiative decay probability ($P_{surr ,\gamma}^{n-fed}$) becomes negligible whereas both surrogate-like total neutron emission and fission probabilities ($P_{surr,total~n}^{n-fed}$ and $P_{surr ,f}^{n-fed}$ respectively) contribute each to half of total decay. In terms of SWFCF behavior as function of energy, we recover on Fig.~\ref{fig:WFCFtransAllPu241*} the customary high energy pattern since each SWFCF tends to unity when the total number of channels involved becomes very large; in practice above $(S_n +1.6)$~MeV. The absence of elastic reaction prevents any usual elastic enhancement and we observe that both radiative and fission decays now endorse the role of the enhanced channels with maximum impact on the $\gamma$ decay channel and up to $+110\%$ of enhancement at $(S_n +200)$~keV. By reciprocity, neutron emission channels are depreciated accordingly to the total amount of reaction rate redistributed. We notice the new role of  neutron emission with residual nucleus in ground state since this channel represents the largest SWFCF flux contributor  ($W_{surr,n_{ground}}$ curve on Fig.~\ref{fig:WFCFtransAllPu241*}) for capture and fission channels. \\   

%In terms of surrogate WFCF behavior as function of excitation energy, we observe on Figure~\ref{fig:WFCFtransAllPu241*} up to about 1 MeV neutron-incident energy no fluctuation correction in any of the decay channels (i.e.; all WFCF=1). This fact can be understood by the fact that WFCF carries the correlation strength between the observed fission channel and total channel width (and indirectly on the other partials since the sum of the partials must be equal to the total amount). Although the number of degrees of freedom of the fission width distribution is rather limited below fission threshold and should manifest in terms of large fluctuations, the quasi-zero width of the fission channel prevent any actual correlations. Above and up to a large total number of involved channels, significant fluctuation corrections show up with neutron emission enhancement as large as fission lowering (about $|5\%|$). In the meantime radiative decay, which magnitude decreases rapidly, appears to carry an enhancement factor larger than the correction due to neutron emission (Fig.~\ref{fig:WFCFtransAllPu241*}).\\

\paragraph{Fissile nuclides}
We refer logically to the $^{240}Pu^*$ case for consistent comparison with pre-cited neutron-incident cross section WFCF behavior. Once more, we support our argumentation with corresponding computed neutron-fed transfer reaction decay probabilities (Fig.~\ref{fig:Pgfi240*}). On contrast to $^{241}Pu^*$, fission decay is now the dominant process above $S_n$ whereas capture decay, representing only one third of total decay, drops regularly and even steeper above neutron 'inelastic' emission threshold energy (the capture cusp is well visible around $8~keV$). Regarding SWFCF pattern as function of excitation energy, we observe on Fig.~\ref{fig:WFCFtransAllPu240*} trends similar to fertiles except in terms of magnitude which variation range remains limited [$_{-10}^{+30}\%$]. This indeed contrasts with the overall picture drawn by a fertile isotope which fission barrier heights lying above neutron emission threshold, constrain strongly the number of fission channels possibly involved and return large amount of width fluctuations at low neutron energy (cf. Fig.~\ref{fig:WFCFtransAllPu241*}).

\begin{figure}[t]
\center{\vspace{1.cm}
\resizebox{0.95\columnwidth}{!}{
\includegraphics[height=5cm,angle=0]{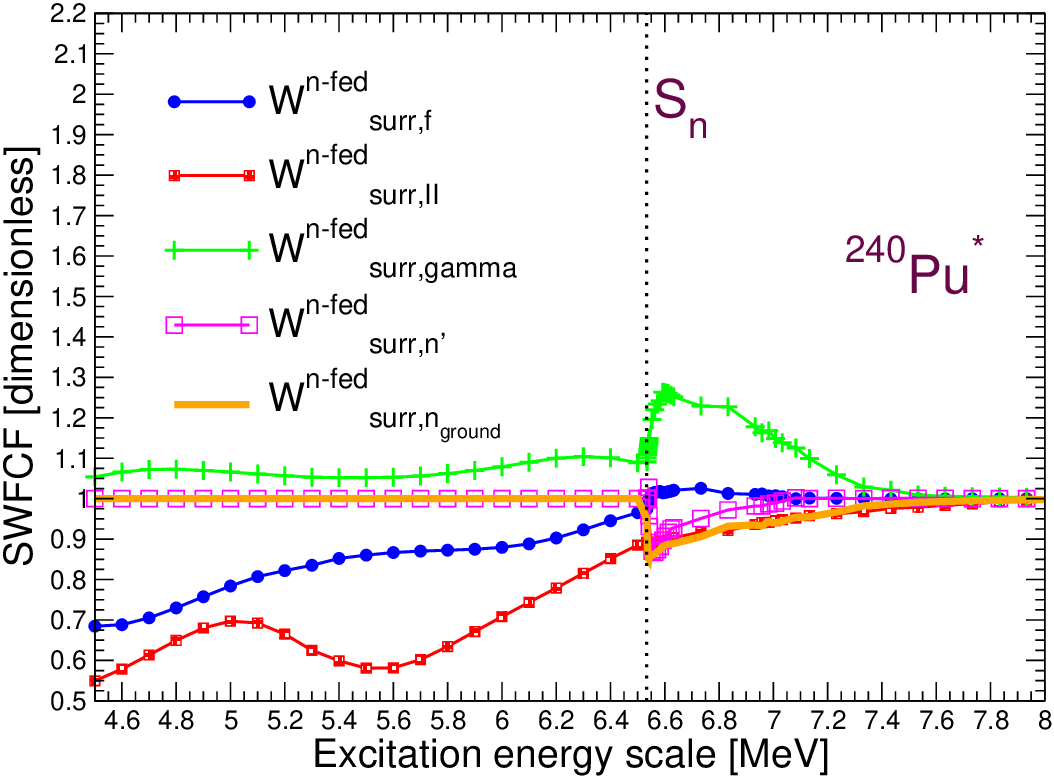}}}
\caption{(Color online) Same as Fig.~\ref{fig:WFCFtransAllPu240*} but over the whole excitation energy range. We materialize immediately the growing impact of $W_{surr,f}$ and $W_{surr,II}$ for decay probability calculations as excitation energy decreases.}%
\label{fig:WFCFtransAllPu240*U}
\end{figure}

\begin{figure}[t]
\center{\vspace{1.cm}
\resizebox{0.95\columnwidth}{!}{
\includegraphics[height=5cm,angle=0]{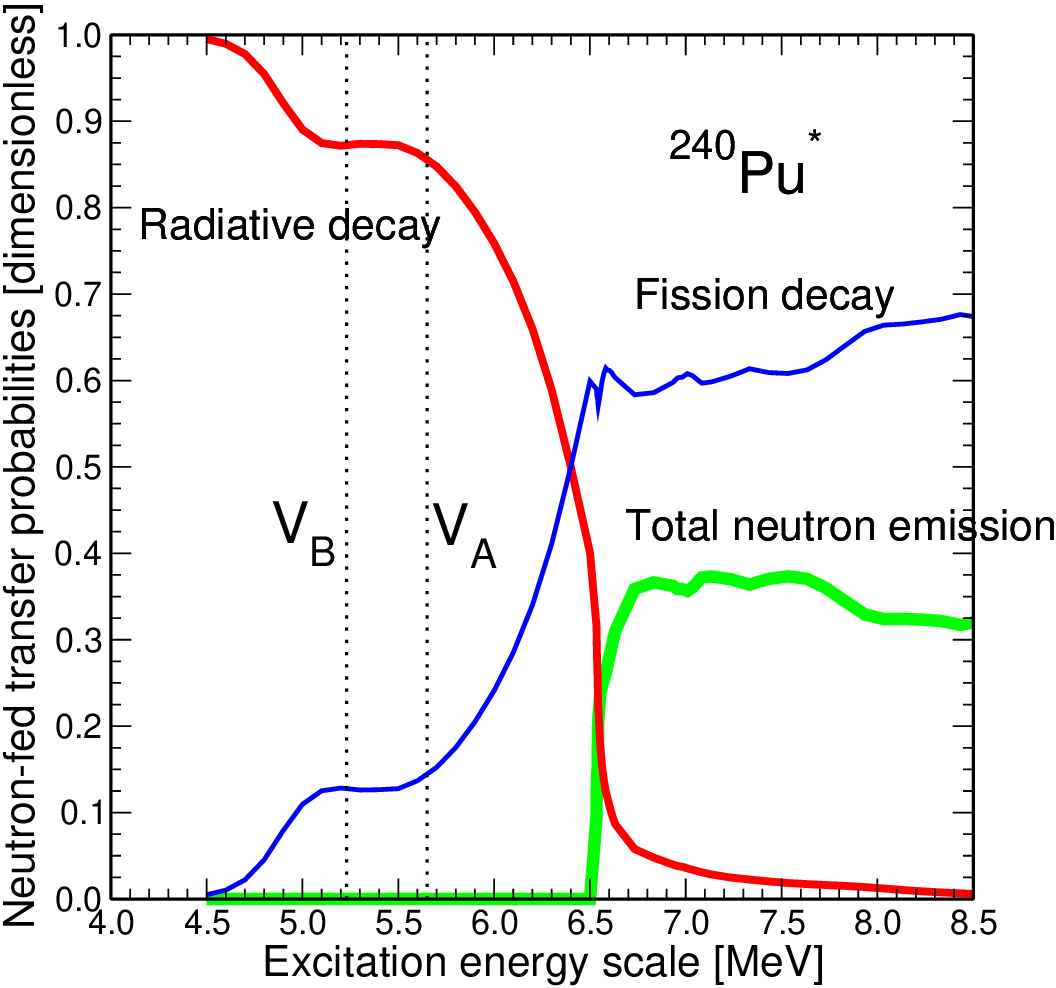}}}
\caption{(Color online) Same as Fig.~\ref{fig:Pgfi240*} but over the whole excitation energy range.}%
\label{fig:Pgfi240*U}
\end{figure}

\subsubsection{Overview of neutron sub-threshold WFCF features}

At this stage, it is interesting to recall that our main objective is best quality simulation of surrogate-like experimental probabilities but SRM does not carry WFCF and $W_{II}$ corrections. %In previous paragraphs, we reviewed the behavior of these latter quantities over the neutron-incident fluctuation energy range. 
As addressed in the introduction, fission probability measurements must be used for extraction of fission barrier heights that lie below neutron energy threshold meaning for isotopes corresponding to the fissile family. We realize that digging under neutron emission threshold requires extension of Fig.~\ref{fig:WFCFtransAllPu240*} down to low excitation energy. As far as we benchmark SWFCF results by supplying to the calculations the neutron cross section incident state population, we observe for $W_{surr,II}$ limited impact over the range $S_n$ to $(S_n+1~MeV)$ but enhanced negative impact as decreases the excitation energy (greater than $30\%$ on Fig.~\ref{fig:WFCFtransAllPu240*U}). On the opposite and since the flux that can be redistributed from the fission channel to the $\gamma$ channel shrinks dramatically (report to Fig.~\ref{fig:Pgfi240*U}), the enhancement on radiative decay remains moderate and constant ($\sim 10\%$). \\

As conclusion to current section, %~\ref{ss:WFCFcomparison}
 we have enlightened SWFCF characteristics as assumed in the WFCF extreme limit (Eq.~\ref{eq:RSRconcept}) that we are not usually accustomed to deal with. We have seen that both radiative and fission decays can now endorse the role of the enhanced channel with positive enhancement as large as $100\%$ (fertile nuclides) right above $S_n$. 
 % --- parametres ajustes meme pour E<Sn peuvent avoir un impact sur l'ahjustement au-dessus 
 %Various above findings suggest significant impact only on neutron-induced capture cross section extrapolation from surrogate applications for neutron reactor applications.  
 %  this finding is expected to impact significantly only neutron-induced capture cross section extrapolation since neutron reactor fission cross sections are either tiny (fertile nuclides) or too large (fissile nuclides) to require very precise SWFCF calculations. 
% or too large (fissile nuclides) to require very precise SWFCF calculations. 
% Fertile grosse correction E>Sn sur fission mais fxs small
% Fissile correction < 5%
\begin{figure}[t]
\center{\vspace{1.cm}
\resizebox{0.95\columnwidth}{!}{
\includegraphics[height=5cm,angle=0]{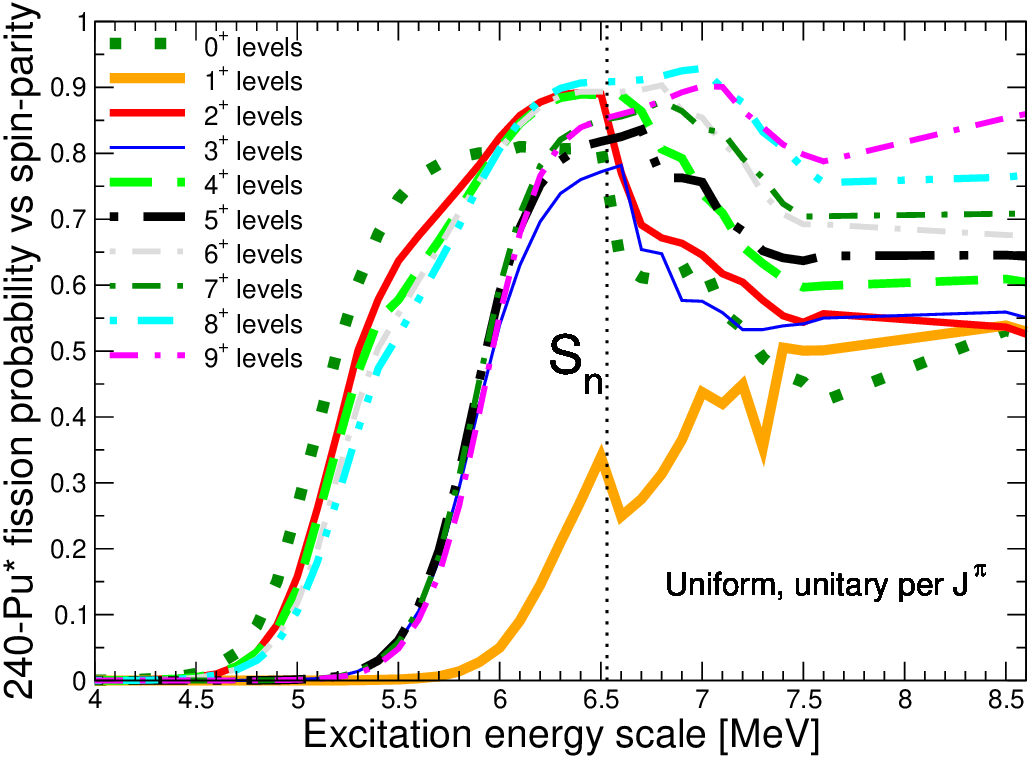}}}
\caption{(Color online)\label{fig:Pfis240*pi+} Monte Carlo $\mathcal{R}$-matrix double-barrier fission surrogate-like probabilities of $^{240}$Pu$^{*}$ as a function of resonance spin (positive parity) and excitation energy up to neutron kinetic energy of 2.1~MeV. The vertical bar at 6.53~MeV materializes neutron emission threshold. Figure above displays in particular the peculiar $1^+$ fission barrier probability (orange thick solid curve) that creates an untypical small contribution to low neutron-incident energy fission cross section. Note: for unbiased illustration, those decay probabilities  according to Eq.~\ref{eq:Pmc}, are fed  with entrance population both uniform and unitary per $J^\pi$.}
\end{figure}

\begin{figure}[t]
\center{\vspace{1.cm}
\resizebox{0.95\columnwidth}{!}{
\includegraphics[height=5cm,angle=0]{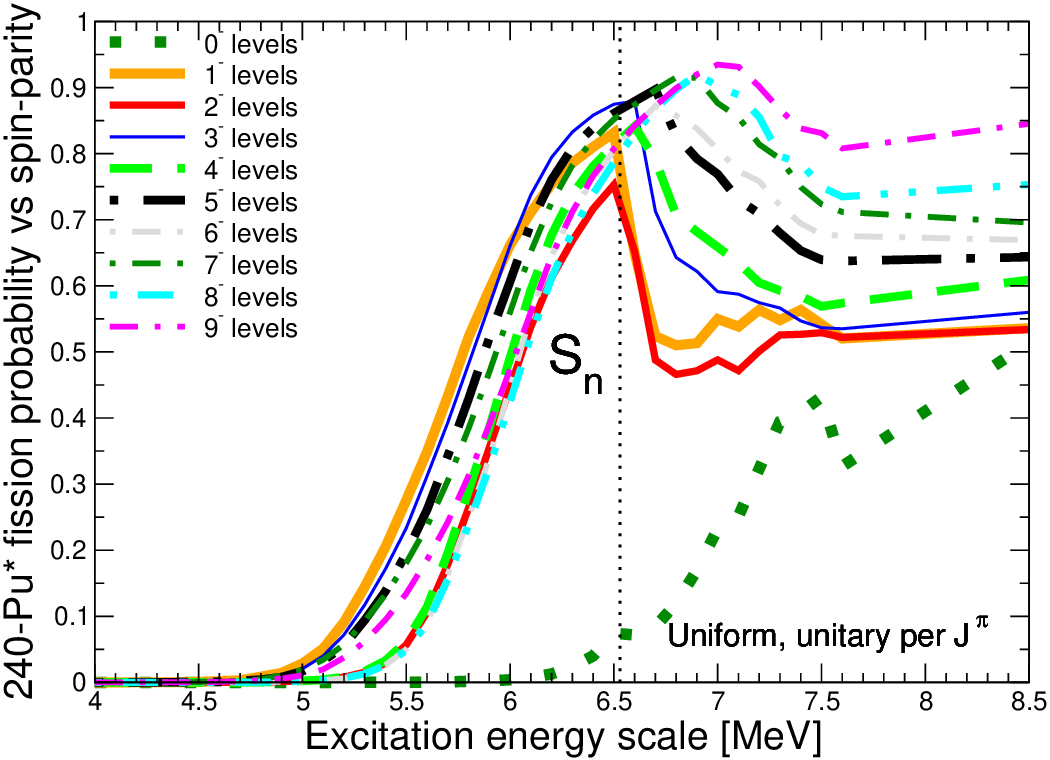}}}
\caption{(Color online)\label{fig:Pfis240*pi-} Same as Fig.~\ref{fig:Pfis240*pi+} but for negative parity Bohr transition states. Figure above shows the $0^-$ probability (green-dotted curve) which singular low excitation energy range contribution results from continuum transition state fission barrier tunnelling only.}
\end{figure}
   
\begin{figure}[t]
\center{\vspace{1.cm}
\resizebox{0.95\columnwidth}{!}{
\includegraphics[height=5cm,angle=0]{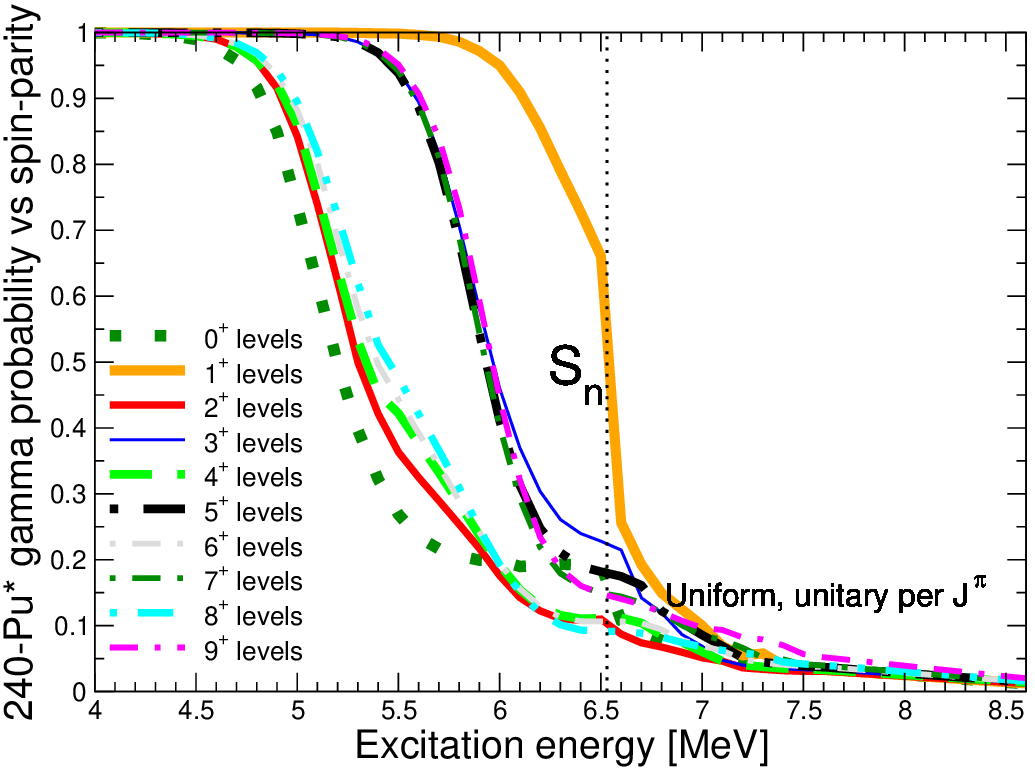}}}
\caption{(Color online)\label{fig:Pgamma240*pi+} Monte Carlo $\mathcal{R}$-matrix $\gamma$ decay surrogate-like probabilities of $^{240}$Pu$^{*}$ as a function of resonance spin (positive parity) and excitation energy up to neutron kinetic energy of 2.1~MeV. The vertical bar at 6.53~MeV materializes neutron emission threshold.}
\end{figure}

\begin{figure}[t]
\center{\vspace{1.cm}
\resizebox{0.95\columnwidth}{!}{
\includegraphics[height=5cm,angle=0]{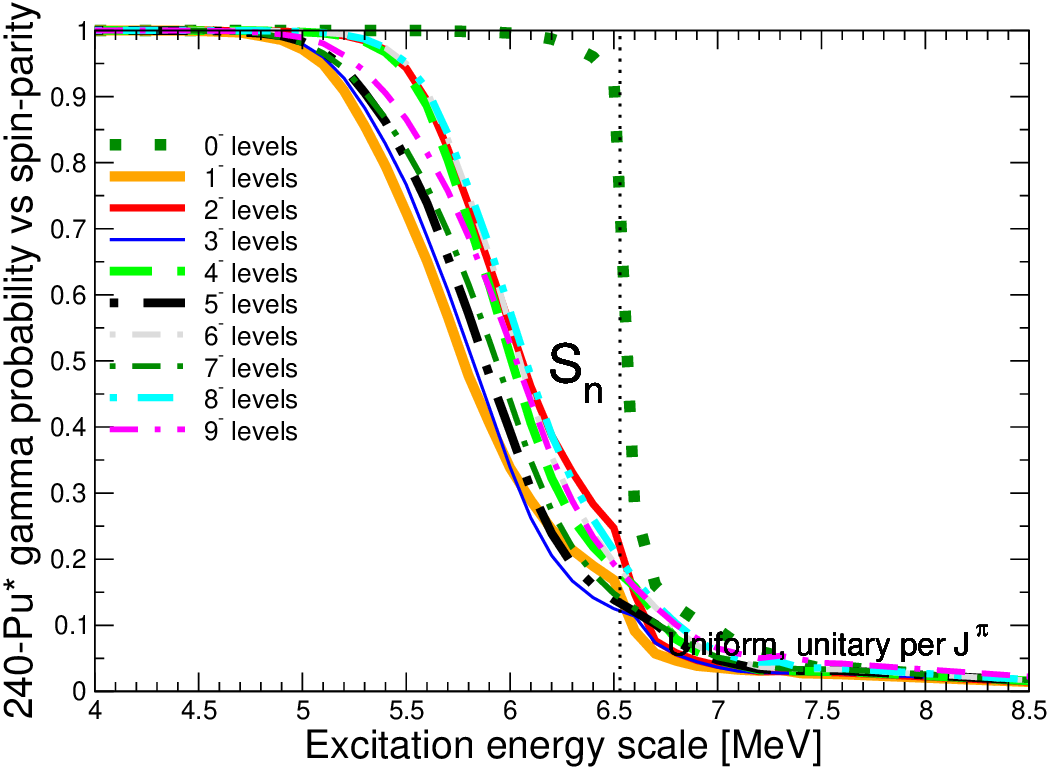}}}
\caption{(Color online)\label{fig:Pgamma240*pi-} Same as Fig.~\ref{fig:Pgamma240*pi+} but for negative parity excited states.}
\end{figure}

\begin{figure}[t]
\center{\vspace{1.cm}
\resizebox{0.95\columnwidth}{!}{
\includegraphics[height=5cm,angle=0]{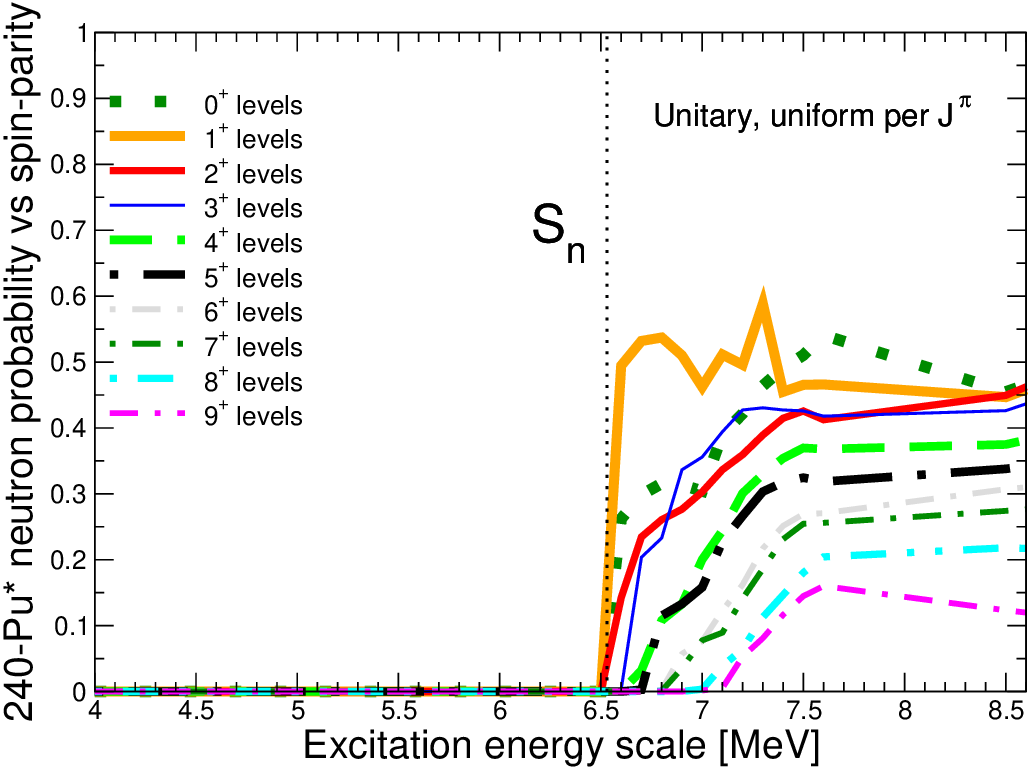}}}
\caption{(Color online)\label{fig:Pn240*pi+} Monte Carlo $\mathcal{R}$-matrix neutron emission decay surrogate-like probabilities of $^{240}$Pu$^{*}$ as a function of resonance spin (positive parity) and excitation energy up to neutron kinetic energy of 2.1~MeV. The vertical bar at 6.53~MeV materializes neutron emission threshold.}
\end{figure}

\begin{figure}[t]
\center{\vspace{1.cm}
\resizebox{0.95\columnwidth}{!}{
\includegraphics[height=5cm,angle=0]{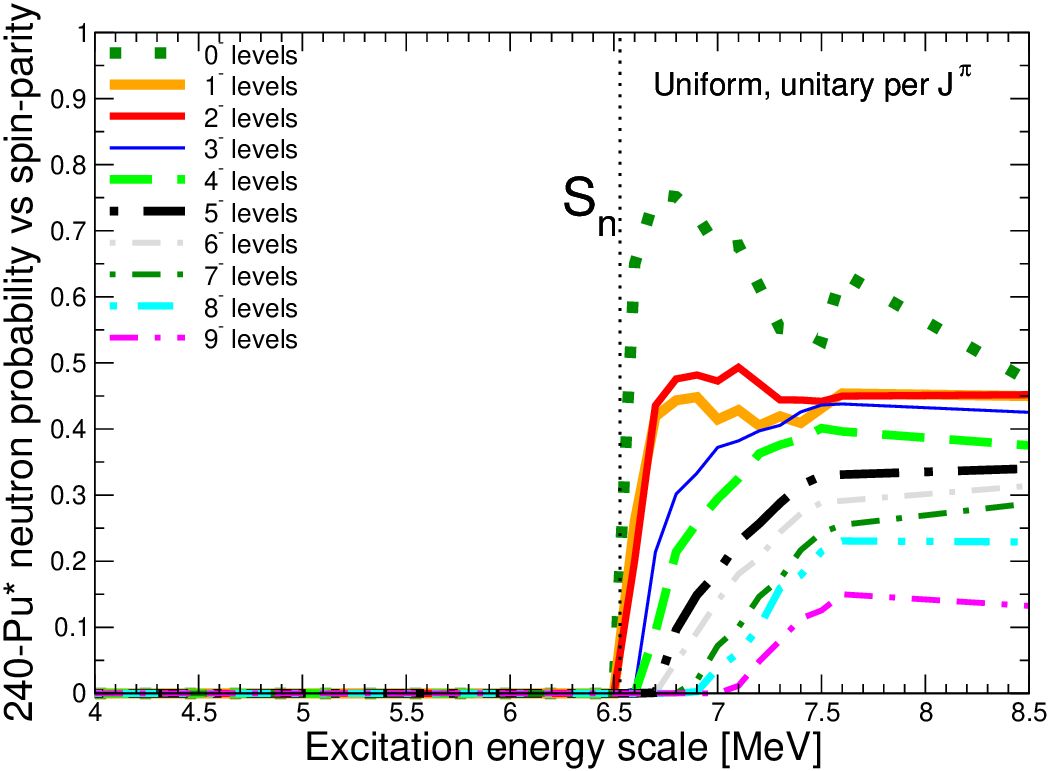}}}
\caption{(Color online)\label{fig:Pn240*pi-} Same as Fig.~\ref{fig:Pn240*pi+} but for negative parity excited states.}
\end{figure}

\subsection{\label{ss:MCdecaytesting}Low excitation energy reaction decay \\ Monte Carlo probabilities}

As recalled in the introduction, SRM relies in particular on the WE assumption of reaction decay probability spin-parity independence. Validity of latter hypothesis has been investigated quite extensively last decade (as in the review by J.E.~Escher \etal~\cite{esc:12} or in the experiments by G.~Kessedjian \etal~\cite{kes:10}, G.~Boutoux \etal~\cite{bout:12} and Q.~Ducasse \etal~\cite{duc:16}). Therefore this section aims revisiting this question on view of present robust formalism. %Now we have got some feedback about how partial reaction width fluctuations modify average partial decay probabilities by using the extreme limit definition of WFCF (Eq.~\ref{eq:RSRconcept}), we can 
We now extend our thinking to the joined effect of {\it SWFCF and subthreshold fission} on transfer reaction probabilities. For this investigation, we keep on studying our two examples namely $^{240}Pu^*$ and $^{241}Pu^*$ representing fissile and fertile categories. 

\paragraph{\label{sss:MCdecaytestingF}Fissile nuclides}
Backdrop for fissile nuclides  below $S_n$ is simpler since only fission and $\gamma$ decay reactions compete across Eq.~\ref{eq:Pmc}. As already specified, in the context of fundamental fission barriers sitting below $S_n$, %estimated respectively to $V_A=-880~keV$ and $V_B=-1300~keV$ according to $^{240}Pu^*$, 
fission occurs mainly across discrete Bohr transition states built solely from pure collective excitations that were carefully generated using the procedure described in ~\cite{bou:13}. Collective vibrations are of several kind, beginning with the {\it zero-vibration  ground state} ($K^\pi=0^+$), then involving low excitation energy collective vibrations such that the {\it $\gamma$-axis vibration} ($K^\pi=2^+$),  the {\it mass-asymmetry} ($K^\pi=0^-$), the {\it bending} ($K^\pi=1^-$) or even the {\it octupole} vibration  ($K^\pi=2^-$); all of them combined or not, supplying bandheads for rotational structure under classical $J^\pi$ rule construction as reminded below   
\begin{eqnarray} 
J^{\pi}&=&
\left\lbrace
\begin{array}{llll}    
K^\pi, (K+1)^\pi, (K+2)^\pi, \cdots&\mbox{for}&K\neq 0\\
0^+, 2^+, 4^+, \cdots&\mbox{for}&K=0^+\\
1^-, 3^-, 5^-, \cdots&\mbox{for}&K=0^-\mbox{ .}\\
\end{array}
\right.
\label{kvalues}
\end{eqnarray}

$^{240}$Pu$^*$ is specific in the way that several $0^+$ transition states are involved at $S_n$ whereas $1^+$ transition states play little role. $1^+$ state building requires at least combination of two collective phonons on top of the inner saddle (viz. bending associated to mass-asymmetry which resulting energy is estimated to $0.7~MeV$ above $S_n$). %  
%suggesting high barrier energies 
%(estimated in our work at $0.7~MeV$ and $-0.7~MeV$ about $S_n$ according to inner and outer barriers respectively)
%(estimated to $+0.7~MeV$ about $S_n$ according to inner barriers). 
$^{240}$Pu$^*$ specificity is properly told by Fig.~\ref{fig:Pfis240*pi+} (and ~\ref{fig:Pfis240*pi-}) which displays surrogate-like individual fission probabilities for transition states of positive parity (resp. negative parities). % according to Eq.~\ref{eq:Pmc} which entrance CN population is chosen uniform and unitary per $J^\pi$ for unbiased feedback. 
We readily imagine that the decay probability spin-parity independence (WE assumption) cannot be truly satisfied at low excitation energy for this isotope since regarding positive parities we observe three groups corresponding respectively to total level spin sequences $J=0,2,4,6,8$, $J=3,5,7,9$ and $J=1$ whereas for negative parities this trend reduces to two groups with first group merging all spins except zero spin. The $0^-$ peculiarity must be granted to the absence of $0^-$ states in the discrete transition state spectrum of even-even compound nuclei  since Eq.~\ref{kvalues} prevents construction of any of those quantum numbers below pair breaking energy.\\

In terms of $\gamma$ decay surrogate-like probabilities, total flux conservation imposes below $S_n$ reciprocal behavior to fission channel decay probabilities. This statement can be verified on respective positive (Fig.~\ref{fig:Pgamma240*pi+}) and negative (Fig.~\ref{fig:Pgamma240*pi-}) parity excited state $\gamma$ decay probability plots where the three and two groups of probability shapes are recovered. Above $S_n$, cusps in $\gamma$- and fission decay probabilities due to neutron emission opening (refer to Figs.~\ref{fig:Pn240*pi+} and~\ref{fig:Pn240*pi-}) are well visible on corresponding figures. Since the entrance population was chosen uniform, neutron emission probability magnitude and energy threshold differences are solely due to $l$ relative orbital momentum dependency from one side and the competition with fission and $\gamma$ decay channels from the other side. The former cause is ruled by Eq.~\ref{tnn} supplemented by our hypothesis of $S_l$ value even-odd dependence. The latter cause explains larger neutron emission probability magnitudes encountered both for $0^-$ (Fig.~\ref{fig:Pn240*pi-}) and $1^+$ (Fig.~\ref{fig:Pn240*pi+}) since corresponding fission decay probabilities at $S_n$ remain much lower than other individual decay probabilities. Finally from pictures above, it is hard to conclude that WE limit is ever reached as excitation energy increases since fission and neutron emission decay probabilities still exhibit large spreading and $\gamma$ decay probabilities collapse. 

%to be ($\mathcal{P}_{surr,\gamma}^{A^*}<0.1$)
%be  encountered values ($\mathcal{P}_{surr,\gamma}^{A^*}<0.1$).
%whereas this issue for  $\gamma$ decay cannot be evaluated because of too small encountered values ($\mathcal{P}_{surr,\gamma}^{A^*}<0.1$).
%To conclude this paragraph regarding fissile compound nuclei, we might bring to present debate that as general trend from various reaction decay probability plots for the highest excitation energy tackled in this paper (8.5 MeV meaning 2.1 MeV above $S_n$), we do not yet verify the validity of the WE assumption of reaction decay probability spin-parity independence since fission and neutron emission decay probability magnitudes still show serious spreading whereas this issue for compound nucleus $\gamma$ decay can not be evaluated because of small calculated values .\\ 

\paragraph{Fertile nuclides} 

\begin{figure}[ht]
\center{\vspace{0.4cm}
\resizebox{0.95\columnwidth}{!}{
\includegraphics[height=5cm,angle=0]{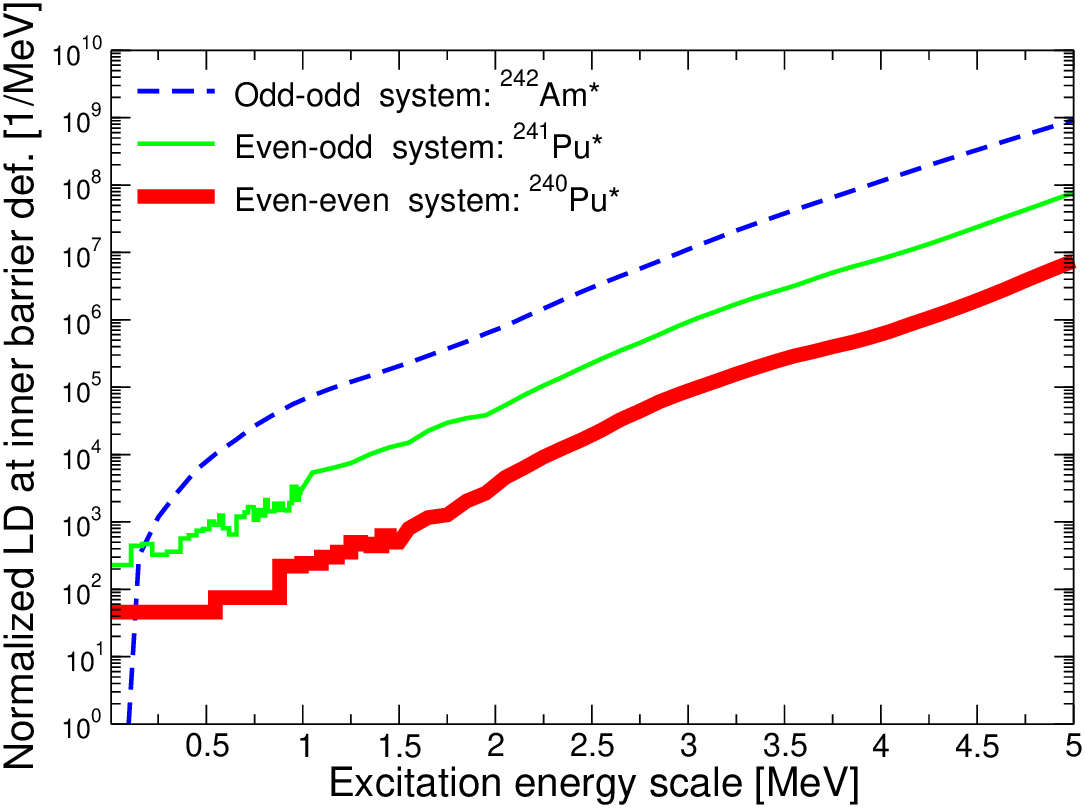}} }
\caption{\label{fig:LDeeVseoVsoo}  (Color online) Combinatorial quasi-particle-vibrational-rotational level density predictions  on top of inner barrier corresponding to even-even ($^{240}$Pu$^*$  - thick solid curve), even-odd ($^{241}$Pu$^*$ - thin solid curve) and odd-odd   ($^{242}$Am$^*$  - dotted curve) nuclei.}
\end{figure}

\begin{figure}[t]
\center{\vspace{1.cm}
\resizebox{0.95\columnwidth}{!}{
\includegraphics[height=5cm,angle=0]{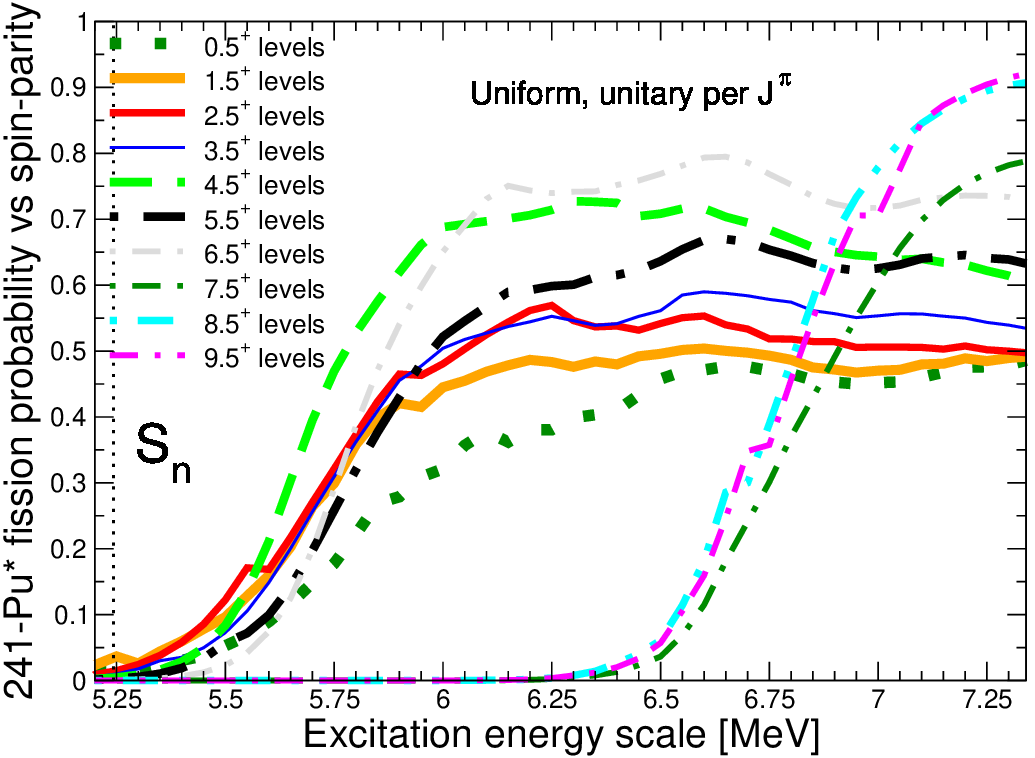}}}
\caption{(Color online)\label{fig:Pfis241*pi+} Monte Carlo $\mathcal{R}$-matrix double-barrier fission surrogate-like probabilities of $^{241}$Pu$^{*}$ as a function of resonance spin (positive parity) and excitation energy up to neutron kinetic energy of 2.1~MeV. The vertical bar at 5.24~MeV materializes neutron emission threshold.}
\end{figure}

\begin{figure}[t]
\center{\vspace{1.cm}
\resizebox{0.95\columnwidth}{!}{
\includegraphics[height=5cm,angle=0]{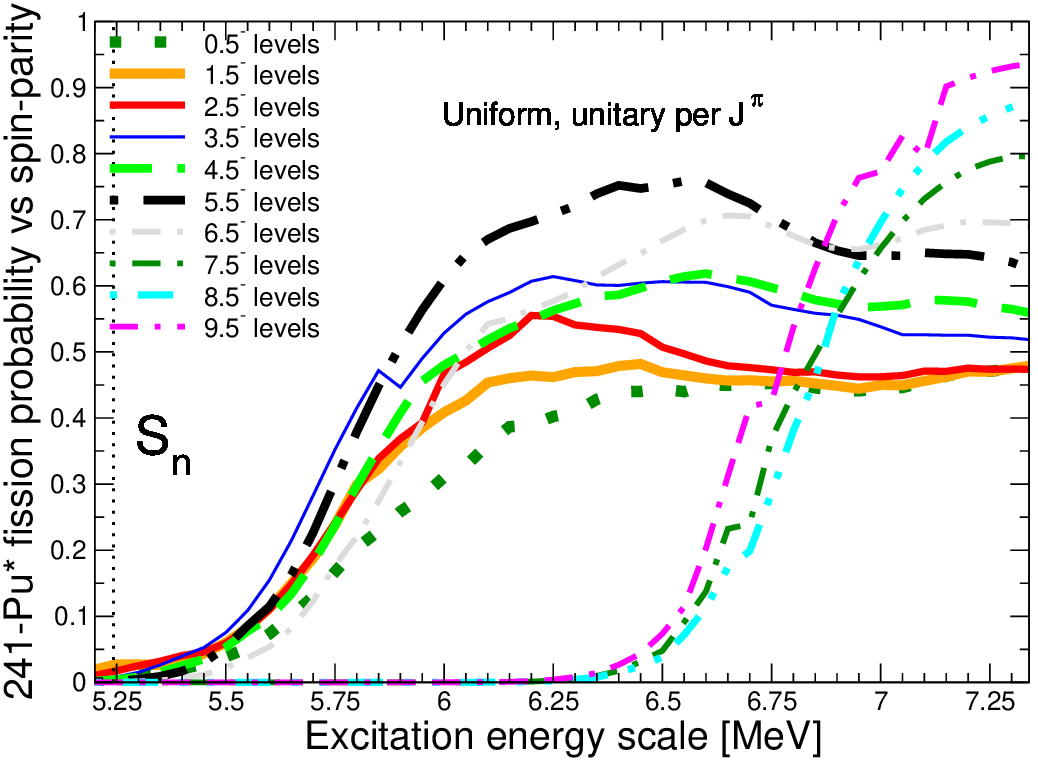}}}
\caption{(Color online)\label{fig:Pfis241*pi-} Same as Fig.~\ref{fig:Pfis241*pi+} but for negative excited  parity states.}
\end{figure}

Odd-neutron isotopes are characterized by low excitation energy combination of single-quasi-neutron states with collective vibrations. We expect fewer transition state fission spectrum oddities because of the dissimilar nature of quasi-particles as well as denser level density right above Fermi energy. Figure~\ref{fig:LDeeVseoVsoo} that illustrates combinatorial QPVR~\cite{bou:13} level densities simulated as a function of nucleus character, returns some hints on WE assumption plausibility. 
%about possible impact of the WE assumption. 
The denser the low energy spectrum is (odd-odd nuclei), the quicker it should tend towards  statistical regime where decay of the nucleus is dominated by statistical level density. Within such regime, 
% to be expected at high excitation energy, 
%spin-parity of the PDEN state prior to fission 
spin-parity features are likely of less importance because of tens of fully-opened transition states for given spin-parity that fission barrier tunneling strengths balance each other (e.g; more than 250 bandheads are counted within [0-1] MeV range above inner barrier according to odd-N fissioning nuclei). %Therefore according to odd-N  configuration context, the trend carried by respective 
%plots of surrogate-like MC partial fission probabilities for transition states of positive (Fig.~\ref
%According to odd-N nuclei, individual positive (Fig.~\ref{fig:Pfis241*pi+}) and negative (Fig.~\ref%{fig:Pfis241*pi-}) parities probabilities, does not come as a surprise since large number of transition states are involved within [0-1] MeV range above fundamental barriers (e.g; counting more than 250 bandheads at inner deformation) and where obviously no special $K=0^-$ bandhead pattern shows up. 
%ruled by Eq.~\ref{kvalues} applies. 
Little parity discrimination (positive on Fig.~\ref{fig:Pfis241*pi+} and negative on Fig.~\ref{fig:Pfis241*pi-}) is observed with only two groups of fission probabilities corresponding respectively to discrete and continuum ($J>6.5\hbar$) energy ranges. Full analysis of the picture commands to explore also $\gamma-$ and neutron-decay probabilities shape patterns. 
%Now,  what can we observe and expect for $\gamma-$ and neutron decay probabilities as a function of excitation energy according to those input fission probability features ?\\

\begin{figure}[t]
\center{\vspace{1.cm}
\resizebox{0.95\columnwidth}{!}{
\includegraphics[height=5cm,angle=0]{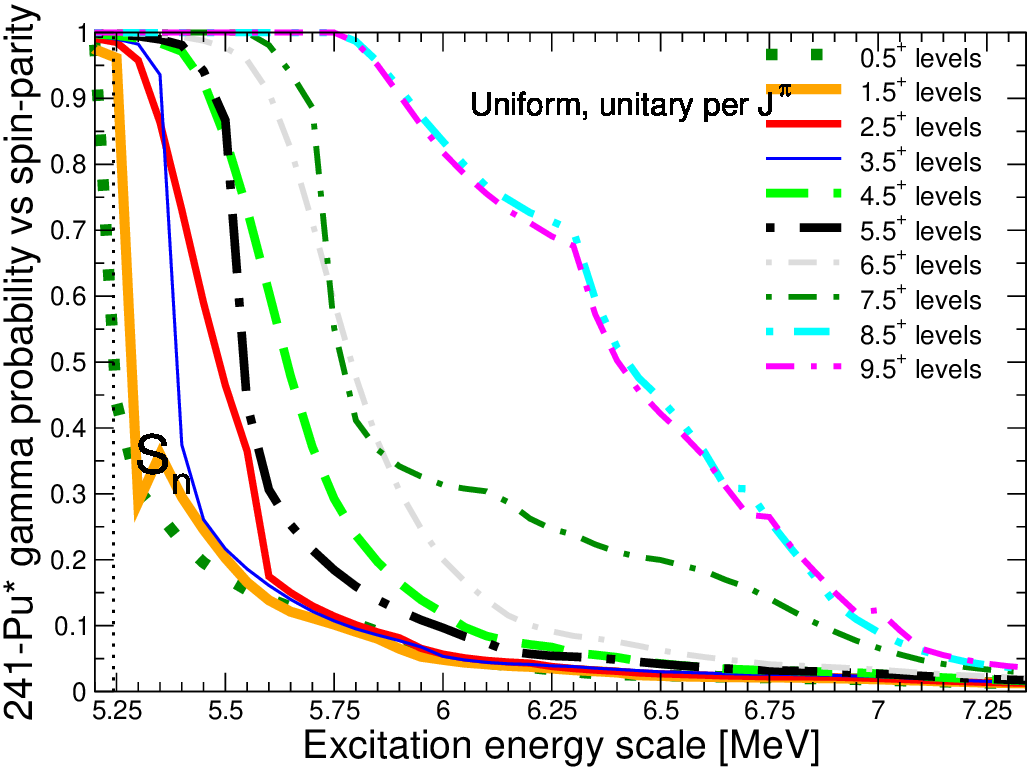}}}
\caption{(Color online) \label{fig:Pgamma241*pi+} Monte Carlo $\mathcal{R}$-matrix double-barrier $\gamma$ decay surrogate-like probabilities of $^{241}$Pu$^{*}$ as a function of resonance spin (positive parity) and excitation energy up to neutron kinetic energy of 2.1~MeV. The vertical bar at 5.24~MeV materializes neutron emission threshold.}
\end{figure}

\begin{figure}[t]
\center{\vspace{1.cm}
\resizebox{0.95\columnwidth}{!}{
\includegraphics[height=5cm,angle=0]{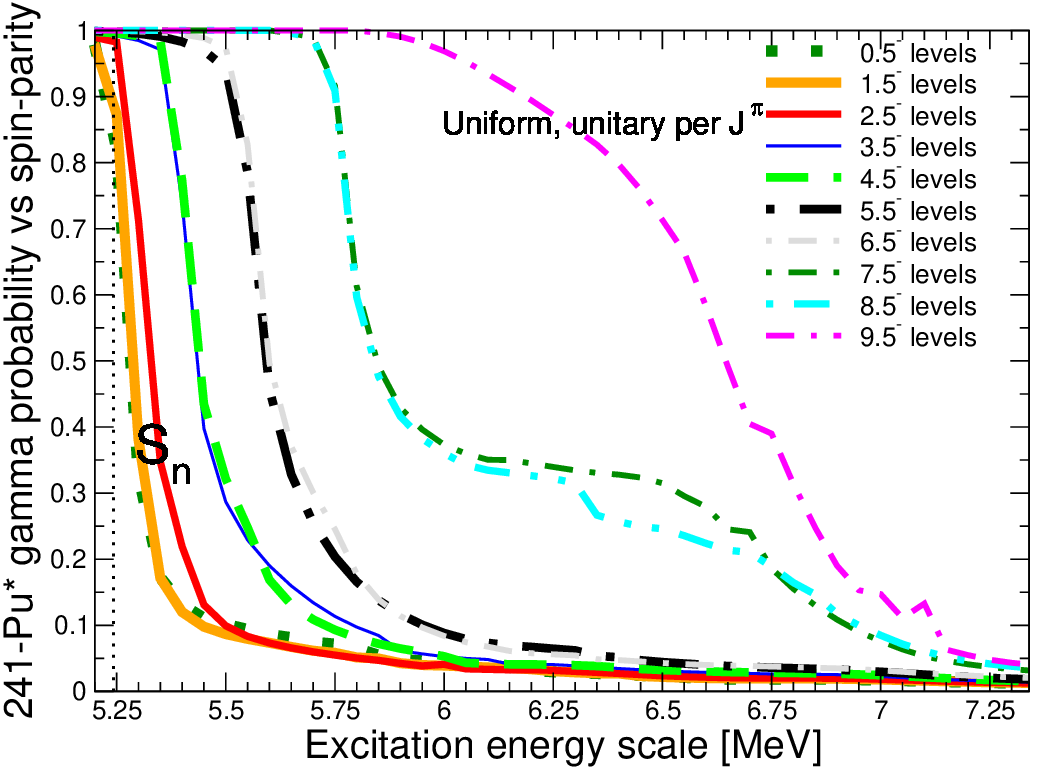}}}
\caption{(Color online) \label{fig:Pgamma241*pi-}Same as Fig.~\ref{fig:Pgamma241*pi+} but for negative parity excited states.}
\end{figure}

\begin{figure}[t]
\center{\vspace{1.cm}
\resizebox{0.95\columnwidth}{!}{
\includegraphics[height=5cm,angle=0]{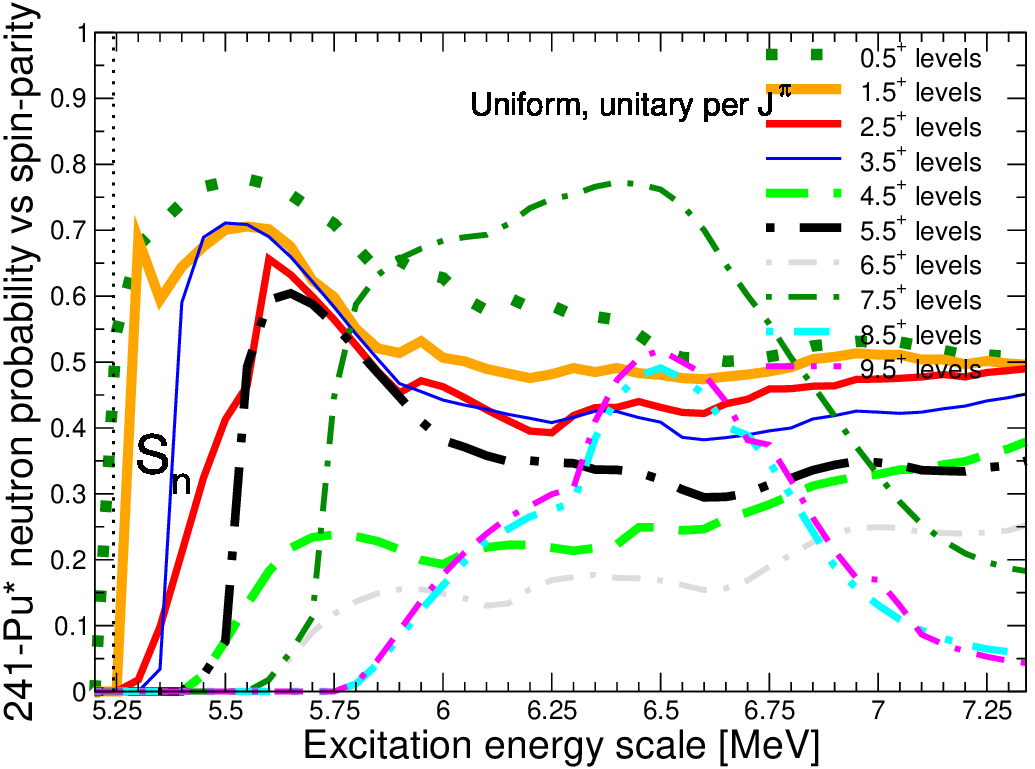}}}
\caption{(Color online) \label{fig:Pn241*pi+} Monte Carlo $\mathcal{R}$-matrix double-barrier neutron emission surrogate-like probabilities of $^{241}$Pu$^{*}$ as a function of resonance spin (positive parity) and excitation energy up to neutron kinetic energy of 2.1~MeV. The vertical bar at 5.24~MeV materializes neutron emission threshold.}
\end{figure}

\begin{figure}[t]
\center{\vspace{1.cm}
\resizebox{0.95\columnwidth}{!}{
\includegraphics[height=5cm,angle=0]{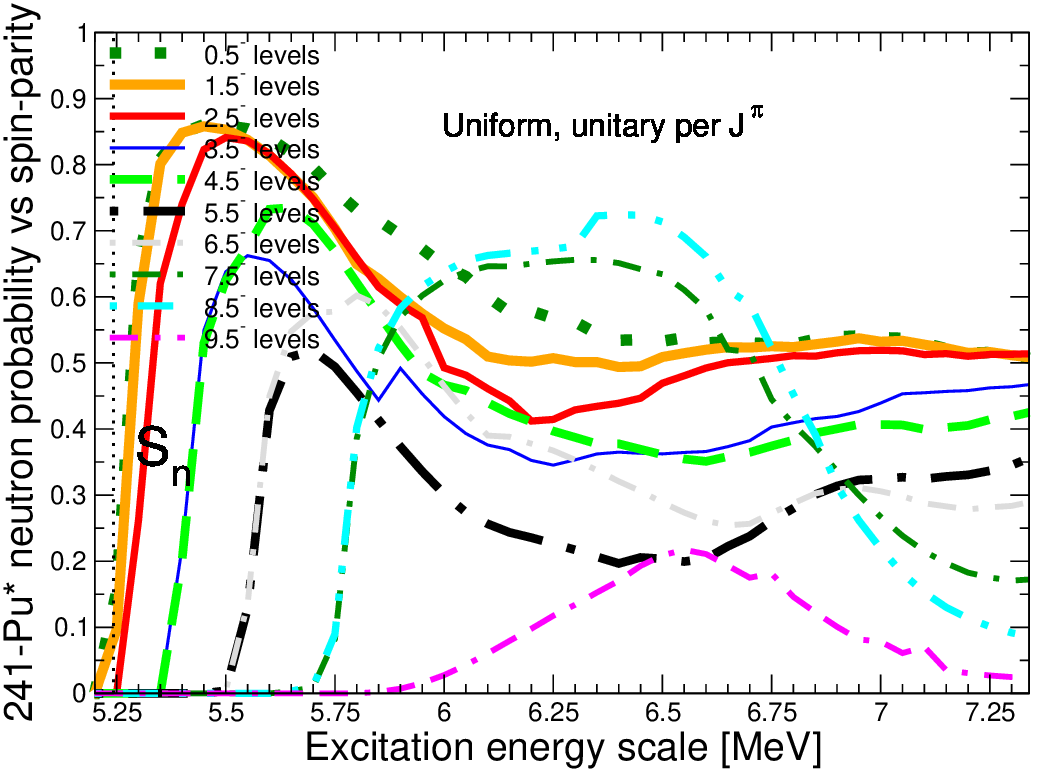}}}
\caption{(Color online) \label{fig:Pn241*pi-} Same as Fig.~\ref{fig:Pn241*pi+} but for negative parity excited states.}
\end{figure}

 We remember that the neutron emission width is related to the reduced width amplitude such that $\Gamma_{n}^{1/2} \equiv \gamma_{n}\sqrt{2P_l}$ with $P_l$, the relative orbital angular momentum dependent centrifugal penetrability. In compliance with Eq.~\ref{eq:jj12}, we address $\overrightarrow J_{A+1}= \overrightarrow I_{A} +  ( \overrightarrow i_{n} + \overrightarrow l_{n})$. Applied  to even-even residual nucleus ($I_{A}=0$), high $J_{A+1}$ can be reached only when $l_n$ increases. Since the centrifugal barrier penetrability decreases as $l_n$ increases, neutron emission is blocked at low energy for large $l_n$ values . This well-known allegation is verified on Figs.~\ref{fig:Pn241*pi+} and ~\ref{fig:Pn241*pi-}. %According to the odd-N nuclide category over present analysis excitation energy range, fission decay, $\gamma$ decay and neutron emission compete with each other. However for 
For high spins, fission decay probabilities are negligible below $6.5$~MeV restricting the problem to a $\gamma$ and neutron emission dual decay competition. In this configuration, the  $\gamma$ decay probability is then exactly the reciprocal of the neutron emission probability (clearly manifest when comparing Figs.~\ref{fig:Pgamma241*pi+} and ~\ref{fig:Pn241*pi+} and,  Figs.~\ref{fig:Pgamma241*pi-} and ~\ref{fig:Pn241*pi-}); which acts as the driver of the $\gamma$ decay channels. When both neutron emission and fission coexist, fission competes strongly with neutron emission as illustrated by the $J^\pi=9.5^+$ curves on Figs.~\ref{fig:Pfis241*pi+} and ~\ref{fig:Pn241*pi+}. Indeed neutron emission reaches its maximum when continuum fission opens at $6.5$~MeV while the $\gamma$ drop is still reinforced.\\
 
%We can summarize this section by the following arguments: 
At final, $\gamma$ decay is much driven by fission and/or neutron emission; this can be justified by the very low sensitivity of the $\gamma$ decay width to the initial state spin-parity. % At those rather high excitation energies ($>5$~MeV), 
Indeed at high excitation energies ($>5$~MeV), level density for the nucleus in the ground state is quite large 
%at normal nucleus potential deformation is quite large 
and there is somehow an equiprobability for $\gamma$ cascade from any spin-parity state so that the $\gamma$ decay probability is being adjusted to external constrains. On the contrary, at about $S_n$ excitation energy neutron emission is absolutely ruled by the centrifugal barrier penetrability %, $l$ dependent,
 while fission is driven essentially by barrier tunneling at large nucleus deformation following Aage Bohr~\cite{boh:53} postulate. 
 \newpage
%Aage Bohr~\cite{boh:53} suggested that a fissioning nucleus, although highly excited, remains cold for excitation energies slightly above saddle ground state because almost the entire excitation energy is tied up by a single degree of freedom (i.e; in terms of potential energy of deformation) leaving at the saddle point a relatively small fraction of energy to share between the other degrees of freedom (kinetic energy of deformation, rotation and vibrations in orthogonal modes, quasi-particle excitations). Then, paradoxically to the very numerous fission fragments emitted and the subsequent very large number of channels expected in fission reaction, we observe an excitation spectrum on top of fission barrier showing level density as low as the one at ground state.\\
\subsection{\label{ss:third}Monte Carlo simulation of \\ surrogate-like experimental probability}    

\begin{figure}[t]
\center{\vspace{1.cm}
\resizebox{0.95\columnwidth}{!}{
\includegraphics[height=5cm,angle=0]{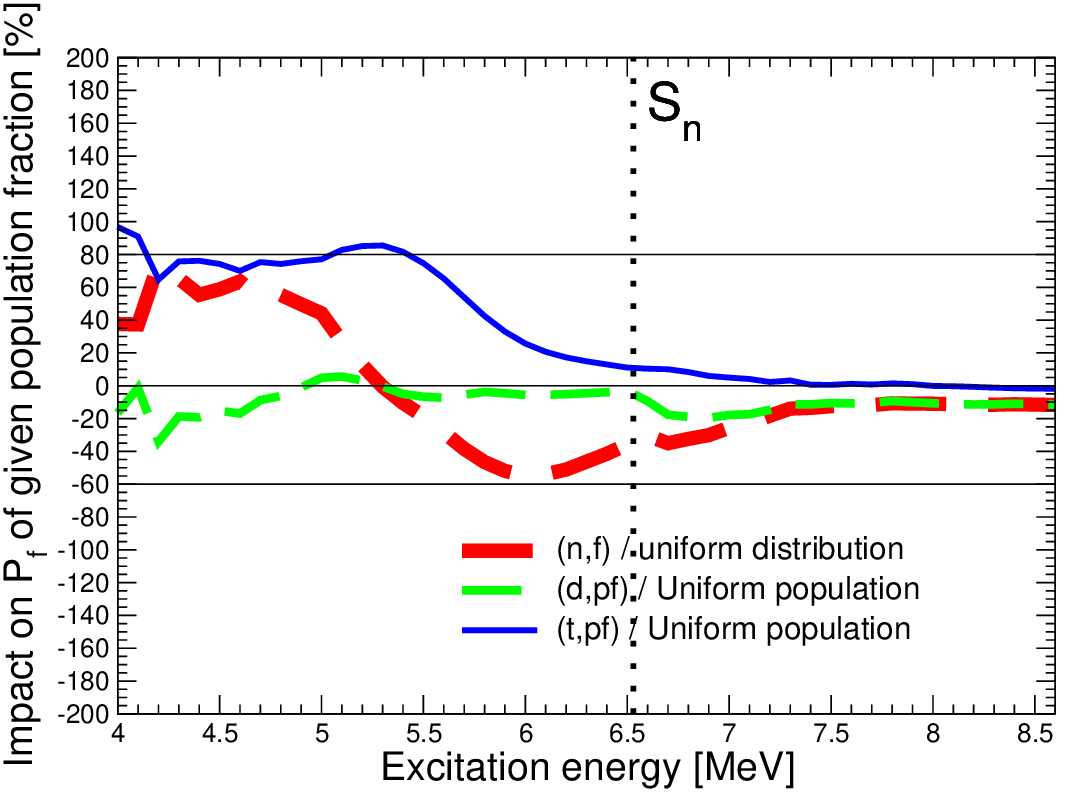}}}
\caption{(Color online)\label{fig:spDist242*} Forseen impact [\%] on $^{240}$Pu$^*$ fission  decay surrogate-like probabilities of $(n,f)$, $(d,p)$ or $(t,p)$ spin-parity distribution relatively to uniform unitary distribution on the ground of Eq.~\ref{eq:unit}. $(d,p)$ and $(t,p)$ distribution information is extracted from Refs.~\cite{and:70} and~\cite{bac:74} respectively whereas neutron-induced fractions are generated by the $\mathcal{LNG}$ code. Neutron-induced distribution regarding excitation energies lower than $S_n$ correspond to the one  at $S_n$ that supports mainly $0^+$ and $1^+$  excited levels by $s$-waves ($24\%$ and $73\%$ respectively of total fraction).}
\label{fig:Pu240*ImpactFjpiPf}
\end{figure}

\begin{figure}[t]
\center{\vspace{1.cm}
\resizebox{0.95\columnwidth}{!}{
\includegraphics[height=5cm,angle=0]{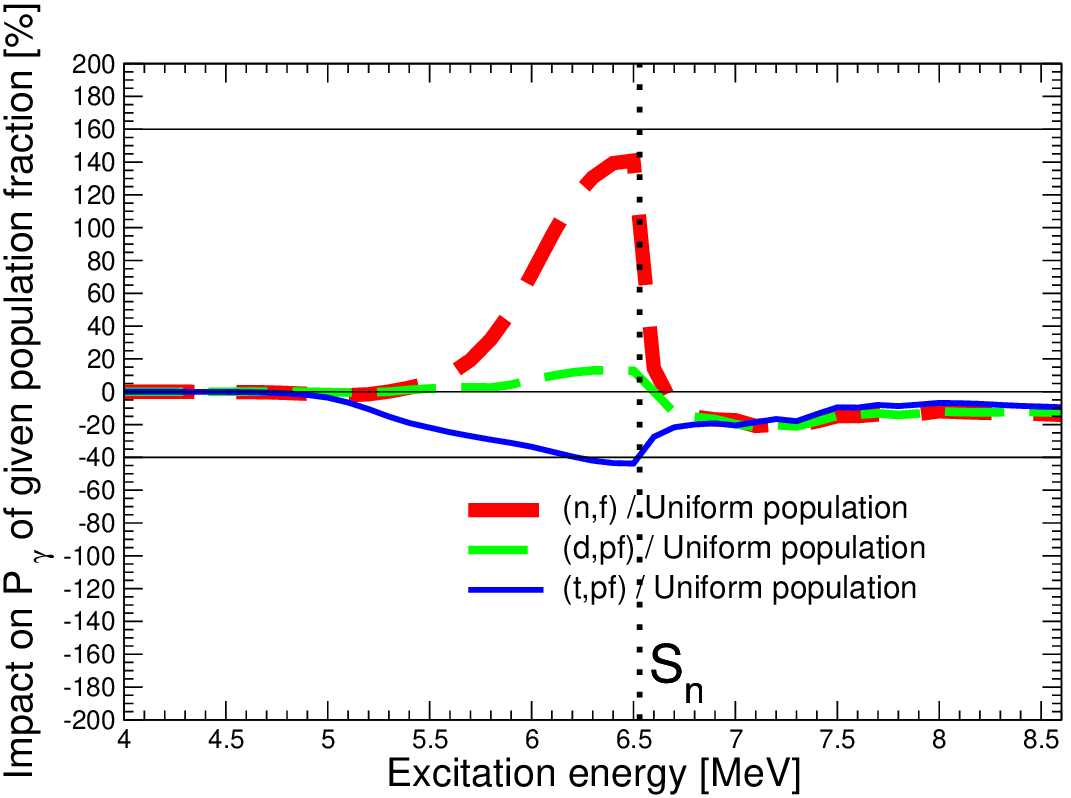}}}
\caption{(Color online)\label{fig:spDist242*} Same as Fig.~\ref{fig:Pu240*ImpactFjpiPf} but according to  total $\gamma$ decay probabilities in $^{240}$Pu$^*$.}
\label{fig:Pu240*ImpactFjpiPg}
\end{figure}

%We remember from the surrogate-like probability analytical formulation (Eq.~\ref{eq:Pdecoupled}) that 
We remember from the analytical formulation of the probability (Eq.~\ref{eq:Pdecoupled}) that
three main ingredients have to be weighted to evaluate the validity of the SRM (Eq.~\ref{eq:hfstdsp}) for neutron-induced cross section reconstruction. In sections above, two out of three ingredients have been examined leaving the consequence of nonuniform PDEN population across $\mathcal{F}_{surr}^{A^*}(E_{x},J,{\pi})$ to be addressed. We have recalled in section~\ref{ss:Compound} that neutron-induced reaction and direct reaction support distinct total angular momentum population distributions. The former $J^\pi$ distribution profile is sharp and centered about $J_{A+1}\approx I_{A}$ (Fig.~\ref{fig:U236*JpiDist}) at low neutron energy~\footnote{We realize that neutron-induced distributions are centered either about values of low spin (case of even-even targets for which $J_{A+1} \approx 0\hbar$)  or about high spin values when  the intrinsic spin carried by the target is high (case of a $^{235}U$ neutron target with $J_{A+1} \approx 3.5\hbar$).} whereas direct reaction distributions are rather broad and centered about high total angular momentum (Fig.~\ref{fig:Pu240*JpiDist_all}) since the transferred relative angular momentum is at least of 3 units. From this general trend, must be distinguished the peculiar case of $(t,p)$ reactions on even-even target nuclei that verify rather single parity per $J$ given (Fig.~\ref{fig:Pu240*JpiDist_tp}). To quantify the error carried by the insecure use of neutron-induced distribution during the retrieval procedure from surrogate experimental data (on the ground of Eq.~\ref{eq:hfstdsp}), we have carried the surrogate-like simulation all way long following Eq.~\ref{eq:Pmc} for three typical spin-parity population distributions according to $^{240}Pu^*$ fissioning nucleus. We expect this nucleus to supply  representative impact of the approximations since it covers excitation energies below $S_n$, excited states preferentially reached with low energy neutrons (close to zero $\hbar$) and discrete transition state spectra on top of barriers that are sparse and built from collective DoF motions only. %In order to quantify the only impact of given spin-parity population distribution across the full MC equation, 
For best illustration, results are benchmarked against uniform distribution, now unitary on ground of the Eq.~\ref{eq:unit}. Although feeding neutron-induced population fractions for excitation energies lower than $S_n$ is meaningless, calculations are made according to distribution at $S_n$ for the purpose of our review. We recall that present debate concerns clearer picture of SRM for fission and $\gamma$ decay data as retrieval procedure to neutron-induced cross section prediction for which it was pointed out that it matches in terms of fission~\cite{kes:10} as good as it is poor in terms of capture reaction~\cite{bout:12,duc:16}. Figs.~\ref{fig:Pu240*ImpactFjpiPf} and~\ref{fig:Pu240*ImpactFjpiPg} are then respectively drawn according to fission and $\gamma$ decay probabilities. We immediately realize that the broadest  entrance distribution, $(d,p)$, returns  logically results close to the reference. %As expected, results show opposite trends in terms of fission and capture below $S_n$ because of total flux conservation across the Hauser-Feshbach equation. 
The peculiar $(t,p)$ population distribution impacts significantly the low excitation energy fission probability with a deviation from the reference (up to $80\%$). It happens that using a $(n,f)$ population rather than a $(d,p)$ population generates maximum absolute error of $60~\%$ below $S_n$ in terms of fission and $160~\%$ in terms of $\gamma$ decay. However above $S_n$, swapping  distributions brings less error than we could have ascribed according to this ingredient ($<40~\%$). 

\section{\label{s:Vibrational}Simulation of surrogate-like probabilities in the region of $\beta$-vibrational resonances}  

For fission probability measurements on fissile isotopes, it is worth to envision occurrence of Vibrational Resonance Structures ($VRS$) at excitations lower than neutron separation energy. Indeed, even with moderately good resolution (25-50~keV), the observation of well-defined peaks at sub-barrier energies % 
%Indeed, the many cases of non-statistical fluctuations observed, even with moderate resolution (25-50~keV), in fission probabilities
 since the end of the sixties that were readily assigned to collective $\beta$-vibrations in second well of fission barrier, suggest the dedication of the Hamiltonian of the compound nuclear system to vibrations along the fissioning nucleus elongation axis (i.e; according to prolate shape vibrations). This reads~\cite{bou:13},
\begin{equation} 
H=H_\beta+H_{int}(\zeta,\beta_{barrier})+H_{c}(\beta, \zeta; \beta_{barrier})\label{eq:fisRmatbeta} \mbox{ }\mbox{ }\mbox{ }\mbox{ }\mbox{ .}
\end{equation}
with $H_{int}$, the intrinsic Hamiltonian and $H_{c}$ the interaction between the  $\beta$-mode and other modes of excitation (collective  and single-particle types).  Literature on experimental fission data reports several types of situations from which two of them reveal a pattern of vibrational structure. The most convincing case is related to the rare observation of resonance of essentially pure vibrational nature as illustrated by the fission cross section measurement of $^{230}$Th around  $720$~keV neutron energy. %Report page828 , column 2 and page808 Lynn80; R-6901 p7). 
This ideal setting is related to specific conditions since we know that the $\bra\lambda_{II}^{(q)}|H_{c}|{\mu}{\nu_{II}}\ket$   matrix elements are responsible for damping of a $\beta$-vibrational state $\nu_{II}$ into the quasi $(q)$ complete set of class-II compound nucleus  states $\lambda_{II}^{(q)}$. In other words, we expect class-II compound nucleus $R$-matrix internal states, expressed as the incomplete expansions in terms of vibrational wave functions localized in the secondary well $X_{\lambda_{II}}^{(II)}=\sum_{\mu\mbox{}\nu}{C^{\lambda_{II}}_{\mu \nu}}\chi_\mu\Phi_{\nu{(\mu)}}^{(II)}$, to restrict themselves to nearly a single class-II state contribution,
\begin{eqnarray} 
X_{\lambda_{II}}^{(II)}&\approx&\chi_0\Phi_{\nu{(0)}}^{(II)} \label{eq:purevib} \mbox{ }\mbox{ with eigenvalue, }\\
E_{\lambda_{II}}&\approx&  \varepsilon_{\nu_{II}} + \epsilon_0 \label{eq:Epurevib} \mbox{ }\mbox{ }\mbox{ }\mbox{ }\mbox{ ,}
\end{eqnarray}
where $\chi_0$ and $\Phi_{\nu{(0)}}^{(II)}$ are respectively the eigenfunctions of $H_{int}(\zeta,\beta_{barrier})$ and  $H_\beta$ and, $\epsilon_0$ and $\varepsilon_{\nu_{II}}$ the associated eigenvalues.
Absence of vibrational damping in second well suggests the existence of a shallow well such that the vibrational excited state lies either within the energy gap for even-even fissioning nuclei or very close to the fission isomeric ground state in odd-A and odd nuclei. According to this, the strength function of the class-II ($\beta$-vibrational) state coupling $directly$ to the class-I compound nucleus states exhibits a resonance peak in the $coupling$ strength function through a specific transition state, $\alpha$,  over the inner barrier. Simultaneously the class-II $fission$ strength function, related to a specific outer barrier transition state, $\mu$, might show a resonance as well; both resonance energies built according to the $\beta$-vibrational state and intrinsic barrier state energies. Assuming Hill-Wheeler~\cite{hil:53} barrier   penetrabilities across inner and outer saddles, namely $P_A$ and $P_B$, supported by Wigner statistical method~\cite{wig:38}, we define the corresponding coupling $\downarrow$ and fission $\uparrow$ widths of an {\it idealized pure} vibrational resonance such that%%Report page808 , 5.9 and 5.10)
% AVXSF utilise des valeurs hwII differente suivant le couplage avec  barA ou barB --> PRC97 Eq 10/11 
\begin{equation} 
\Gamma_{vib_{II\mbox{ }\downarrow}} =\frac{\hbar\omega_{II}}{2\pi}P_A \mbox{ and } \Gamma_{vib_{II\mbox{ }\uparrow}} =\frac{\hbar\omega_{II}}{2\pi}P_B\mbox{ ,}
\label{eq:Gvibs}
\end{equation}  
with $\hbar\omega_{II}$, the quantum of vibrational energy in second well. Occurrence of substructures over an IS resonance due to the strength fragmentation into  more complex class-II excitations for this 'pure' vibrational case, is not expected. Origin of observed substructures are then associated to the %unavoidable 
existence of a rotational band built on the intrinsic vibrational state $\mu\nu_{II}$.

 %See notes Eric Treatment of beta-vibrational resonances
However in most VRS observations, we face to a relatively deep secondary well so that the vibrational configuration corresponds to excitation energy above the energy gap sitting on top of the zero-point vibration energy such that the $\beta$-vibration fission strength be spread at least weakly or moderately among a pretty dense class-II compound nucleus state resonance structure. This second configuration, which still manifests a characteristic peak in the fission strength function but broader, is described as a $damped$ vibrational resonance; which damping strength is characterized by its damping width $\Gamma_{vib_{II\mbox{ }D}}$.  %See aere R6901 page 6 
In the extreme limit where the secondary well becomes very deep, the principal fission mode $\chi_0\Phi_{\nu{(0)}}^{(II)}$ across the outer barrier saddle is strongly spread over the class-II CN states with consequence of observed monotonically rising fission cross section after coupling of these states  with the class-I states assuming standard experimental resolution.\\

To cope with the modeling of the various vibrational resonance configurations, we supplement our computer code (presented in Section II-C of Ref.~\cite{bou:13}) with an option putting a sequence~\footnote{In present simulation, the number of of resonance terms is set to 20  to guaranty a negligible residual transmission coefficient.} of resonance terms in the transmission function for either or both inner and outer saddles (respectively $T_A(\alpha)$ and $T_B(\mu)$). In the following we assume no overlapping of class-II resonances (i,e; allowing the use of narrow resonance approximation) that constitute the damped vibrational resonance. According to uniform (picket-fence) model approximation, the distribution of the strength of a $single$ vibrational state among the class-II states will borrow the usual form of a Lorentzian profile such that the coupling  width of a class-II state ($\lambda_{II}$) across an inner barrier transition state $\alpha$, with the hypothesis $\Gamma_{\lambda_{II\mbox{ }\downarrow}}\gg\Gamma_{\lambda_{II\mbox{ }\uparrow}}$, be equal to \\ %See Lynn 1980 Eq 5.78 p830 and p829 last paragraph. See also Lynn 1980 Eq 5.51 p823. See also Wagemans page 90, eq (89)
\begin{eqnarray} 
\Gamma_{\lambda_{II\mbox{ }\downarrow}}(\alpha)  = \frac{D_{II}}{2\pi} \frac{\Gamma_{vib_{II\mbox{ }\downarrow}}(\alpha) \mbox{ . } \Gamma_{vib_{II\mbox{ }D}}(\alpha)}{\small{( \epsilon_{0(\alpha)} +
E_{vib_{II}(\alpha)}-E_{\lambda_{II}})^2+\Gamma^2_{\lambda_{vib_{II\mbox{ }D}}}(\alpha)/4 }} \mbox{ .}\mbox{ }\mbox{ }\mbox{ }
\label{eq:lorentGc}
\end{eqnarray}
The class-II fission widths, $\Gamma_{\lambda_{II\mbox{ }\uparrow}}$, across given outer barrier transition state, labeled  $\mu$, are obtained by a similar formula. Both formulae assume that $\Gamma_{vib_{II\mbox{ }\uparrow}}$ and $\Gamma_{vib_{II\mbox{ }\downarrow}}$ add a negligible amount to the damping of the vibrational state. By contrast when the vibrational state coupling width to the fission continuum is  stronger than the vibrational coupling width across the inner barrier,  an extra spreading to the vibrational state must be considered in the derivation of the class-II compound state fission width. This can be reinterpreted as a sequential process where first a mixing of the vibrational state with the fission continuum is made and then with the class-II compound states. Latter statement leads to  
\begin{eqnarray} 
\Gamma_{\lambda_{II\mbox{ }\uparrow}}(\mu)  = \frac{D_{II}}{2\pi} \frac{\Gamma_{vib_{II\mbox{ } \uparrow}}(\mu) \mbox{ . } \Delta_{vib_{II}}(\mu)}{(\small{ \epsilon_{0(\mu)} + E_{vib_{II}(\mu)}-E_{\lambda_{II}})^2+\Delta_{vib_{II}}^2(\mu)/4 }} \mbox{ ,}\mbox{ }\mbox{ }\mbox{ }
\label{eq:lorentGf}
\end{eqnarray}
with $\Delta_{vib_{II}}\simeq \Gamma_{vib_{II\mbox{ }D}}+\Gamma_{vib_{II\uparrow}}\mbox{ .}$
The above result relies on $S$-matrix theory~\cite{lyn:68} that considers the continuum in its foundation and resonance occurrence as a singularity of the collision matrix when extended to the complex energy plane (i.e; the imaginary component of the singularity supplies the width of the resonance). By supplying an option to manage resonance peaks in the class-II fission width (Eq.~\ref{eq:lorentGf}), we acknowledge the existence of $\beta$-vibrational levels in a possible tertiary well which expected shallow minimum suggests damped or pure class-III vibrational state features. Recent work testing tertiary well impact has been presented in~\cite{bou:18}.\\

Obviously Eqs.~\ref{eq:lorentGc} and ~\ref{eq:lorentGf} do not restrain themselves to a single vibrational state  (neither to the lowest intrinsic state at saddle, $\epsilon_0$) but a sequence that assumes for instance harmonic oscillator modeling.  Right-hand sides of Eqs.~\ref{eq:lorentGc} and ~\ref{eq:lorentGf} must then be substituted by a sum of $n$ transmission coefficients according to each vibrational resonance $i$ with $E_{vib_{i}(\mu)}$, the corresponding energy as $E_{vib_{i}(\mu)} = E_{vib_{1}(\mu)}+(i-1)\hbar\omega_{vib_{II}(\mu)}$. We assume here exponential dependence of the vibrational state damping width that is calibrated on the value of the width, labeled $\Gamma_{0(\mu)}$, expected at the energy corresponding to the second well ground state. This reads
% '{\it effective} ground state'. 
% Eric assigned finally to the isomer (2.95 as EII) The energy of the latter is usually chosen much above actual second well zero-point vibration energy, as for instance the lowest observed giant resonance, to insure validity of the damping width single energy dependence model.  This reads
\begin{eqnarray} 
\Gamma_{vib_{II\mbox{ }D}}(E)= \Gamma_{0(\mu)}\exp{[\kappa_D(E-E_{II,eff})]}\mbox{ ,}\mbox{ }\mbox{ }\mbox{ }
\label{eq:Gdvariation}
\end{eqnarray}
with $\kappa_D$, the vibrational damping constant.\\

Likewise class-I and -II coupling modes treated by Monte Carlo simulation in our $\mathcal{LNG}$ program, we consider complete/full damping of a vibrational state regarding class-II structure above an arbitrary limit set to four vibrational state spacings; meaning $\Gamma_{vib_{II\mbox{ }D}}\geq 4\hbar\omega_{vib_{II}}$. From this cutoff, the calculation switches back to strong coupling formulae (i.e; the simple Hill-Wheeler penetrability according given transition state energy). Assuming $\Gamma_{0}=50$~keV, $E_{II,eff}=4.4$~MeV, $\kappa_D=1.5$ and $\hbar\omega_{vib_{II}}=800$~keV, full damping treatment does not occur below $7.20$~MeV excitation energy. 

%damping width of $\Gamma_{0}=50$~keV according to $4.4$~MeV 'effective' isomeric ground state and vibrational damping constant of $\kappa_D=1.5$, full damping treatment does not occur below $7.20$~MeV excitation energy (assuming $\hbar\omega_{vib_{II}}=800$~keV).

\section{\label{s:application}Surrogate-like probability simulations applied to the Pu fissile isotopes over the [4-8] MeV range}  

%In preceding sections, we have described both formulations and approximations related to the various quantities involved in the simulation of surrogate-like probabilities (Eq.~\ref{eq:Pdecoupled}). Hence 
%We are now eligible 
Above substantial framework allows confident application to the Pu fissile isotopes over the [4-8] MeV range; isotopes which  fissility character makes them adequate for subthreshold neutron emission energy investigation. Present study complements neutron-induced cross section simulations published in Ref.~\cite{bou:13}. 
%We recall that complementary fission and capture~\footnote{Erratum: we must underline that a graphic substitution has occurred in Ref.~\cite{bou:13} regarding Fig.~18. Indeed the latter is related to the $^{241}$Pu  capture cross section on the contrary to what is stated by associated text and caption in the publication. Relevant Figure is displayed in present appendix for completeness.} neutron-induced cross sections have been published in a previous Ref.~\cite{bou:13}.
Latest calculations remain consistent with initial parameter extraction except when specifically mentioned. Table~\ref{tab:surrogateData} gives the list of fissile Pu isotopes studied in this work and associated surrogate data simulated for comparison using our computer code. Present experimental data base includes stripping $(d,p)$ reactions, ($^3He,d$) and ($^3He,t$) charge-exchange reactions and, two-neutron transfer $(t,p)$ reactions only. For easier comparison with published calculations, we simply feed the  code with entrance state population quoted in literature according to the direct reaction vector unless otherwise stated.  %Details according the simulations are reviewed thereafter.      

\subsection{\label{ss:application237}$^{237}$Pu$^*$ compound nucleus}
\begin{figure}[h]
\center{\vspace{1.cm}
\resizebox{0.95\columnwidth}{!}{
\includegraphics[height=5cm,angle=0]{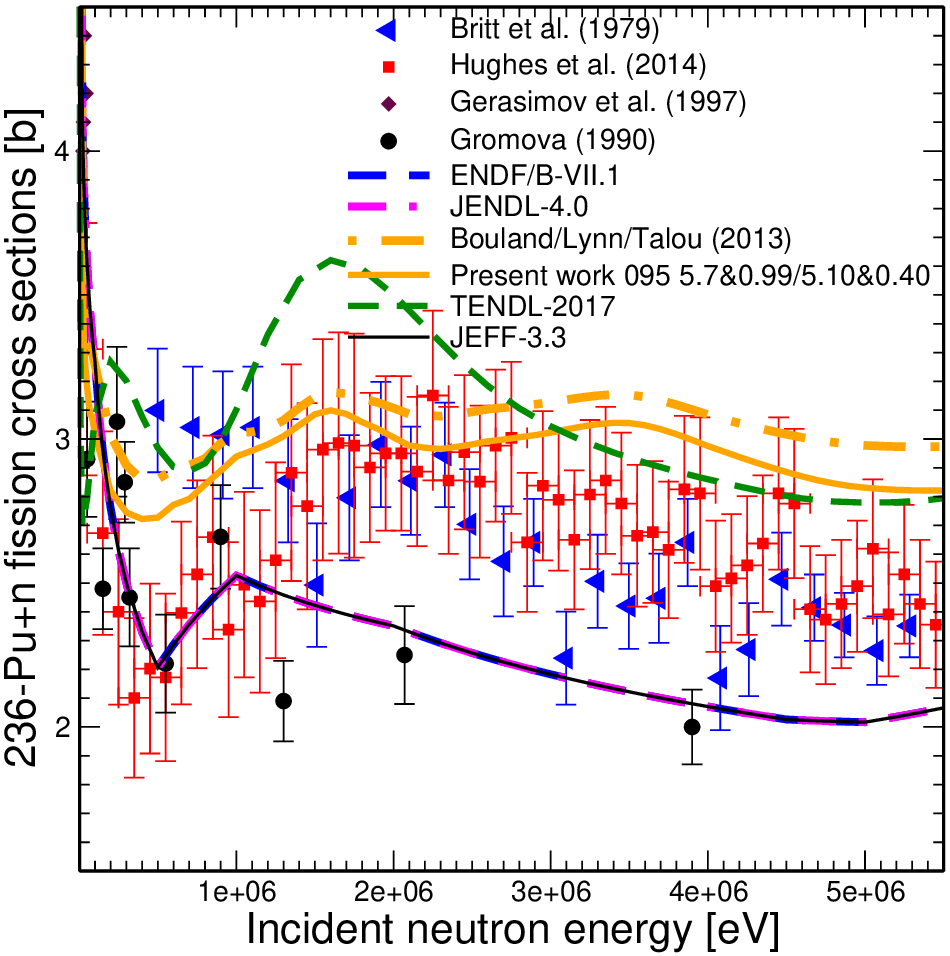}} }
\caption{\label{fig:pu237*xsfz1}(Color online) $^{236}$Pu  neutron-induced fission cross section (orange-solid curve) computed with $\mathcal{LNG}$ and compared to some evaluated data (ENDF/B-VII.1, JEFF-3.3,  JENDL-4.0 and TENDL-2017) and to either NPS or surrogate data. }
\end{figure}

\begin{figure}[h]
\center{\vspace{1.cm}
\resizebox{0.95\columnwidth}{!}{
\includegraphics[height=5cm,angle=0]{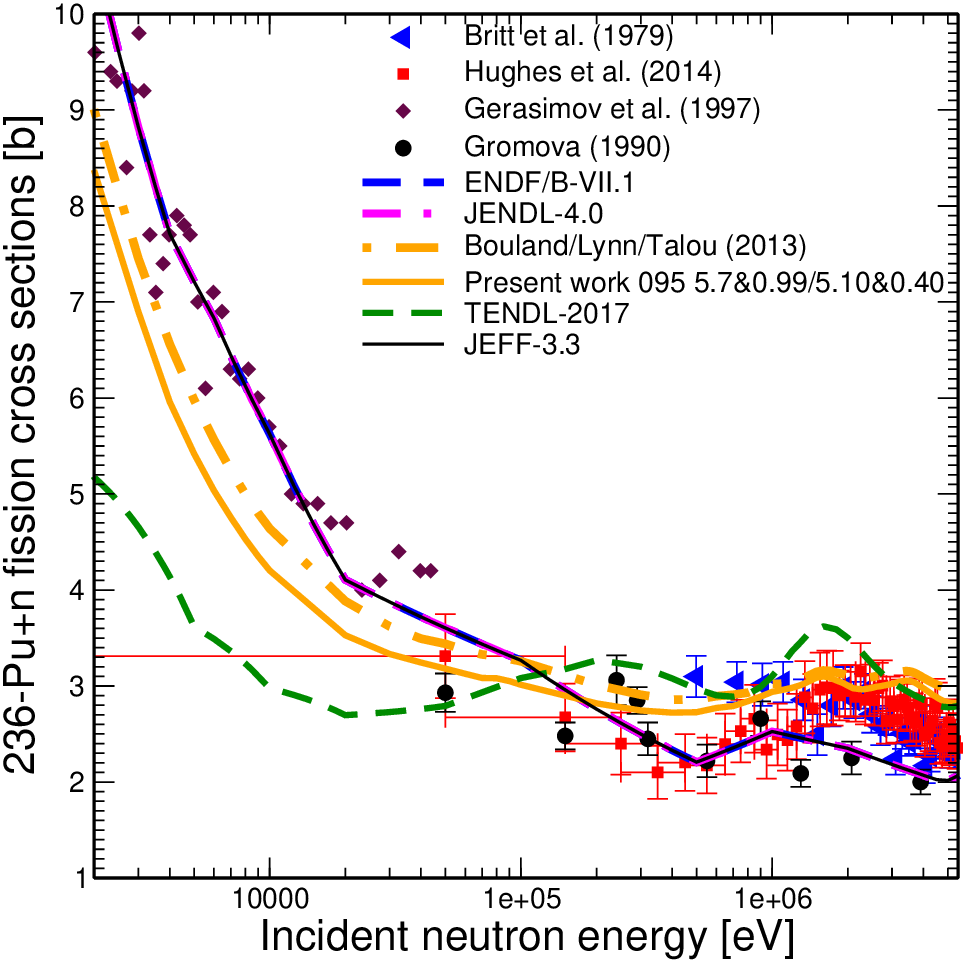}} }
\caption{\label{fig:pu237*xsf}(Color online) Same as Fig.~\ref{fig:pu237*xsf} but using a logarithmic  abscissa energy scale.}
\end{figure}

Surrogate data investigation dedicated to $^{237}$Pu$^*$ appears to be of major interest since neutron-induced experimental data of the short-lived target nuclide $^{236}$Pu ($\tau_{1/2}$=2.85y) remain very sparse. The latter are limited in the resonance region to a single differential measurement that used lead slowing-down spectrometer of inherent low energy resolution (V.F.~Gerasimov \etal~(1997) EXFOR \#41369) and at higher energy to measurements with monoenergetic neutrons (E.F.~Gromova \etal~(1990) EXFOR \#41064). Those NPS data are displayed on Figs.~\ref{fig:pu237*xsfz1} and~\ref{fig:pu237*xsf} against present simulation and recent experimental surrogate data by~Hughes \etal~(2014) (EXFOR \#14396) who used SRM. However we expect the latter $^{236}$Pu$(n,f)$ derived from ratio of cross sections $^{235}$U$(p,tf)$/ $^{239}$Pu$(p,tf)$ to be more reliable than neutron cross section extrapolation using the standard SRM since both direct reaction measurement vectors are of similar nature. Present work (orange solid curve) is closer to SRM feedback than to the evaluated cross sections (all identical except TENDL-2017) that follow above 1.5~MeV neutron energy the lower trend carried by Gromova \etal.\\

%confirms returned surrogate data feedback on evaluated files (except for TENDL-2015) to be consistently low above 1.5~MeV neutron energy (Fig.~\ref{fig:pu237*xsfz1}). Within the energy range [100~keV-1~MeV] (Fig.~\ref{fig:pu237*xsf}) the disagreement with Hughes \etal~is tangible whereas a better agreement is obtained with the neutron cross section extrapolation (EXFOR \#14229.030) proposed in 1979 by Britt \etal~\cite{bri:79} from $^{237}$Np($^{3}$He,$tf$) surrogate data  that includes nonetheless a straightforward use of the WE approximation.\\ 

\begin{figure}[t]
\center{\vspace{1.cm}
\resizebox{0.95\columnwidth}{!}{
\includegraphics[height=5cm,angle=0]{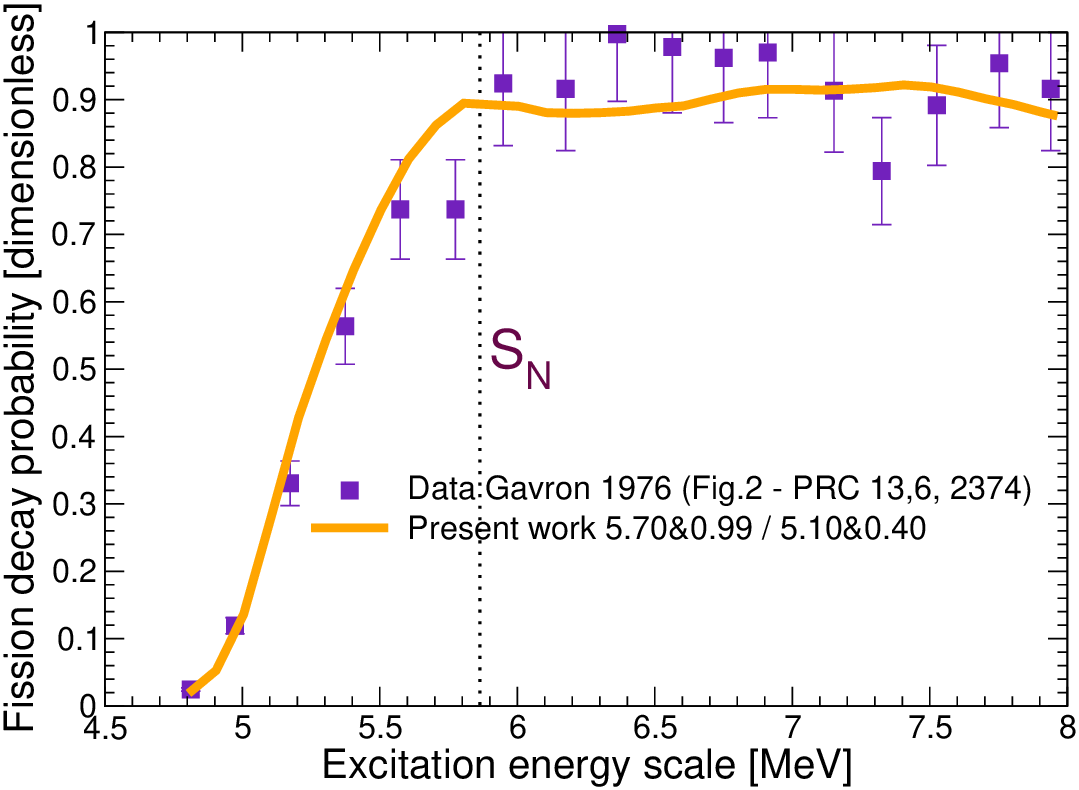}} }
\caption{\label{fig:pu237*pf}(Color online) $^{237}$Pu$^*$  fission surrogate-like probabilities (orange-solid curve) computed with $\mathcal{LNG}$ and compared with $^{237}$Np($^{3}$He,$tf$) surrogate data from  A. Gavron \etal~\cite{gav:76}.}
\end{figure}

Using our code, information on fission barrier heights can be actually obtained from the $^{237}$Np($^{3}$He,$tf$) surrogate measurement by  A.~Gavron~\etal~\cite{gav:76}. Quoting the authors, typical systematic experimental errors in the determination of the absolute fission probabilities are less than 10\%. Unfortunately, no information is available concerning accuracy of the $^{237}$Pu excitation energy scale or energy resolution. Since there is no evidence from the rather sparsely separated data points ($\sim 0.2$~MeV) of vibrational resonance structure, the $\mathcal{LNG}$ simulation was carried out assuming full $\beta$-vibrational damping. 
The present $V_A=5.70$~MeV, $V_B=5.10$~MeV barrier heights, addressing the curve shown on Fig.~\ref{fig:pu237*pf}, are slightly higher that our initial guess (Tab.~\ref{tab:surrogateData}) but remain both lower than neutron emission threshold; preserving the fissile nucleus feature of $^{237}$Pu$^*$. In terms of spin-parity barrier curvature sensitivity of the even-odd fissioning nuclides, we retained for $^{237}$Pu$^*$ previously estimated values; meaning at inner saddle a spin-dependent sequence ($\hbar \omega_{A}^{J=1/2->13/2}=0.99/0.95/0.90/0.85/0.80/0.60/0.55$~MeV) and a spin-independent sequence at outer saddle  ($\hbar \omega_{B}^{J}=0.40$~MeV).\\

\begin{figure}[t]
\center{\vspace{1.cm}
\resizebox{0.95\columnwidth}{!}{
\includegraphics[height=5cm,angle=0]{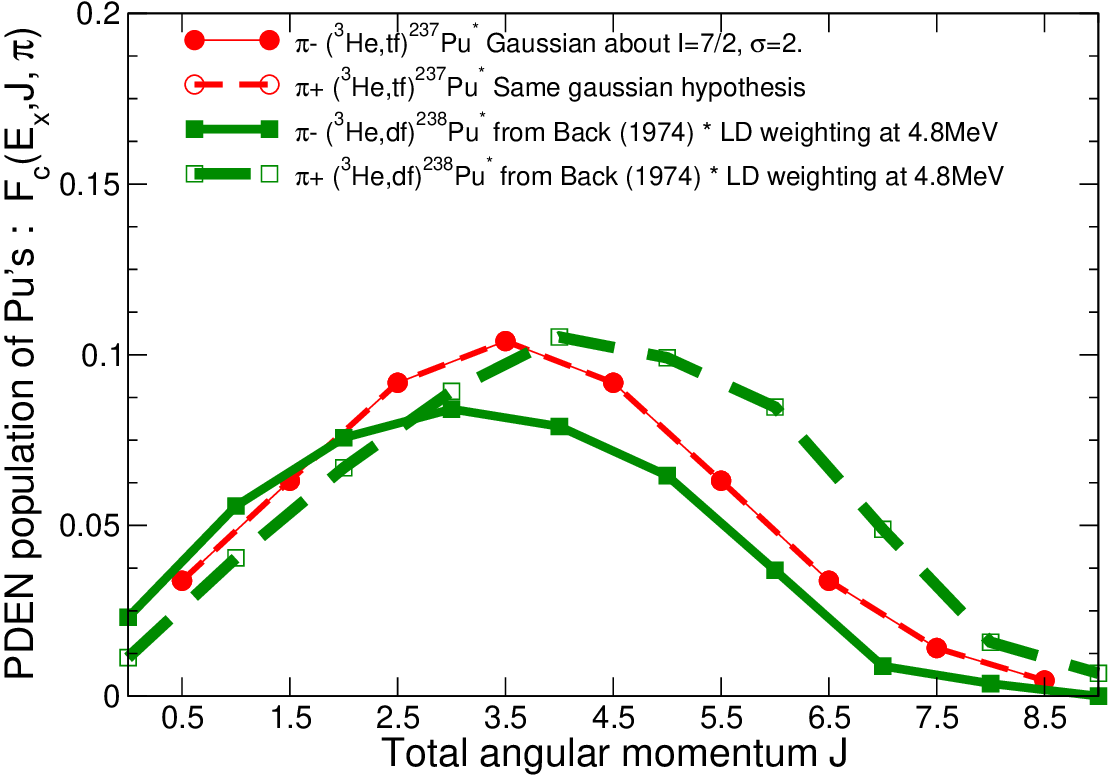}}}
\caption{(Color online) Distributions of total angular momenta corresponding to $^{237}Pu^*$ and $^{238}Pu^*$  excited states formed respectively in ($^3He,t$) and ($^3He,d$) charge-exchange reactions. For the former, we assumed a truncated gaussian distribution centered about $J=7/2$ and a dispersion $\sigma=2$. The latter distribution is converted from Ref.~\cite{bac:74} that supplies  the distribution of orbital angular momenta according to the transfer of a particle (separated out of the light incident projectile) into a single-particle shell of the target nucleus. Lines between symbols are drawn to guide the eye. Close symbols address negative $\pi$ (respectively open symbols for positive $\pi$). ($^3He,d$) population level density weighting (thick curves and Eq.~\ref{eq:jpiCN}) is made in agreement with results obtained in Ref.~\cite{bou:13}.}
\label{fig:Pu237*JpiDist_all}
\end{figure}

A.~Gavron~\etal~\cite{gav:76} do not supply information about spin-parity surrogate population distribution to feed our decay calculation in the case of ($^{3}$He,$t$) reactions. Hence we have chosen  a truncated gaussian parity equiprobable spin-parity distribution centered about $J=7/2$ with dispersion $\sigma=2$. The lack of information and high angular momenta of surrogate measurements 
%See Boutoux thesis p186
suggest both the gaussian assumption and the choice of $J=7/2$; the latter choice according to the intrinsic spin of the $^{237}$Pu$^*$ PDEN since quoting G.~Boutoux \etal~\cite{bout:12} {\it 'The $J^\pi$ distribution populated in a ($^{3}$He,$p$) transfer reaction is close to the one populated in the photon-induced reaction'}. % meaning close to the intrinsic spin of prior-to-fission compound nucleus. 
Figure~\ref{fig:Pu237*JpiDist_all} displays selected spin-parity distribution feeding the simulation for reproduction of $^{237}$Np($^{3}$He,$tf$) surrogate data. Fig.~\ref{fig:pu237*pf} compares our calculated fission probability curve with A.~Gavron~\etal~\cite{gav:76} data. The agreement is quite good in view of the quoted 10\% systematic errors on the experimental fission probabilities.   

%early direct calculations made by  Andersen \etal~\cite{and:70} for the particular case of the $^{239}Pu(d,pf)$ one-particle stripping reaction and by Back \etal~\cite{bac:74} for the remaining $(d,p)$, ($^3He,d$) and  $(t,p)$ surrogate reactions. We have used for calculating Eq.~\ref{eq:hfstdspA} the $J^\pi$  entrance fractions as a function of excitation energy supplied by Andersen \etal~\cite{and:70} in graphic form

\subsection{\label{ss:application238}$^{238}$Pu$^*$ compound nucleus}

\begin{figure}[h]
\center{\vspace{1.cm}
\resizebox{0.95\columnwidth}{!}{
\includegraphics[height=5cm,angle=0]{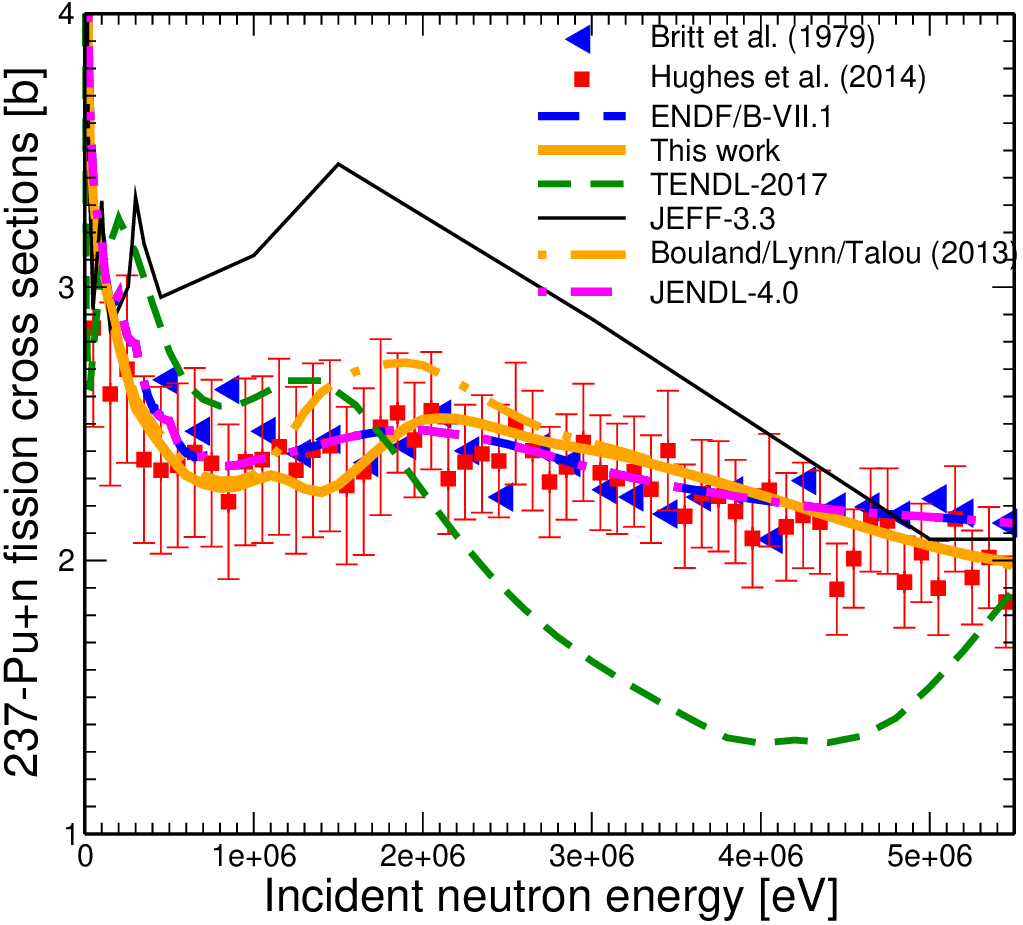}} }
\caption{\label{fig:pu238*xsf}(Color online) $^{237}$Pu  neutron-induced fission cross section (orange-solid curve) computed with $\mathcal{LNG}$ and compared to some evaluated data (ENDF/B-VII.1, JEFF-3.3,  JENDL-4.0 and TENDL-2017) and to extrapolated neutron data from experimental surrogate measurements respectively by Britt \etal~\cite{bri:79} and R.O.~Hughes \etal~\cite{hug:14}.}
\end{figure}
  
\begin{figure}[t]
\center{\vspace{1.cm}
\resizebox{0.95\columnwidth}{!}{
\includegraphics[height=5cm,angle=0]{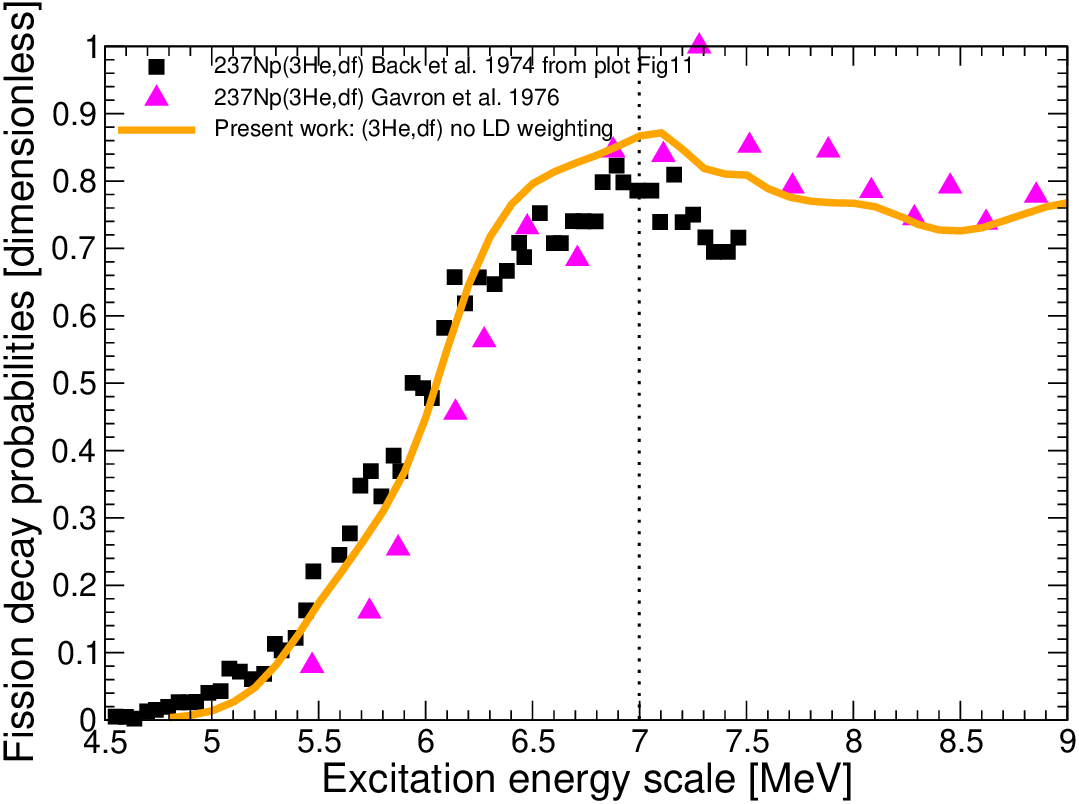}} }
\caption{\label{fig:pu238*pf}(Color online) $^{238}$Pu$^*$  surrogate-like fission probability (solid curve) with $\mathcal{LNG}$ and directly comparable with $^{237}$Np($^{3}$He,$df$) surrogate data from Back \etal~\cite{bac:74} and A. Gavron \etal~\cite{gav:76}.}
\end{figure}

There are no neutron-induced fission cross section measurement on $^{237}$Pu ($\tau_{1/2}$=45.1 days) but two extrapolated data  sets from surrogate measurements are released in EXFOR. Consistently with the $^{237}$Pu$^*$ experiment, Britt \etal~\cite{bri:79} extrapolated neutron cross sections (EXFOR \#14229 displayed on Fig.~\ref{fig:pu238*xsf}) but by using the $^{237}$Np($^{3}$He,$df$) canal. Once more, R.O.~Hughes \etal~\cite{hug:14} bring information from recent surrogate data material relying on the $^{235}$U$(p,tf)$/ $^{239}$Pu$(p,tf)$ ratio technique (EXFOR \#14396.003). The computed $^{237}$Pu  neutron-induced fission cross section (orange-solid curve) shows pretty good agreement with both surrogate data extrapolations (Fig.~\ref{fig:pu238*xsf}). Our cross section calculation is also supported by joint analyses of $^{237}$Np($^{3}$He,$df$) surrogate data from Back \etal~\cite{bac:74} and A. Gavron \etal~\cite{gav:76} that suggest  barrier heights close to $V_A=5.65$~MeV, $V_B=5.45$~MeV (i.e; unchanged from Ref.~\cite{bou:13}) and much lower than $S_n=7.0$~MeV. Latter statement suggests less fission cross section sensitivity on barrier curvature spin-parities for even-even than for even-odd fissioning nuclides. Hence no dependence was applied on barrier curvatures which assigned values are $\hbar \omega_{A}=1.05$~MeV and $\hbar \omega_{B}=0.60$~MeV. Final simulation of both fission probability data, fed with spin-parity surrogate population distribution reconstructed from \cite{bac:74}, is satisfactory as exemplified on Fig.~\ref{fig:pu238*pf}. The $^{238}$Pu$^*$  simulation was carried out assuming full $\beta$-vibrational damping.

\subsection{\label{ss:application240}$^{240}$Pu$^*$ compound nucleus}

Evaluating $^{240}$Pu$^*$ features is much more challenging than those above 'exotic' compound nuclei since we have to address high-quality neutron-induced resonant and continuum cross sections for standard and advanced reactor fuels. $^{239}$Pu is a fissile nucleus owing to the fission barrier of $^{240}$Pu$^*$ lying well below its neutron separation energy. Therefore our first requirement is an independent estimate of barrier heights that can indeed be achieved using surrogate spectroscopy investigating excitation energies below $S_n$. Literature provides many theoretical investigations of fission potential surfaces, but most seem to carry uncertainties in their absolute value of the order of 0.5 to 1 MeV. However there is general agreement that the outer barrier of the plutonium isotopes ($A^*<245$) is some few to several hundred keV lower than the inner barrier. As example, Refs.~\cite{Mol:09a} and~\cite{Deb:09} quote barrier pairs respectively of ($6.0$; $4.9$)~MeV and ($7.0$; $4.5$)~MeV  according to $^{240}$Pu$^*$ first and second saddle heights. %Surrogate reactions are especially suited for barrier height estimates below neutron emission threshold as suggested by the numerous measurements performed in this field. 
Fig.~\ref{fig:pu240*norm} enlightens  surrogate experimental data relevant to present work according to $^{240}$Pu$^*$. The main difficulties are two-fold: one is that there is a clear evidence for a pronounced 'resonance' structure at about $5$~MeV in the fission probability function that needs to be modeled; the other is possible systematic error in the normalization of those experimental data. 

\subsubsection{\label{sss:norm}Suitable normalization correction}
\begin{figure}[h]
\center{\vspace{0.4cm}
\resizebox{0.95\columnwidth}{!}{
\includegraphics[height=5cm,angle=0]{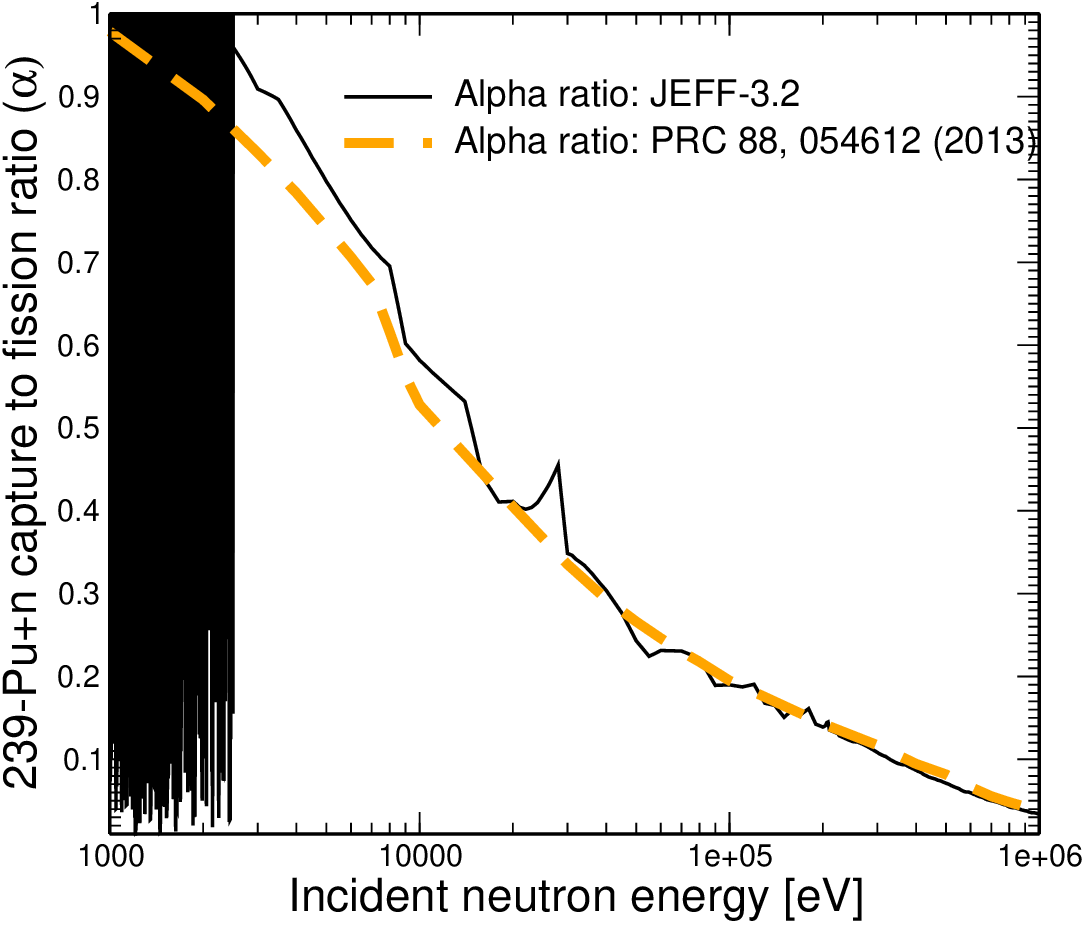}} }
\caption{\label{fig:pu240*alpha}(Color online) $^{239}$Pu capture to fission cross section ratio (so-called $\alpha$) shape as a function of neutron incident energy. For reference, the JEFF-3.2 recommendation  (black-solid curve) is compared to the simulation (orange-dashed curve). The large discontinuity observed in the JEFF-3.2 ratio right below $30$~keV is due to rough data transition between the URR and continuum energy ranges.}
\end{figure}

\begin{figure}[ht]
\center{\vspace{0.1cm}
\resizebox{0.95\columnwidth}{!}{
\includegraphics[height=5cm,angle=0]{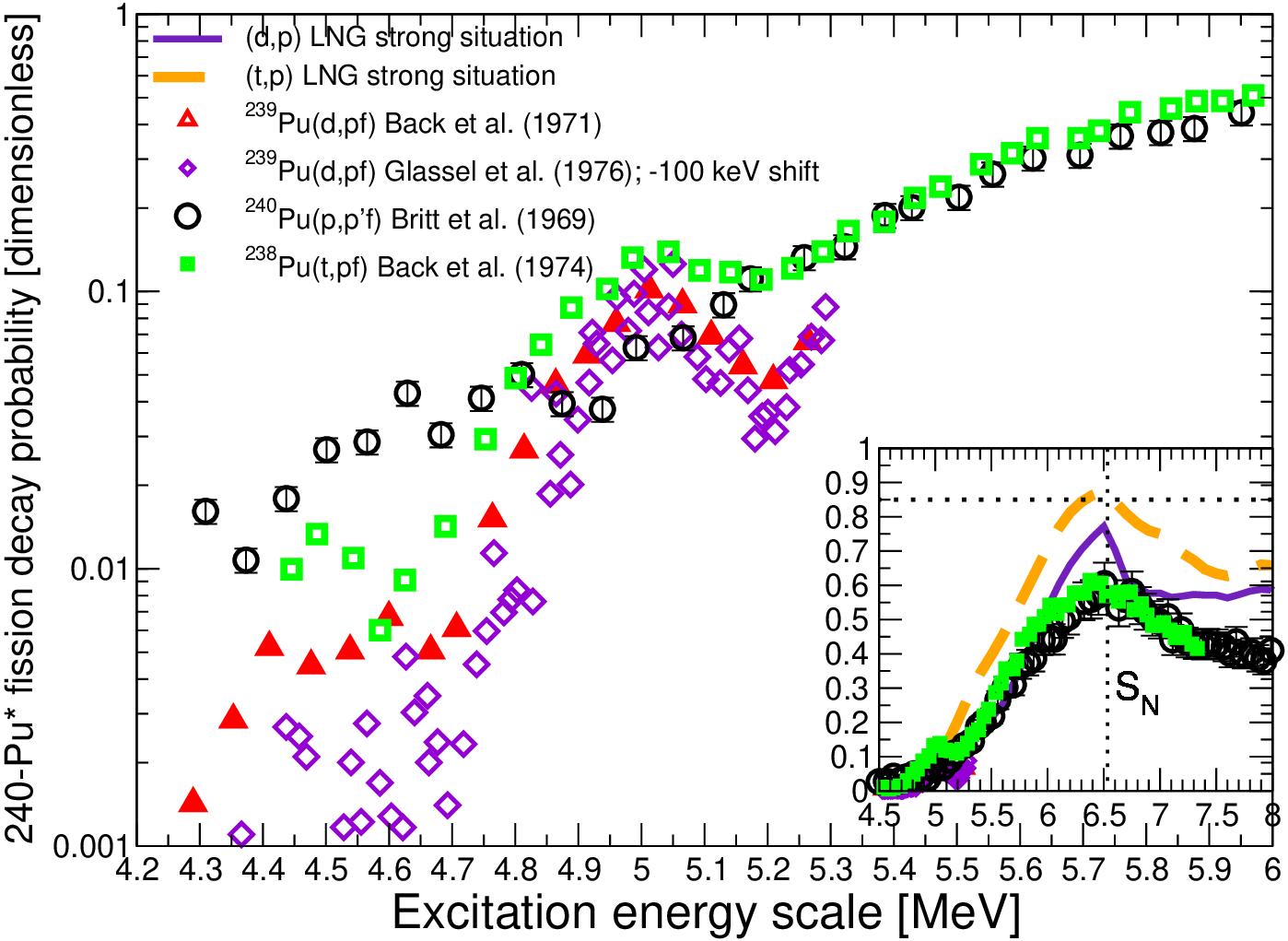}} }
\caption{\label{fig:pu240*norm}(Color online) Some measured fission probabilities according to $^{240}$Pu$^*$ as a function of both excitation energy and direct-reaction entrance vector: $^{238}$Pu($t,pf$)~\cite{cra:70a},  $^{239}$Pu($d,pf$)~\cite{gla:76}~\cite{bac:71} and $^{240}$Pu($p,p'f$)~\cite{bri:69}. The inset image displays the same  experimental data sets over the whole excitation energy range as well as present  simulations under full $\beta$-vibrations damping. The latter calculations are fed with either ($t,p$) or ($d,p$) direct reaction population distribution.}
\end{figure}

Systematic error in the normalization of $(t,p)$ experimental data covering the neutron emission threshold can be estimated  from neutron-induced $\alpha$-ratio (i.e; the ratio of the capture to fission cross sections) which trend is foreseen either from a recommended database (as JEFF-3.2) or from our $\mathcal{LNG}$ results~\cite{bou:13}. As displayed on Fig.~\ref{fig:pu240*alpha}, the $\alpha$-ratio is about 0.9 below neutron energies of a few keV. Above $5$~kev the latter falls rapidly to reach 0.1 in the $200-300$~keV region. The reason of this fall comes from the predominance of $1^+$ resonances in the compound nucleus over the $s$-wave region (Fig.\ref{fig:Pu240*JpiDist_all}) and its prompt decreasing importance with the onset of $p$-wave and $d$-wave absorption. This is still amplified by the absence of low-lying $1^+$ transition state at the inner barrier in the $^{240}$Pu$^*$. %voir (Paragraph \ref{sss:MCdecaytestingF}).    
%In addition $^{240}$Pu$^*$ is specific in the way that several Bohr fission transition channels are accessible at $S_n$ for (one-step) $0^+$ fission whereas $1^+$ fission occurs across a single transition state; the latter spin/parity feature  requiring at least the combination of two collective phonons (viz. bending coupled to mass-asymmetry DoF) which combined energy is therefore expected to be much higher than zero-vibration energy (about $1.5$~MeV and $0.8$~MeV above inner and outer barriers respectively). 
This fact makes the $1^+$ resonance contribution untypically small to low energy neutron fission cross section.  Referring to Eq.~\ref{eq:hfwccp}, the fission probability is equivalent at $S_N$ to the ratio $\frac{T_f}{T_f+T_\gamma}$; meaning $\frac{1}{1+\alpha}$. Since the neutron-induced $\alpha$-ratio within the $200-300$~keV region does not occur across $1^+$ states similarly to $(t,p)$-induced reactions (as justified  by  Fig~\ref{fig:Pu240*JpiDist_tp}; green open  circle), we can expect at $S_N$ a value of about $\frac{1}{1+0.15}$ meaning between 0.85 to 0.9 for the fission probability in contrast with the low value of $P_f(S_n)\approx0.60$ claimed by the authors of the $^{238}$Pu($t,pf$) measurement~\cite{bac:74} (see inset of Fig.~\ref{fig:pu240*norm}). However since the spin distribution of the states excited in the ($t,p$) reaction deviates significantly from the $^{239}$Pu($n,f$) reaction population, peaking at about $J=3-5$ rather than about $1-2$ (Section~\ref{s:Dedicated}), % as well as fluctuation factor formulations (refer to Eq.~\ref{eq:Pdecoupled}), <-- correction prise en compte
our final argument lies in our  surrogate-like fission probability simulation fed with the expected ($t,p$) excited state population (Fig.~\ref{fig:Pu240*JpiDist_tp}) that is based on our most reasonable set of evaluated resonance and nuclear structure parameters fitting reasonably neutron-induced cross sections~\cite{bou:13}. Resulting  fission probability, shown on Fig.~\ref{fig:pu240*norm} (inset; orange-dashed curve), confirms our expectation with a $P_f$ value around to $0.9$ at $S_N$. Therefore it seems necessary, for accurate model fitting of the fission probability to $(t,pf)$ data, to raise the latter by at least $30\%$. This renormalization is larger than maximal systematic uncertainty ($20\%$) claimed by the authors~\cite{bac:74} on the experiment but we can also assume $10\%$ error in our calculations due to various parameter-related uncertainties. 

\subsubsection{\label{sss:structure}Vibrational resonance structures modeling}
\begin{figure}[t]
\center{\vspace{1.cm}
\resizebox{0.95\columnwidth}{!}{
\includegraphics[height=5cm,angle=0]{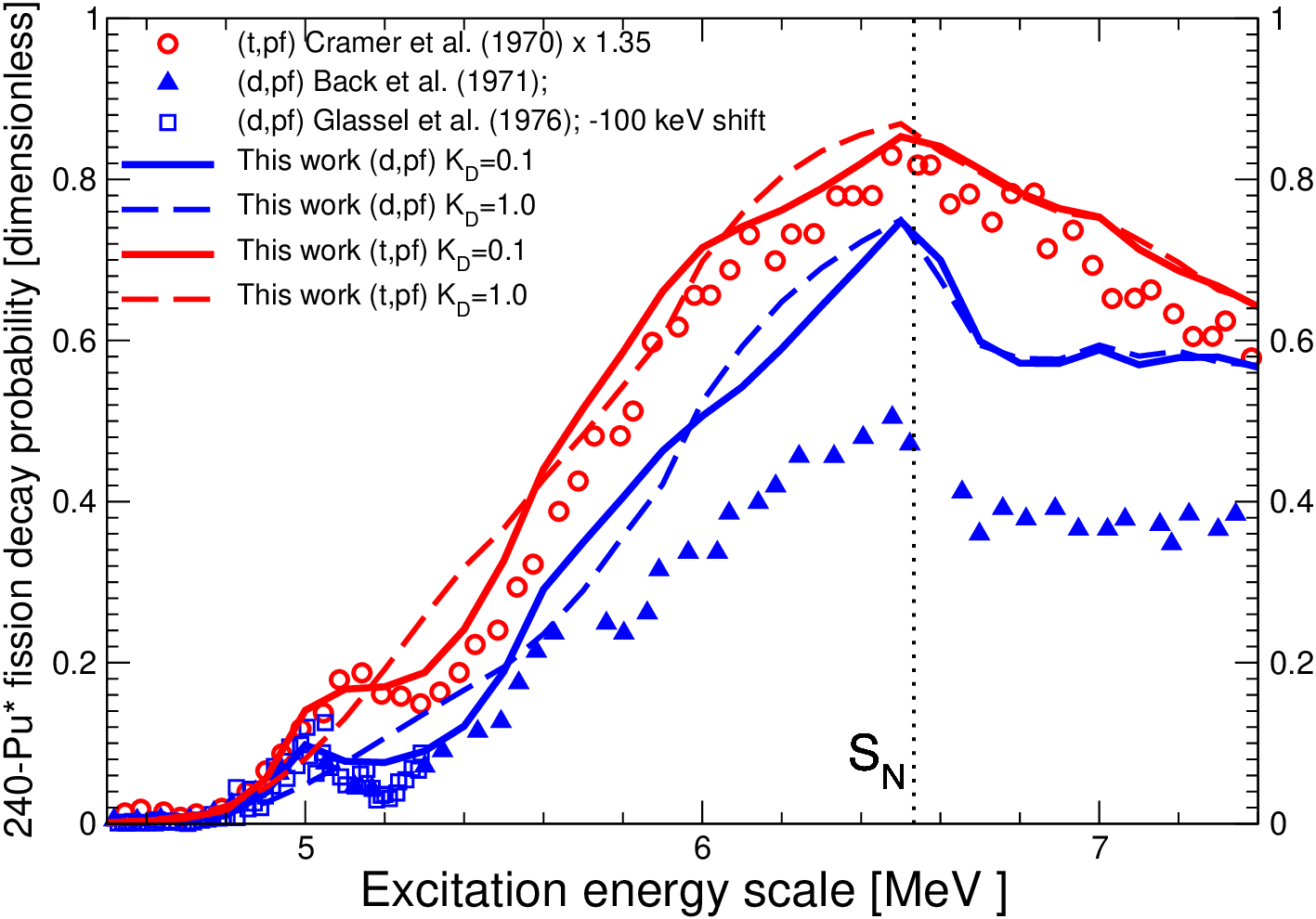}} }
\caption{\label{fig:Pu240*giant}(Color online) $^{238}$Pu($t,p$)~\cite{cra:70a} and $^{239}$Pu($d,p$)~\cite{gla:76}~\cite{bac:71} measurements and corresponding calculations according to $^{240}$Pu$^*$ as a function of  excitation energy. The simulations are performed under incompletely-damped second well $\beta$-vibrations involving an oscillator harmonic sequence of $K^\pi=0^+$ $\beta$ levels respectively at 2.95, 3.95, 4.95, 5.95, 6.95~MeV, etc. whose width increases exponentially with the excitation energy. Agreement between theory and experimental data around 5.0~MeV suggested to shift the $^{239}$Pu($d,p$) data from Glassel \etal~towards lower energies by 100 keV (doing thus the exact opposite of procedure carried in Ref.~\cite{gla:76}).}      
\end{figure}

$^{239}$Pu($d,p$) reactions have been widely used to investigate experimentally the well-known resonances at $5.0$~MeV (gross structure) and  $4.5$~MeV (weaker structure) excitation energies. Among the two ($d,p$) experimental data sets presented on Fig.~\ref{fig:pu240*norm}, the most accurate is the one from Glassel \etal~\cite{gla:76} whose measurement carries an energy resolution of $3$~keV (FWHM). In addition the authors tested quantitatively various hypotheses on the nature of those structures and their projection onto the barrier deformation channel states. This work is not aiming to mimic this baseline study but rather to include explicitly and with some degree of confidence observed intermediate structures from fission isomer excitation energy  until full damping, for best estimates of fission barrier heights when comparing fission probability simulation with experimental data. ($d,p$) experimental data that can help assessing gross structure resonances are likely similarly impacted as ($t,p$)  by possible systematic normalization error  since dating back to the same era. Using again the  simulation as support but fed with a dedicated ($d,p$) excited state population (as pictured on Fig.~\ref{fig:Pu240*JpiDist_all}), resulting calculations suggest a value closer to $0.75$ rather than $0.5$ for $P_f$ at $S_N$ (Fig.~\ref{fig:pu240*norm}; inset; purple solid curve). However break-up correction issues mainly related to deuteron incident particles in the field of the heavy target nucleus, seem to be exonerated to be the cause of this disagreement since deuteron breakup is expected to be sizable only above neutron emission threshold. %dixit Back
Because of the difficulty to correct the ($d,p$) data from experimental artifacts, those data have been used as a double-check for vibrational state assignment in second well only below $5.6$~MeV excitation energy. Table XXXVI of Ref.~\cite{bjo:80} reviews the level of knowledge on vibrational states in second well of the $^{240}$Pu$^*$ even-even fissioning nucleus. A first structure at $(4.0\pm0.10)$~MeV is mentioned with no spin-parity assignment, followed by two levels at $4.5$ and $(4.65\pm0.05)$~MeV respectively, which assumed $K^\pi$ values are $0^+$ and $0^-$. The VRS sequence is completed by a $K^\pi=0^+$ bandhead at $(5.05\pm0.02)$~MeV.  
% subjectif et E>Vb ...
%whereas the resonance-like structure around $5.7$~MeV (Fig.~\ref{fig:Pu240*giant}) is likely the highest excitation energy observed incompletely-damped structure as the damping of the vibrational states is expected to increase with energy. 
% Redite Equation~\ref{eq:Gdvariation} is dedicated to spreading of the vibrational strength over  broader energy intervals. The additional cause for the class-II $\beta$-vibrational structure smoothing with increasing energy is related to the exponentially increasing fission probability because of the opening of the various fission channels. 
Our VRS sequence is assumed to project fully on both inner and outer barrier deformation channel states %ruled by Eqs.~\ref{eq:lorentGc} and \ref{eq:lorentGf} but only low $J^\pi$ transition state values provide sizable penetrabilities  for second well $\beta$-vibrational states located below $V_B$. %.5$~MeV.
and aims in particular to reproduce strong resonances observed below the lowest of the two barriers (meaning here below $V_B=5.23$~MeV).   
On the spirit of QPVR calculations (Ref~\cite{bou:13}; Chapter III), the combination of zero-phonon $\beta$-vibration with given intrinsic state at barrier does not modify the original spin and parity of the latter. We recall that a $\beta$-vibration owns $K^\pi=0^+$ spin projection over fissioning nucleus elongation axis and is not present in the Bohr transition spectrum above saddles by definition. Same rule applies whenever  multiphonon $\beta$ combination with given transition state is involved. According to our best incompletely-damped $\beta$-vibrational fission probability modeling (Fig.~\ref{fig:Pu240*giant}), the isomeric ground state energy was set up at $2.95$~MeV excitation energy~\cite{sin:02}.  
%an effective isomeric ground state was set up at $4.50$~MeV excitation energy  (i.e; $2.10$~MeV above the actual second well ground state estimated energy). The latter, interpreted as a $\beta$-vibration of one phonon less than the well-known $5.0$~MeV vibration, 
The latter supplies the calibration for the phonon energy width exponential dependence (Eq.~\ref{eq:Gdvariation}) with a damping width of $\Gamma_{vib_{II\mbox{ }D}}(0^+)\approx 100$~keV, consistent with literature~\cite{gla:76}, and a damping coefficient of $\kappa_D=0.1$~MeV$^{-1}$. Table II of Ref.~\cite{bou:18} reports assumed properties of collective states in the determination of the level spectra along fission path.
% according to formula~\ref{eq:Gdvariation}, were found to be appropriate for experimental fission probability modeling .\\ 

\begin{figure}[ht]
\center{\vspace{1.cm}
\resizebox{0.95\columnwidth}{!}{
\includegraphics[height=5cm,angle=0]{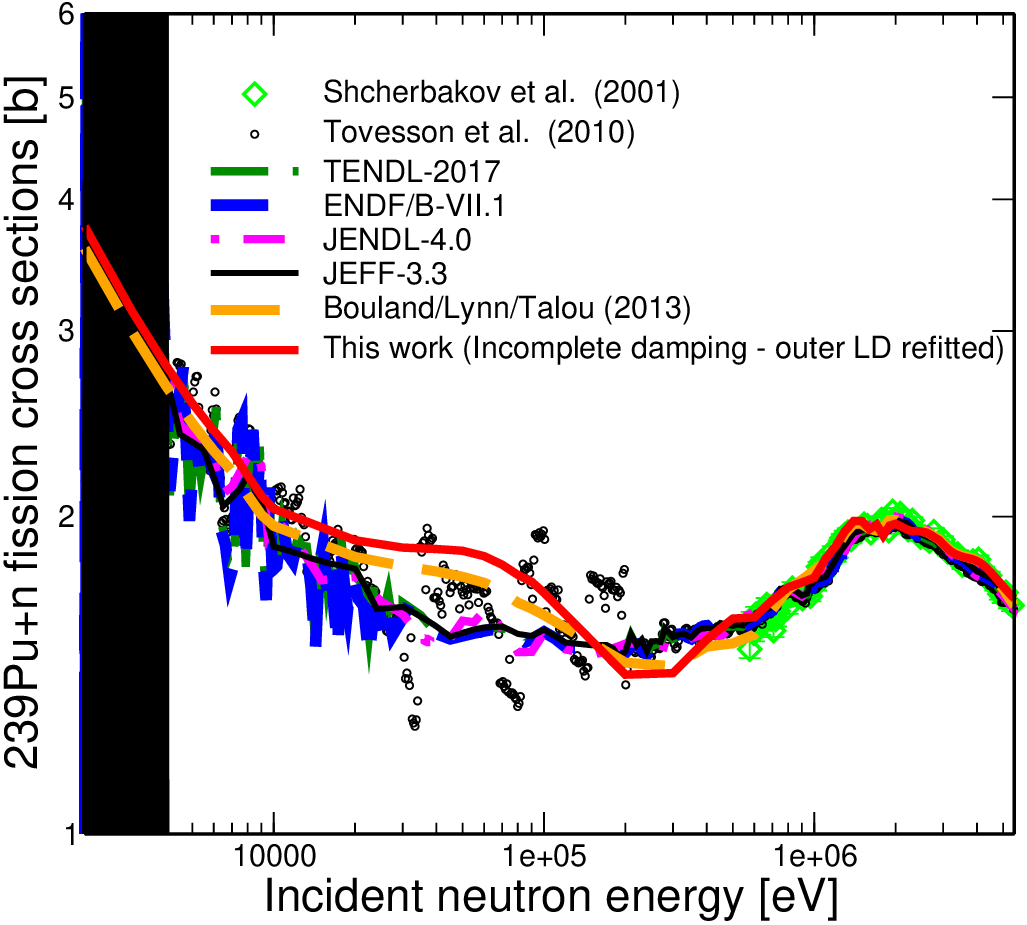}} }
\caption{\label{fig:pu240*fis}(Color online) $^{239}$Pu  fission cross section computed with $\mathcal{LNG}$ using the 'incomplete damping' (red solid curve) hypothesis. For reference, the orange dash curve recalls previous work~\cite{bou:13} that used the 'strong damping' assumption. Present work differs from the latter     also by re-adjustment of the outer barrier multiphase-temperature level density parameters. Results are benchmarked against evaluated data (ENDF/B-VII.1, JEFF-3.3,  JENDL-4.0 and TENDL-2017) and  measurements  performed by Tovesson \etal~\cite{tov:10} and Shcherbakov \etal~\cite{shc:01}.} 
\end{figure}

Figure~\ref{fig:Pu240*giant} shows reasonable representation of the observed VRS that is not unique and can be improved with better knowledge on the sequence of $\beta$-vibrations in second well. The incomplete damping treatment impacts also the fission cross section in the  neutron incident energy range until full damping situation occurs; meaning up to 3 MeV above $S_N$ for $\kappa_D=0.1$~MeV$^{-1}$ or retrained to 1 MeV if $\kappa_D=1.0$~MeV$^{-1}$. Present work (Fig.~\ref{fig:pu240*fis}) differs from~\cite{bou:13} by the incomplete damping treatment but also by re-adjustment of the outer barrier multiphase-temperature level density parameters. The damping mode selected according to fission barrier transmission in the calculation impacts moderately the other partial compound nucleus cross sections but as far as capture is concerned here, the relative effect on the latter remains affordable (Fig.~\ref{fig:pu240*xsg}).\textcolor{red}{}
  
\subsubsection{\label{sss:heights}Barrier height estimates} 
\begin{figure}[ht]
\center{\vspace{0.5cm}
\resizebox{0.95\columnwidth}{!}{
\includegraphics[height=5cm,angle=0]{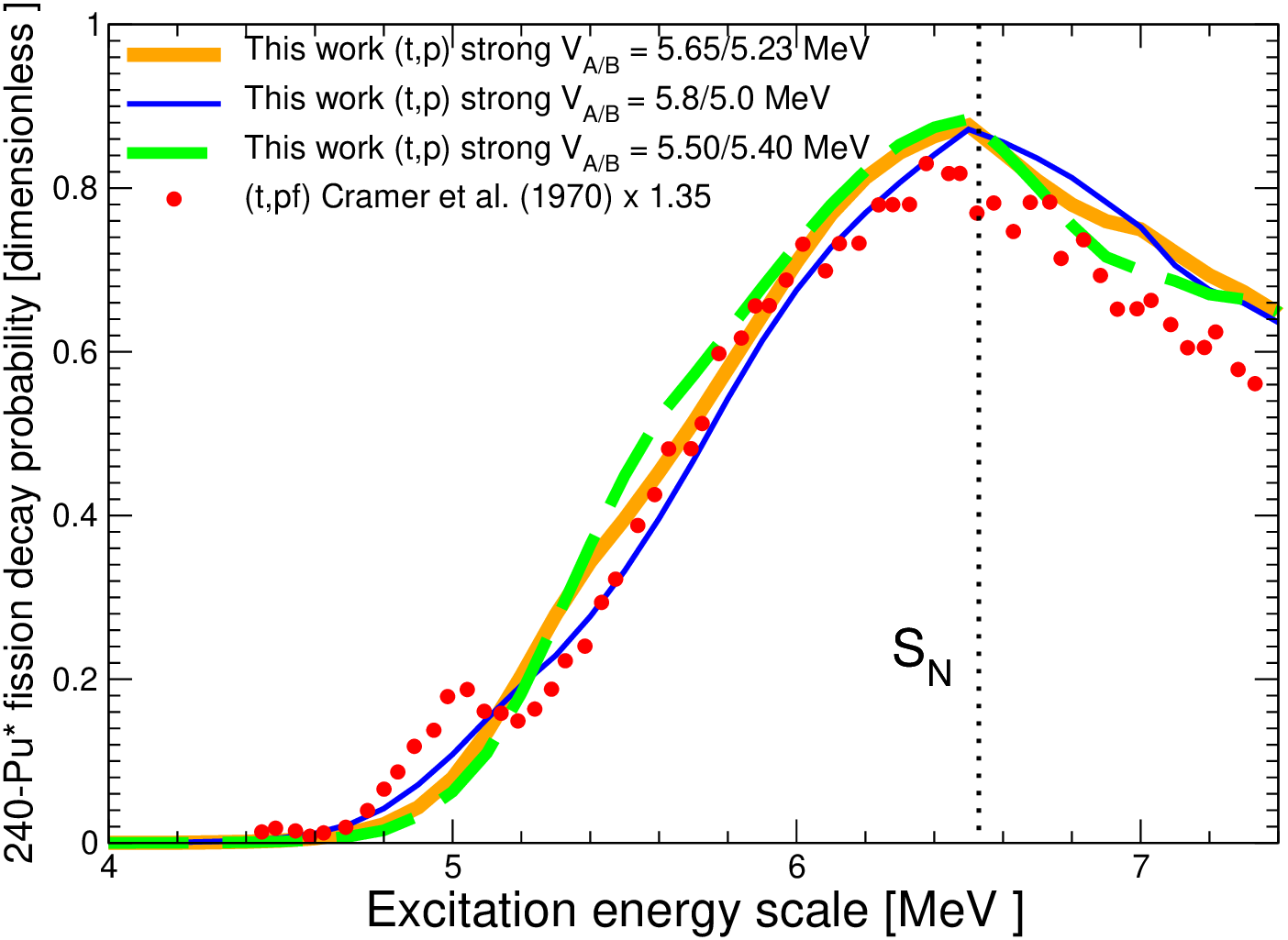}}}
\caption{\label{fig:pu240*heightsPf}(Color online) Fission probability simulation under strong coupling hypothesis according to $^{240}$Pu$^*$ as a function of both excitation energy and eligible  barrier heights. Distant ($V_A=5.8$, $V_B=5.0$~Mev), close ($V_A=5.5$, $V_B=5.4$~Mev) or present choice  ($V_A= 5.65$, $V_B=5.23$~Mev) inner and outer barrier pairs are considered. Renormalized $^{238}$Pu($t,pf$) data from Cramer and Britt~\cite{cra:70a} are displayed for comparison.}
\end{figure}

The range of reasonable fits for the barrier heights covers the parameter sets from $V_A=5.8$, $V_B=5.0$ to $V_A=5.5$, $V_B=5.4$~Mev as shown on Fig.~\ref{fig:pu240*heightsPf} on the ground of the strong coupling model, in which vibrational resonances are completely dissolved into the class-II compound states. Among those, the pair $V_A= 5.65$, $V_B=5.23$~Mev was adopted consistently with neutron-induced fission cross section feedback as reported in Ref.~\cite{bou:13}. This is also in line with theoretical prediction of an outer barrier a few hundred or several hundred keV below the inner barrier. We remember that barrier heights are inversely correlated to level densities~\footnote{As far as the $^{240}$Pu* fissioning nucleus is concerned, level densities on barrier tops start respectively 1.5 and 1.3~MeV above fundamental values; meaning respectively at $S_N$+600~keV and $S_N$ according to barrier height proposal of Ref.~\cite{bou:13} (Table II).} at barrier deformation which knowledge remains quite flimsy. %During our fit of measured neutron-induced fission cross section, we used a parametrized combinatorial level density for the inner barrier and an empirical one for the outer barrier (Fig.10 of Ref.~\cite{bou:13}). 
We notice that predicted fission probability at $S_N$ sounds not to be much dependent on the choice of barrier heights and so, tend to confirm large renormalization ($\geq 30\%$) on investigated old experimental data.

\begin{figure}[t]
\center{\vspace{1.cm}
\resizebox{0.95\columnwidth}{!}{
\includegraphics[height=5cm,angle=0]{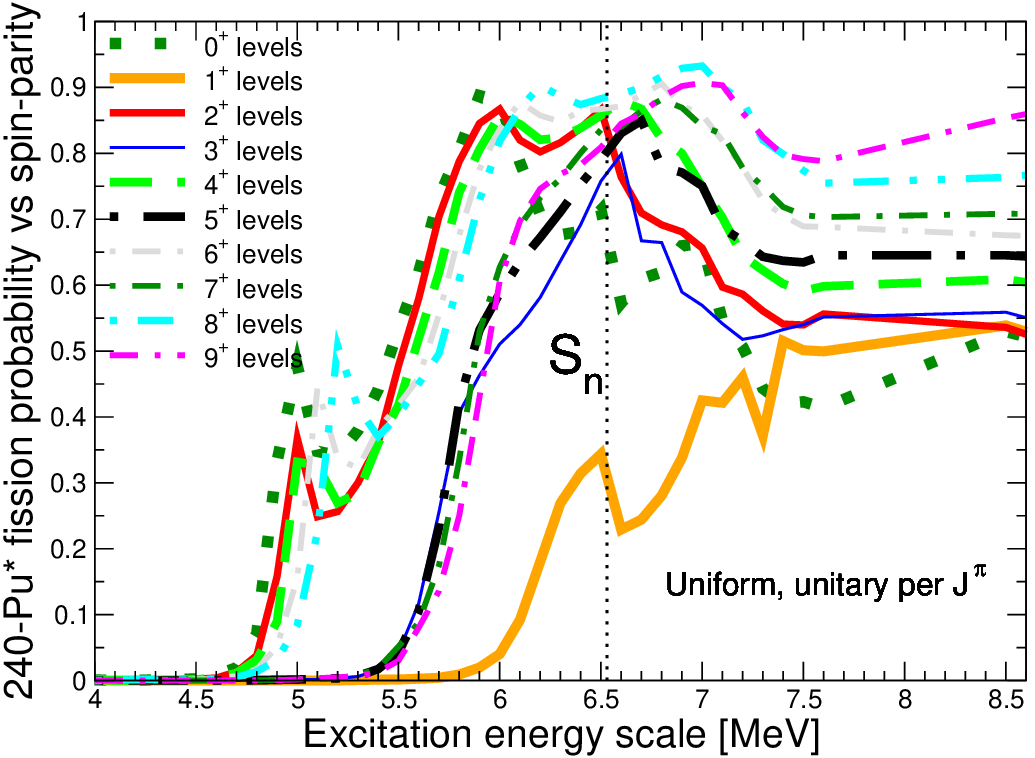}}}
\caption{(Color online)\label{fig:Pfisinc240*pi+} Monte Carlo $\mathcal{R}$-matrix double-humped fission barrier surrogate-like probabilities of $^{240}$Pu$^{*}$ as a function of resonance spin (positive parity) and excitation energy up to neutron kinetic energy of 2.1~MeV. The vertical bar at 6.53~MeV materializes neutron emission threshold. Present calculation differs from Fig.~\ref{fig:Pfis240*pi+} by explicit treatment of the second well $\beta$-vibrations.}
\end{figure}
\begin{figure}[t]
\center{\vspace{1.cm}
\resizebox{0.95\columnwidth}{!}{
\includegraphics[height=5cm,angle=0]{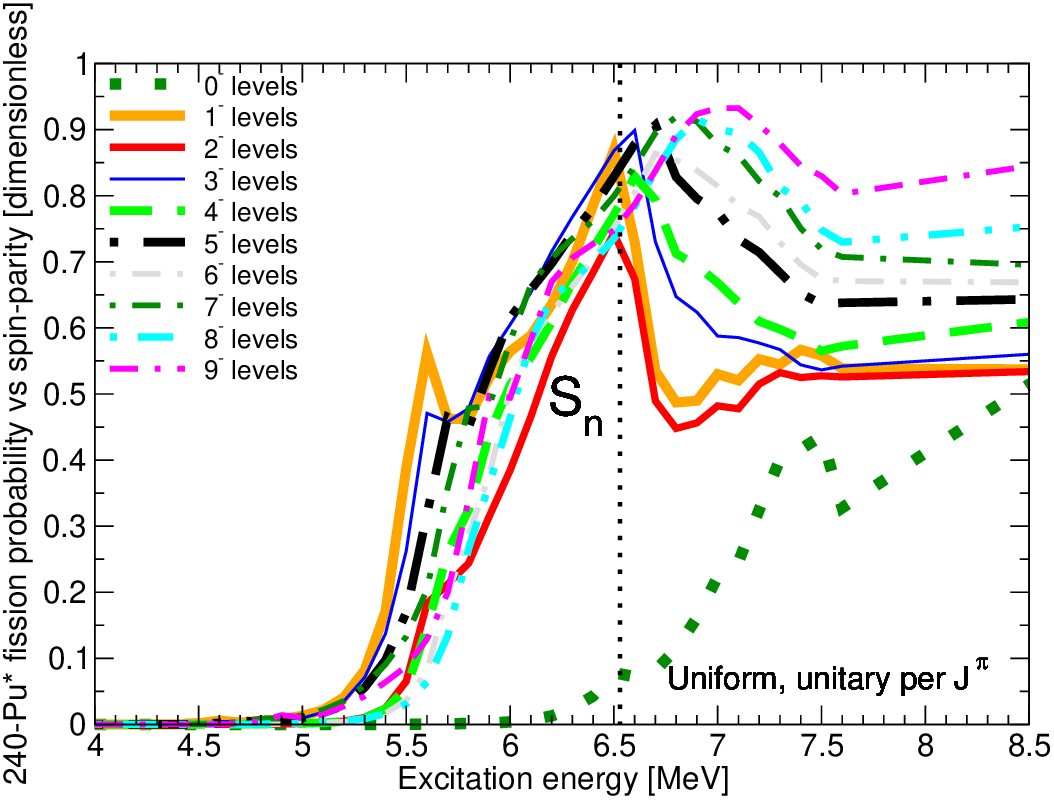}}}
\caption{(Color online)\label{fig:Pfisinc240*pi-} Same as Fig.~\ref{fig:Pfisinc240*pi+} but for negative parity Bohr transition states. Present calculation differs from Fig.~\ref{fig:Pfis240*pi-} by explicit treatment of the second well $\beta$-vibrations.}
\end{figure}
\subsubsection{\label{ss:con}Explicit VRS treatment in fission probability calculation}

We argumented in section~\ref{ss:MCdecaytesting} on surrogate-like fission probability shapes according to $^{240}$Pu$^{*}$ in case of strong damping situation. The incomplete dissolution of $\beta$-vibrational resonances into the quasi complete set of class-II compound states must show up in reaction probabilities and so, preventing smooth increase of Hill-Wheeler fission barrier penetrabilities. Associated penetrabilities are displayed on Figs.~\ref{fig:Pfisinc240*pi+} and~\ref{fig:Pfisinc240*pi-} respectively for  positive and negative parities. We immediately visualize successive picks in the $0^+$ state probability at $4.95$ and $5.95$  resulting respectively from the coupling of two- and three-phonons $\beta$-vibrations with the lowest intrinsic state at both barriers ($K^\pi=0^+$). Corresponding rotational level contributions $2^+,4^+,6^+,8^+$ are build on those bandheads. The rise in the $1^+$ fission penetrability below neutron emission energy, is become steeper because of the adjunction at $6.3$~MeV of a structure made from the admixture of two-phonons $\beta$-vibrations with one-phonon bending and mass-asymmetry vibrations ($K^\pi=1^+$). Structures show up as well in negative parity fission probabilities (Fig.~\ref{fig:Pfisinc240*pi-}). Among them, the structure at $5.60$~MeV is made from the coupling of the two-phonons $\beta$-vibrations with the one-phonon bending-vibration ($K^\pi=1^-$). Subsequent rotational level contributions $1^-,3^-,5^-,9^+$ are build on this bandhead. 

\subsection{\label{ss:application}$^{242}$Pu$^*$ compound nucleus}

\begin{figure}[t]
\center{\vspace{1.cm}
\resizebox{0.95\columnwidth}{!}{
\includegraphics[height=5cm,angle=0]{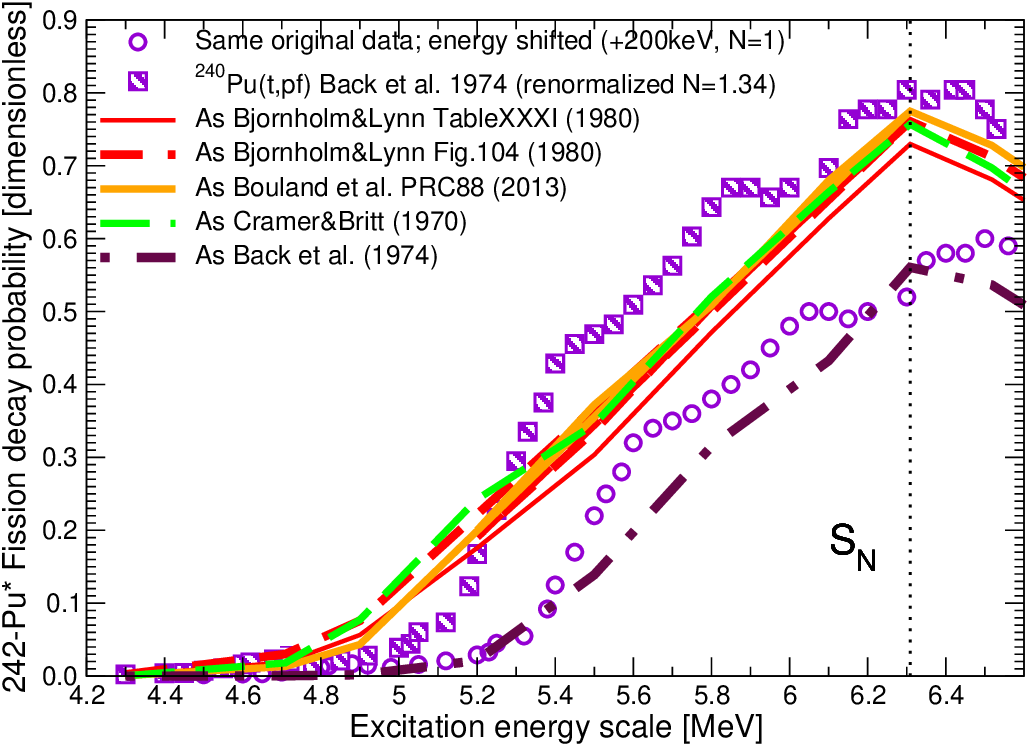}} }
\caption{\label{fig:Pu242*strong}(Color online) $^{240}$Pu($t,p$)~\cite{bac:74} experimental (both original data energy shifted for display and renormalized data) as a function of  excitation energy and corresponding  fission probability estimates. The latter are performed under completely-damped (i.e; strong)  $\beta$-vibrational resonance hypothesis and tested against several pairs of barrier parameters quoted in literature (Tab.~\ref{tab:surrogateData}). }
\end{figure}

Similarly to the $^{238}$Pu($t,pf$)$^{240}$Pu$^*$ data, the magnitude of the experimental fission probability around neutron separation energy is not consistent with evaluated neutron cross section feedback. The latter satisfactory reproduced by our prior estimate of barrier parameters~\cite{bou:13}, suggests at $S_N$ a fission probability approaching 0.8, whereas the measurement peaks at $0.6$. We have therefore renormalized the $^{240}$Pu($t,pf$)$^{242}$Pu$^*$ by a factor $1.34$. As a confirmation to our feeling, we tested various pairs of barrier parameters quoted in literature (Tab~\ref{tab:surrogateData}) to simulate both fission probability ($P_f$) - Fig.~\ref{fig:Pu242*strong} - and cross section. Among those, all sets but one (Back \etal~\cite{bac:74}), return similar $P_f$  trends with a highest probability value lying between $0.75$ and $0.8$ whereas the singular set  was  clearly tuned to reproduce the author's original data magnitude. %The resulting probability simulations (report to Fig.~\ref{fig:Pu242*strong}), under strong $\beta$-vibrational damping, have to be put into perspective with corresponding fission cross section calculations (Fig.~\ref{fig:pu242*strongXs}). 
The correlated fission cross section, drawn on Fig.~\ref{fig:pu242*incXs} in regards to some evaluated curves, %, representative of the bulk of fission measurements, 
confirms the low confidence to put on measurement original normalization.\\ %of the $^{240}$Pu($t,pf$)$^{242}$Pu$^*$. %(Fig.~\ref{fig:pu242*strongXs}; dot double-dashed purple curve).\\

\begin{figure}[t]
\center{\vspace{1.cm}
\resizebox{0.95\columnwidth}{!}{
\includegraphics[height=5cm,angle=0]{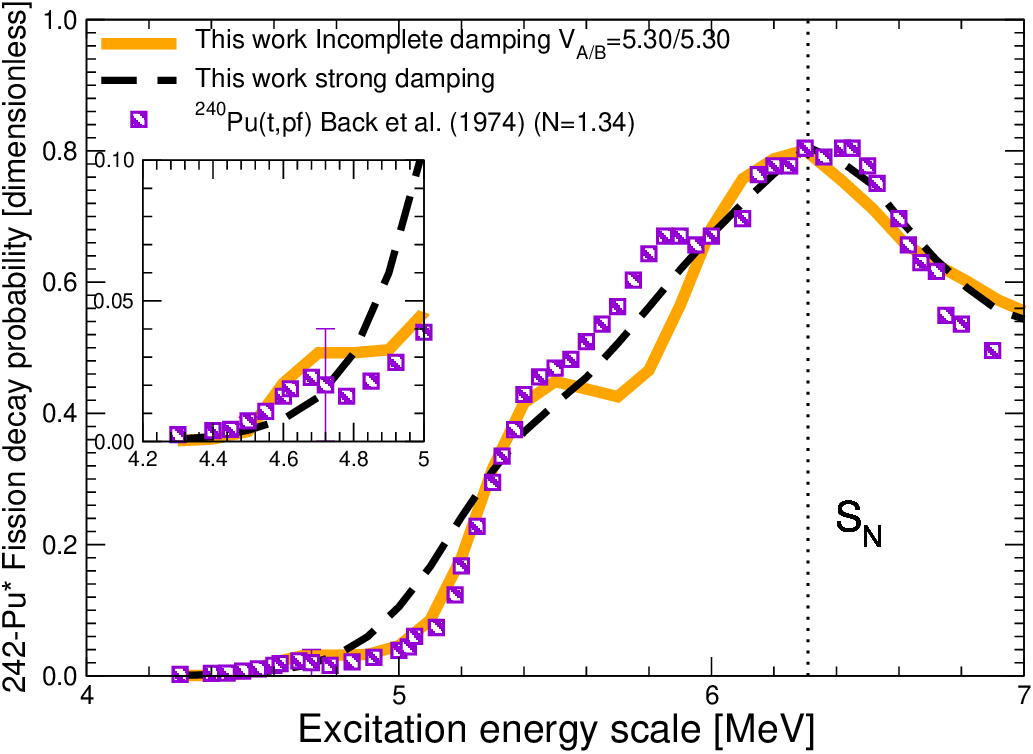}} }
\caption{\label{fig:Pu242*inc}(Color online) Renormalized $^{240}$Pu($t,p$)~\cite{bac:74} experimental data ($+34\%$) as a function of  excitation energy that are compared to strong- and incomplete-damping fission probability  simulations. Both results suggest slight decrease of the inner barrier height postulated in Ref.~\cite{bou:13} (from $V_{A/B}=5.40/5.30$~MeV to $V_{A/B}=5.30/5.30$~MeV). The inset graphic, enlarging the $4$ to $5$ MeV energy zone, reveals the vibrational structure observed around $4.65$~MeV.}
\end{figure}

\begin{figure}[t]
\center{\vspace{0.5cm}
\resizebox{0.95\columnwidth}{!}{
\includegraphics[height=5cm,angle=0]{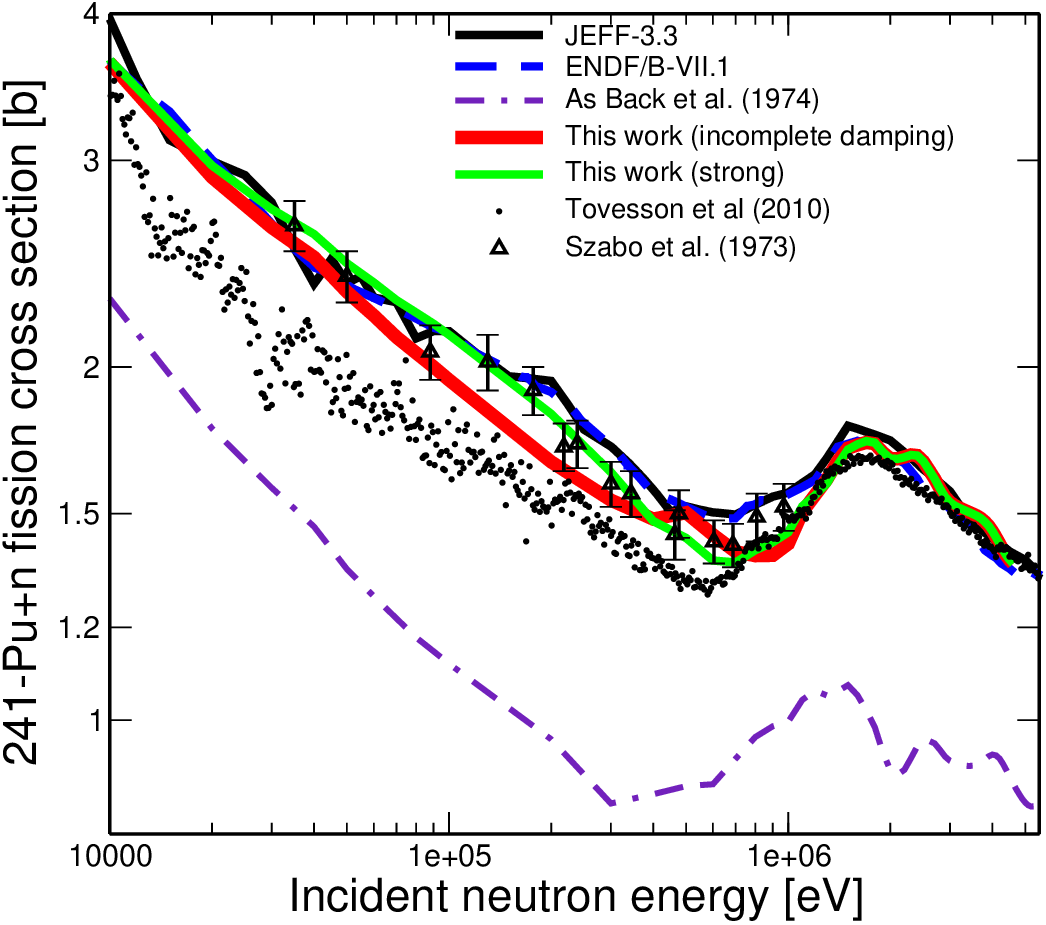}} }
\caption{\label{fig:pu242*incXs}(Color online) $^{241}$Pu  fission cross section computed under strong or incomplete damping hypothesis as a function of  excitation energy. Present
 calculations suggesting slight lowering of the inner barrier are compared to the two  measurements by Tovesson \etal~\cite{tov:10} and Szabo \etal~\cite{sza:73}, and with some standard evaluated files (ENDF/B-VII.1 and JEFF-3.3).The dot double-dash purple curve, based on the parameters determined by Back \etal~\cite{bac:74} to fit their own data, shows a trend far from the other simulations and so, confirms significant bias (larger than $30\%$) on the normalization of the original surrogate measurement.}
\end{figure}

%The $^{240}$Pu($t,pf$)$^{242}$Pu$^*$ experimental data drawn versus energy (Fig.~\ref{fig:Pu242*strong}) exhibit clearly structure that has been interpreted in literature as incompletely damped vibrational structure. 
Reference~\cite{bjo:80} quotes the observation of a single VRS owning $K^\pi=0^+$ at ($4.65\pm0.05$)~MeV that justifies present use of incompletely damped resonance formalism.  Figure~\ref{fig:Pu242*inc} displays a simulation based on this modeling that includes  sequences of $\beta$-vibrational resonances with $\hbar\omega_\beta$=800~keV  built on the fission isomer at $2.2$~MeV~~\cite{sin:02}. This choice of numerical values imply the reproduction of the VRS by the coupling of a three-phonons $\beta$-vibration with the lowest intrinsic state on both barriers. Beyond this, sequences of $\beta$-vibrations are constructed on the same way as $^{240}$Pu$^*$.  %coupled with  mass-asymmetry, bending and $\gamma$ vibrational modes sitting respectively at $0.6$, $0.65$ and $0.8$~MeV higher than the isomer state are also built to serve as additional rotational bandheads whereas more complex DoF combinations are involved above in excitation energy ($>E_{II}+1.35$~MeV). The damping width of these vibrational states among class-II compound states starts at $100$~keV and increases with energy with a damping constant of $\kappa_D=0.1$~MeV$^{-1}$. 
The several barrier pairs tested suggest to lower slightly  the inner barrier height of Ref.~\cite{bou:13} from $V_{A/B}=5.40/5.30$~MeV down to $V_{A/B}=5.30/5.30$~MeV and thus, confirm very close barrier height values for the $^{242}$Pu$^*$ in the framework of one dimension double-humped parabolic barrier representation. Several attempts to better fit the energy zone between $5.4$~MeV and $5.9$~MeV have remained unsuccessful. However the authors of the measurements~\cite {bac:74} reported trouble fitting the energy zone above $5.3$~MeV, acknowledging weakness in experimental data single events correction.     \\

In view of the above $^{240}$Pu($t,pf$)$^{242}$Pu$^*$ data analysis feedback, we have recalculated the neutron-induced fission cross section of $^{241}$Pu using the slightly modified barrier heights according either the strong or the incomplete damping hypothesis. Figure~\ref{fig:pu242*incXs} plots both calculated curves to be justapoxed with  two  measurements by Tovesson \etal~\cite{tov:10} and Szabo \etal~\cite{sza:73}, and finally with some evaluated files (ENDF/B-VII.1 and JEFF-3.3). Our $strong$ calculation is in good agreement with  recommended fission cross sections and the measurement by Szabo \etal~whereas our {\it best} calculation is slightly worsens by current choice of incomplete damping parameters.  High-precision data from Tovesson \etal~\cite{tov:10},  although resulting from a well-tested technique,  are 20-30$\%$ lower than the evaluations to be put into perspective with the global uncertainty quoted by the authors (ranging from $\pm3\%$ to  $\pm6\%$ below $1.5$~MeV).

%I am not too sure why you seem so concerned, Patrick. I believe that the evaluations are soundly based on good measurements for four cases; Pu239, 240, 241 and 242. Our calculations are in quite good overall agreement with the evaluations for the odd targets, but Tovesson's new data for 241 are running considerably below the evaluations. If Tovesson's data are found to be totally sound (in normalization etc.) then we would need to adjust our barrier LD's at some future date. On the even targets we are underrunning the 240 and (to a lesser extent) overshooting the 242 at about 100 keV and below, but this is well below the barrier energy. We could start adjusting things like ordering of low-lying Nilsson states at the barrier deformations and barrier penetrability parameters to get better fits,


\subsection{\label{ss:application}$^{244}$Pu$^*$ compound nucleus}

\begin{figure}[t]
\center{\vspace{1.cm}
\resizebox{0.95\columnwidth}{!}{
\includegraphics[height=5cm,angle=0]{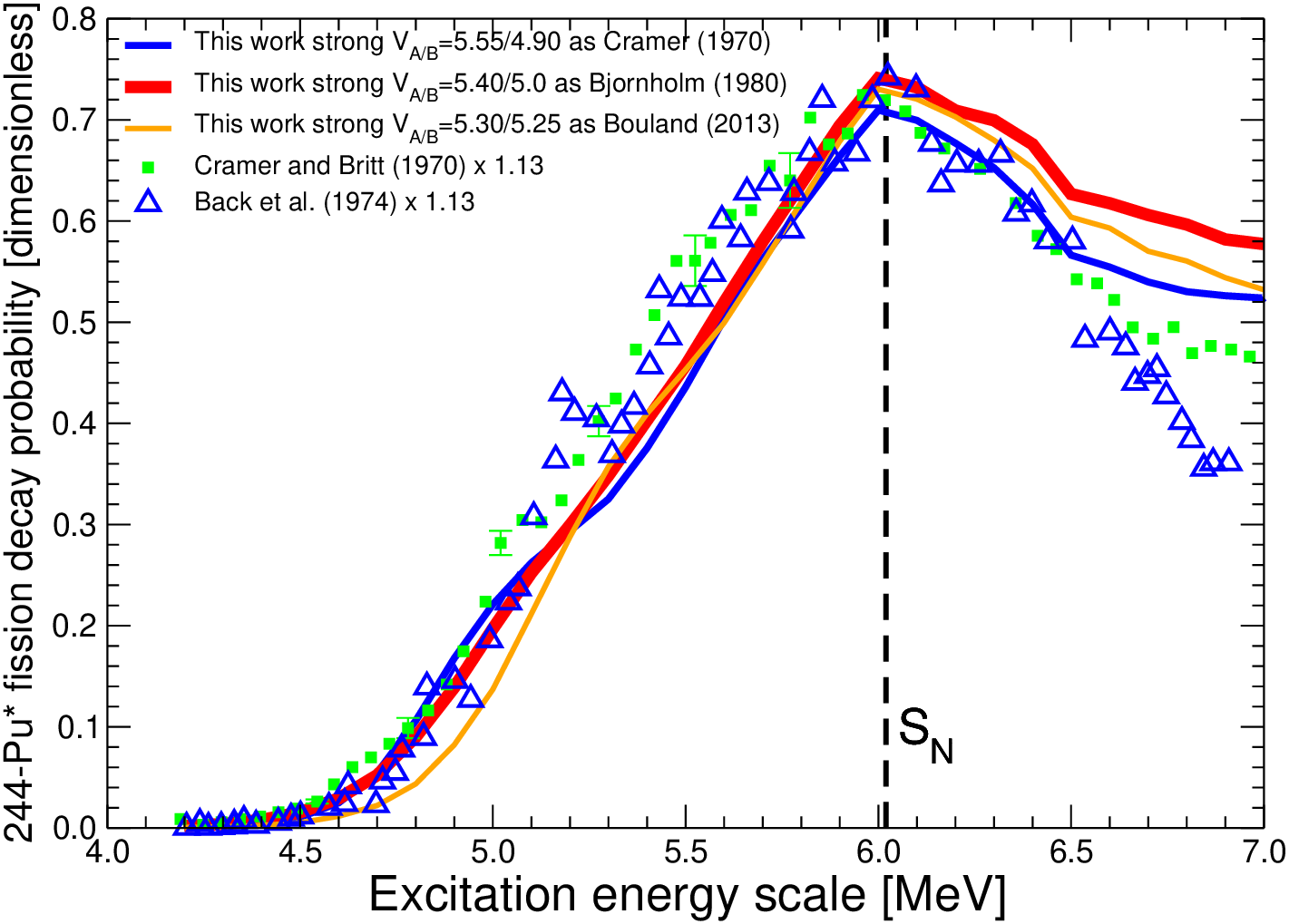}} }
\caption{\label{fig:Pu244*strong}(Color online) $^{242}$Pu($t,p$)~\cite{cra:70a}~\cite{bac:74} renormalized experimental data and fission probability calculations according to $^{244}$Pu$^*$ as a function of  excitation energy. The simulations are performed under completely-damped (i.e; strong)  $\beta$-vibrational resonance hypothesis. The origin of the departure between calculations and experimental data above 6.5~MeV comes from full opening of the inelastic channels and the moderate confidence level we can put in present inelastic level density without any calibration on NPS measurements.}
\end{figure}

Because of the dearth of neutron-induced cross section measurements for $^{243}$Pu, review and analysis of existing of surrogate-induced fission probabilities of $^{244}$Pu$^{*}$ is of considerable interest for neutron cross section evaluation work. The surrogate data bring in particular some support for choosing representative barrier heights but the robustness of our data interpolation for the non-measured isotope is also strongly ensured by present approach and feedback from the study of the other isotopes of the family~\cite{bou:13}. Our surrogate-type reconstruction (Fig.~\ref{fig:Pu244*strong}) suggests for this isotope more reasonable experimental data renormalization, meaning within the experimental uncertainties. The  observed maximal probability is enhanced from $0.65$ to $0.73$ (i.e; $+13\%$). This is in agreement with the simulations made using barrier parameter set proposed either by Cramer and Britt~\cite{cra:70a} or Bjornholm and Lynn~\cite{bjo:80}. Our feedback is reinforced by a $^{244}$Pu$^{*}$ neutron emission threshold lower than other fissile plutonium isotopes (Tab~\ref{tab:surrogateData}) relatively to fission barrier heights which prevents the fission probability to rise as high as the others. Figure~\ref{fig:Pu244*strong} shows several  simulations performed under completely-damped $\beta$-vibrational resonance hypothesis. Present work sticks on the choice of barrier parameters ($5.30$/$5.25$~MeV) used in Ref.~\cite{bou:13} which close values differ from the remote pairs of barrier heights selected by other authors.\\ 
%Present choice is determined by expected impact of VRS formalism that carves the calculated fission probability in between resonances.\\

\begin{figure}[t]
\center{\vspace{1.cm}
\resizebox{0.95\columnwidth}{!}{
\includegraphics[height=5cm,angle=0]{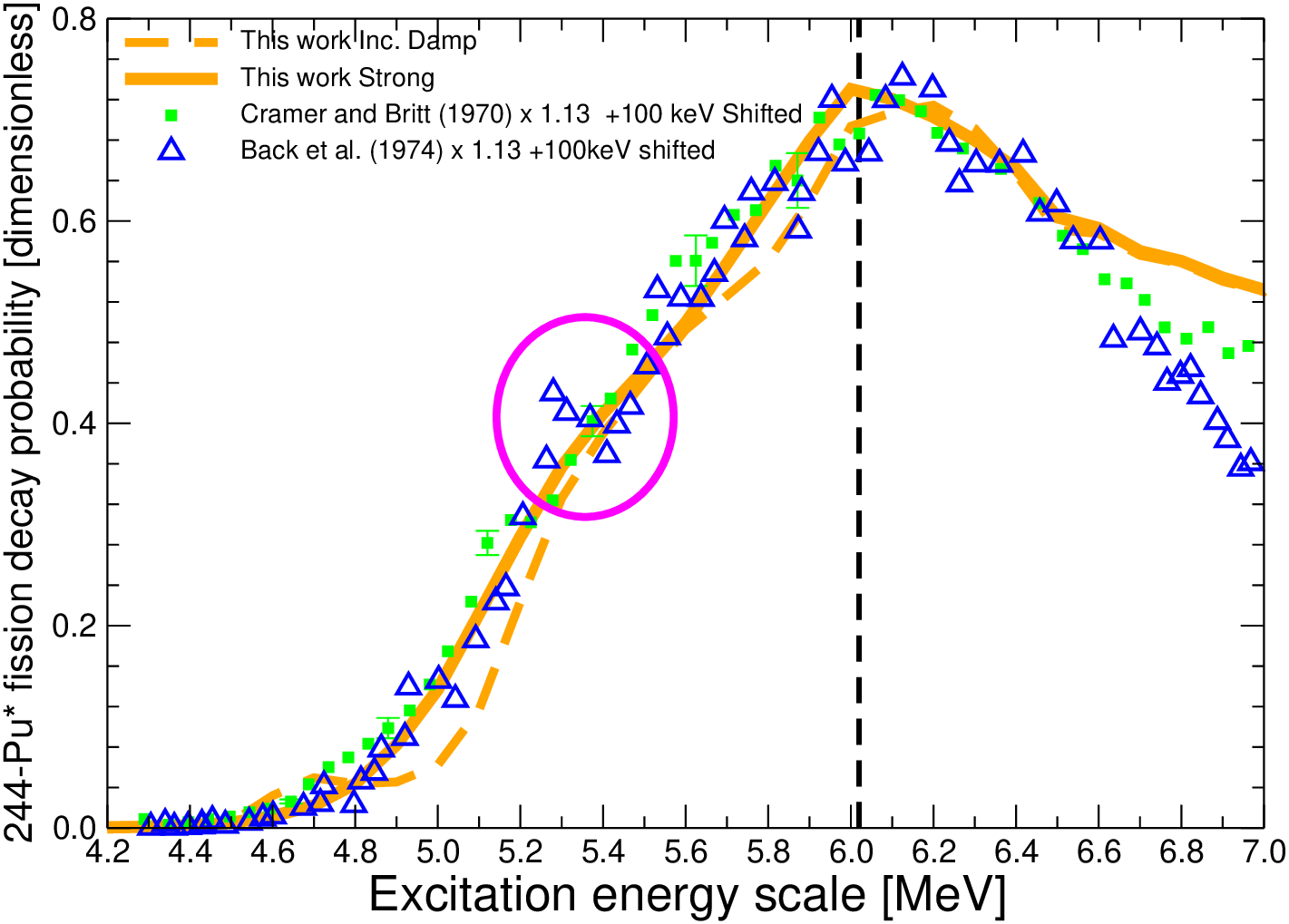}} }
\caption{\label{fig:Pu244*giant}(Color online) $^{242}$Pu($t,p$)~\cite{cra:70a}~\cite{bac:74} renormalized experimental data and corresponding fission probability calculations according to $^{244}$Pu$^*$ as a function of  excitation energy. The simulations are performed under either completely- (i.e; strong; orange solid curve) or incompletely-damped $\beta$-vibrational resonance hypothesis (orange dash curve).}% The latter involves an oscillator harmonic sequence of $K^\pi=0^+$ $\beta$ levels respectively at 4.10, 4.60, 5.10~MeV, etc. (pink medium-thicked solid curve) or 4.10, 5.10, 6.10~MeV, etc. (purple-dashed curve). The damping width of these vibrational states starts at 0.065 MeV and increases exponentially with the excitation energy with the proportionality coefficient 0.5.} %Reasonable representation of the observed giant resonances below $6.0$~MeV either in terms of ($t,p$) or  ($d,p$) entrance reaction is obtained using the $\mathcal{LNG}$ approach based in particular on an identical set of nuclear structure parameters that fits also well neutron-induced cross sections.}
\end{figure}

Consistency is manifest in between the two sets of ($t,pf$) data relevant to $^{244}$Pu$^*$ as documented in literature (Tab.~\ref{tab:surrogateData}) although resonance structure pattern is more glaring in the measurement by Back~\etal~\cite{bac:74} suggesting as for as $^{240}$Pu$^*$ and $^{242}$Pu$^*$, possible use of incompletely damped VRS formalism. Very few information is available in literature on this point except the quotation of a $K^\pi=0^+$ $\beta$-structure at about $4.6$~MeV above normal deformation ground state. If we endorse the latter structure and the pronounced structure at about $5.2$~MeV (Fig.~\ref{fig:Pu244*giant}; purple ellipse) assuming for both $K^\pi=0^+$ character, this would imply a quite low value for the vibrational quantum of about $0.5$~MeV. More probable is the superposition of the $\beta$-vibrational state with another intrinsic excitation of low-energy (other than the lowest intrinsic state) since the expected order of magnitude for the $\beta$-vibration spacing around this excitation energy is $1$~MeV. Referring to Table II of Ref.~\cite{bou:18} that reports low-energy intrinsic mode phonons according to second well, we can notice that a spacing of $0.6$~MeV is carried by a mass-asymmetric phonon of $K^\pi=0^-$ character. On this ground, we have set up a sequence of $\beta$-vibrational resonances combined with low energy intrinsic states that is built on an assumed fission isomer at $2.6$~MeV with $\hbar\omega_\beta$=1~MeV. Figure~\ref{fig:Pu244*giant} shows  simulations corresponding to two alternatives : with strong (solid curve) and incomplete damping (dash curve) of the vibrational resonances into the quasi complete set of class-II compound nucleus states. We take note of the good agreement between the strong calculation and the experimental data, suggesting the non mandatory use of the 'best' theory that carves greatly the calculated fission probability in between resonance structures. We must then endorse the extreme difficulty to determine unambiguously, from the angular-integrated experimental probabilities alone, the mixing of the $\beta$ vibrations with the underlying spectrum of intrinsic modes and corresponding phonon energies. Table~\ref{tab:surrogateData} shows well the various 'best' sets of barrier parameters that fit the $^{242}$Pu($t,pf$)$^{244}$Pu$^*$ fission probability experimental data.

%assume that a structure is observed around this energy (Fig.~\ref{fig:Pu244*giant}; inset, red circle) and we also acknowledge the more pronounced structure at about $5.2$~MeV, of $K^\pi=0^+$ character, this would imply a quite low value for the vibrational quantum of about $0.5$~MeV. More probable is the fragmentation of the $\beta$-vibrational strength over $1$~MeV; the expected order of magnitude for the $\beta$-vibration spacing around this excitation energy. Figure~\ref{fig:Pu244*giant} shows the $\mathcal{LNG}$ simulations corresponding to the two alternatives, meaning $\hbar \omega_\beta=0.5$~MeV (simulating a fragmentation of equal strength) or $1.0$~MeV for the $\beta$-vibrational ($K^\pi=0^+$, incompletely damped) resonance spacing. 
%A first  sequence of $\beta$-vibrational resonances is built from an assumed 'isomer' state at $4.10$~MeV. Sequences of $\beta$-vibrations coupled with other DoF as mass-assymetry, $\gamma$ and bending modes (sitting respectively at $0.8$~MeV, $0.1$~MeV and $1.2$~MeV higher than the isomer state relatively to the inner barrier or $0.1$~MeV, $0.8$~MeV and $0.6$~MeV relatively to the outer barrier) are also built to serve as rotational bandheads. The damping width of these vibrational states among class-II compound states starts at $65$~keV and increases with energy with the proportionality coefficient of 1.0 (Eq.~\ref{eq:Gdvariation}). 
%Between the two alternatives of vibrational quantum, preference is given to the value of $1.0$~MeV from the visual inspection of Fig.~\ref{fig:Pu244*giant} (purple-dashed curve).\\      

\begin{figure}[t]
\center{\vspace{0.5cm}
\resizebox{0.95\columnwidth}{!}{
\includegraphics[height=5cm,angle=0]{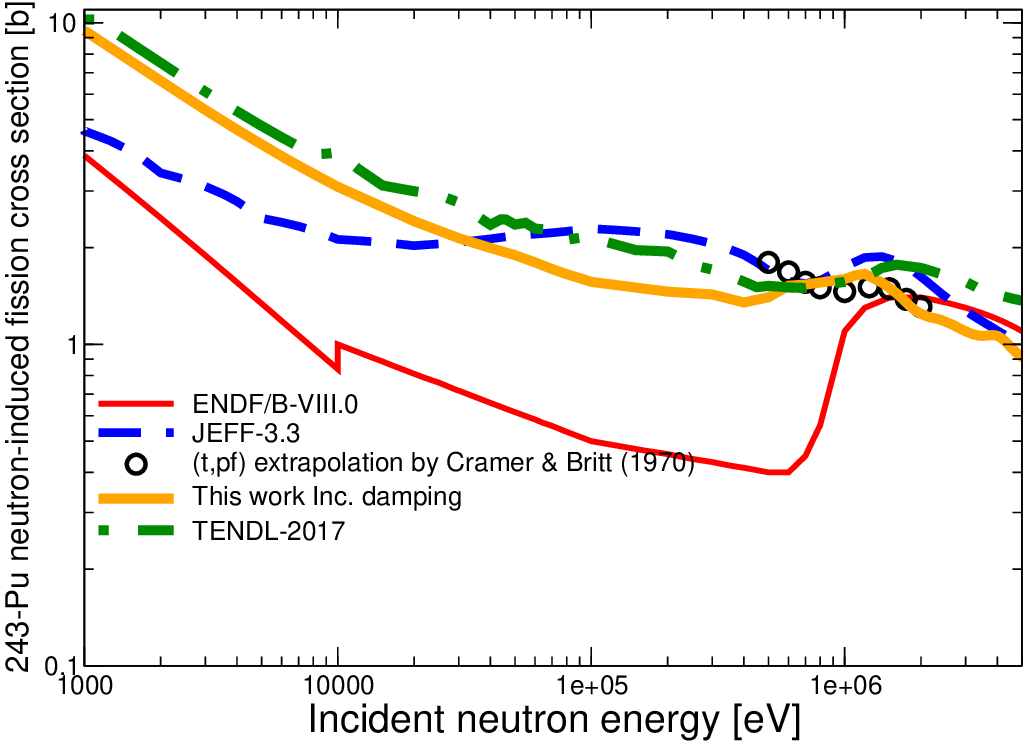}} }
\caption{\label{fig:pu244*fis}(Color online) $^{243}$Pu neutron-induced fission cross section computed with $\mathcal{LNG}$ using incompletely-damped $\beta$-vibrational resonance hypothesis (see details in the text). Present results are compared with major evaluated data files (ENDF/B-VIII.0, JEFF-3.3 and TENDL-2017) and with the surrogate data extrapolation by Cramer and Britt~\cite{cra:70a}. The latter returns similar trend as present work. No recommendations are made in the JENDL-4.0 data library for this isotope. The ENDF/B-VIII.0 evaluation still counts $^{243}$Pu as a member of the fertile family.}    
%The impact of present very  distant barrier height pair ($5.45$/$4.80$~MeV) can be compared to our own past choice involving neighboring values ($5.30$/$5.25$~MeV)~\cite{bou:13}.}
\end{figure}

Since the use of VRS formalism can be claimed for this isotope, we must verify its impact on the calculated neutron-induced fission cross section. No dramatic effect is encountered even at low neutron incident energy (Fig.~\ref{fig:pu244*fis}) but it must be remembered that classically the price to pay using better theory is usually a change in barrier heights or level densities; strongly model-dependent.\\\\\\

\begin{figure}[t]
\center{\vspace{1.0cm}
\resizebox{0.95\columnwidth}{!}{
\includegraphics[height=5cm,angle=0]{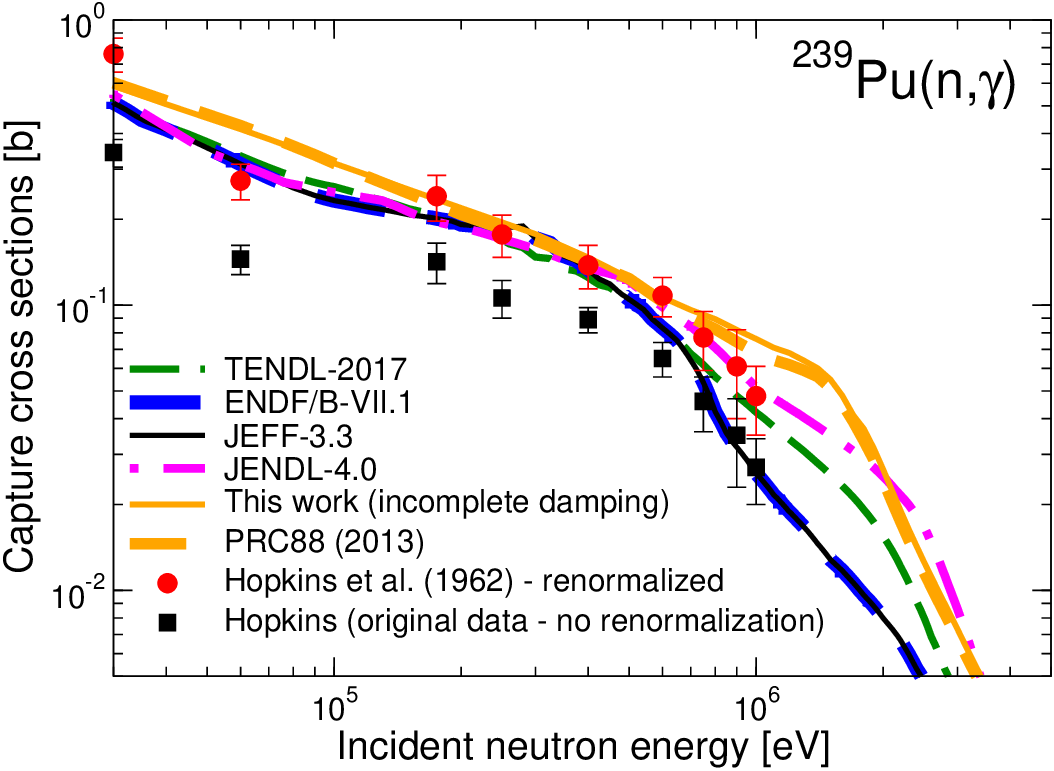}} }
\caption{\label{fig:pu240*xsg}(Color online) $^{239}$Pu  capture cross section computed with $\mathcal{LNG}$ as a function of neutron incident energy and compared to some evaluated data (ENDF/B-VII.1, JEFF-3.3,  JENDL-4.0 and TENDL-2017) and the old  measurement by Hopkins et al.~\cite{hop:61}. Present result (orange solid curve) differs slightly from previous calculation~\cite{bou:13} (orange dash curve).}
\end{figure}

\section{\label{s:Conclusion}Summary and perspectives}

Along this paper, we have enlightened the actual possibility to carry one-dimensional fission barrier extended $R$-matrix Monte Carlo simulations of neutron-induced cross sections jointly with  surrogate-like decay probabilities. The application to the Pu fissile isotope family over the 4 to 8 MeV excitation energy range in terms of simulated probabilities has shown good agreement for the $^{237~and~238}$Pu$^*$ and reasonable agreement for the $^{240, 242~and~244}$Pu$^*$ isotopes with experimental data assuming for the latter sound data renormalization ($<35\%$). Literature review of fission experimental data bringing situations where an obvious pattern of vibrational structure is revealed, we supplemented our calculations with an option putting a sequence of vibrational resonance terms in the transmission function for  both inner and outer saddle barriers to still be more in agreement with the shape of the surrogate data. %Actual impact on the calculated neutron fission cross section of the sub-neutron emission threshold incompletely-damped VRS has been observed  in the case of the $^{240~and~242}$Pu$^*$ below $1$~MeV neutron energy. 
If we omit the question of $\gamma$-decay probabilities to be discussed in a future publication, %was tackled in this work because of lack of old experimental data with respect to the Pu isotope family, 
it is manifest that evaluating simultaneously experimental neutron-induced cross sections and fission decay probabilities will help assessing nuclear parameters for fissile nuclides which fission threshold is not accessible by NPS techniques and for target material with unsuitable lifetimes or with high radio-toxicity. The latter statement has been frequently put forward in literature but also strongly questioned because of the technique used (SRM) in analyzing surrogate data. Present Monte Carlo approach does not suffer from such controversial choice which should be avoided in next generation evaluations. However we are aware of the need to pursue efforts on direct reaction modeling which reliability remains a genuine challenge.\\ 

Beyond Monte Carlo approach, a simpler comparison with the more conventional analytical formulation of decay probabilities has raised the main biases brought by SRM hypotheses; namely on the excited nucleus spin-parity state population distribution, on the in-out going channel width fluctuation correction factor and finally on the shape of individual decay probabilities as far as heavy isotopes are concerned. In terms of population distribution, replacement of peculiar $(t,p)$ or $(d,p)$ direct distribution by a neutron distribution can generate large biases on calculated probabilities below neutron emission threshold. However above $S_n$ the impact seems to be strongly reduced. Calculated differences between  common WFCF formulation and SWFCF surrogate variant remain limited (a few tens of percent) when fissile actinides are concerned whereas the impact of the correction reaches up to $100\%$ above neutron 'inelastic' threshold on both radiative and fission decays which have now endorsed the role of the enhanced channels. Regarding WE that suggests no spin and parity BR dependency, no dramatic fission decay shape differences are observed for actinides except when peculiarity exists in discrete transition state spectrum as for the $1^+$ state in $^{240}$Pu$^*$. In any event, failure of the WE hypothesis regarding $\gamma$-decay surrogate-like probabilities imposes equal conclusion with the fission decay probabilities below $S_n$ since total flux is preserved. Above $S_n$, the picture is fundamentally different because neutron emission probability acts as driver of the $\gamma$-decay reaction. For both reactions, the WE hypothesis does not apply at all. As conclusion, we can say that validity of SRM in surrogate data analyses relies on the overall bias brought by the three main ingredients of Eq.~\ref{eq:Pdecoupled} which definitive impact can not be evaluated by common decoupling as acknowledged from analytical formulae. This reinforces present surrogate simulations based on efficient Monte Carlo sampling of surrogate-like decay probabilities fed by sound spin-parity excited state population distributions.\\

Further work will concern the demonstration of present method validity for experimental surrogate $\gamma$-decay probabilities jointly analyzed with measured fission decay probabilities and neutron-induced cross sections. Foreseen simultaneous $\gamma$- and fission-decay probabilities measurements on heavy targets will bring the opportunity to full validation of present MC technique. We expect more confidence on experimental data normalization by reference to older data sets because of the possibility to control the unitary of $\gamma$-decay probability below both neutron emission and fission thresholds.  Although present method accuracy is a step forward standard inclusion of surrogate data in neutron-induced cross section evaluation procedure, impact of  common  simplifications made in terms of fission path calculation (one-dimension, Hill-Wheeler transmission coefficients, etc.) will have to be quantified and more rigorous treatment envisioned.

%-----------------------------------------------------------------------------------------------------------------------------
%--Appendix
%-----------------------------------------------------------------------------------------------------------------------------
\appendix
\section{Erratum - $^{239}$Pu capture cross section\label{A:erratum}}

A graphic substitution has occurred in Ref.~\cite{bou:13} regarding Fig.~18. Latter caption refers to the $^{239}$Pu neutron capture cross section whereas the graphic displays the $^{241}$Pu neutron capture cross section. To correct this unfortunate exchange, Fig.~\ref{fig:pu240*xsg} displays the foreseen Pu isotope cross section.  The associated experimental database above the resonance range [0-2.25~keV] is rather poor and we rely on the unique and old average measurement by Hopkins et al.~\cite{hop:61}. We observe that main evaluations differ above 700 keV where no experimental data are released and where the very small value of the capture cross section complicates any  NPS measurement. Present calculation deviates from the smooth compound nucleus formation shape at about 1.5 MeV. Investigation of this Wigner-cusp-type inflection shows that it comes from the strong competition by other open channels.

\newpage

%%%%%%%%%%%%%%%%% Acknowledgments
\acknowledgments
{\footnotesize One of the authors (O. Bouland) expresses his deep gratitude to Eric J. Lynn and Patrick Talou from LANL for the many fruitful discussions about the AVXSF code and related physics.}
%\acknowledgments
%{\footnotesize The LANL contribution is carried out under the auspices of the National Nuclear Security Administration of the U. S. Department of Energy at Los Alamos National Laboratory under Contract No.\ DE-AC52-06NA25396.}
%\newpage  

%------------------------------------- MASTER TABLE -- 
\begin{table}[htb]
\centering
 \rotatebox{90}{%
   \begin{varwidth}{\textheight}
\caption{\label{tab:surrogateData}  List of fissile Pu compound isotopes studied in present work and corresponding surrogate data that were fitted using the $\mathcal{LNG}$ code. Present but also older sets of fundamental barrier heights and curvatures are also listed. We recall that these parameters are far from being unique to fit the data but supply reasonable results within one dimensional Hill-Wheeler double-humped fission barrier frame. Neutron emission threshold ($S_n$) values quoted provide the NPS baseline.}      
\begin{tabular}{c|c|cc|c|c|ccccc}
\hline
Compound  & $S_n$ & Reactions & Ref.& Assumed population  & Neutron-induced exp. data &  $V_A$ &$\hbar \omega_A$ & $V_B$ & $\hbar \omega_B$ & Ref.\\
 nucleus  &[MeV] & \multicolumn{2}{c|}{}& distribution &  (EXFOR\#) or data origin&\multicolumn{4}{c}{[MeV]}&\\
\hline
%See also Lynn80 page 882 Table XXXI
$^{237}$Pu$^*$ & 5.86& $^{237}$Np($^{3}$He,$tf$)  &\cite{gav:76}& Gaussian & 14229.030 14386.002 41064 41369&5.70 & 0.99 & 5.10 & 0.40  &Present\\
&&&& &&5.60 & 0.99 & 4.95 & 0.40  &\cite{bou:13}\\
&&&& &&5.90 & 0.80 & 5.20 & 0.52  &\cite{bjo:80};Table XXXL\\
$^{238}$Pu$^*$ & 7.00& $^{237}$Np($^{3}$He,$df$) &\cite{bac:74} \cite{gav:76}&Reconstructed from \cite{bac:74}&14229.031 14396.003&5.65  &1.05 & 5.45 & 0.60  &Present as~\cite{bou:13}\\
&&&& &&5.50 & 1.04 & 5.00 & 0.60  &\cite{bjo:80};Table XXXL\\
$^{240}$Pu$^*$ & 6.53 &$^{239}$Pu($d,pf$) & \cite{gla:76}& As published by \cite{and:70} & 41455.012 14271.002 14271.004&    5.65 & 1.05 & 5.23 & 0.60  &Present as~\cite{bou:13} \\
& &$^{238}$Pu(t,pf) &\cite{cra:70a}&Reconstructed from \cite{bac:74} && 5.57 & 1.04 & 5.07 & 0.60  &\cite{bjo:80}\\
 & &$^{240}$Pu($p,p'$) & \cite{bri:69}& Not analyzed here &&  &&  & & \\
&&&& &&\\
$^{242}$Pu$^*$ & 6.31 &$^{240}$Pu($t,pf$)&\cite{bac:74}&Reconstructed from \cite{bac:74}&14271.007 20567.004& 5.30 & 1.05 & 5.30 & 0.60 & Present\\
&&&& &&5.60 & 1.04 & 5.10 & 0.60  &\cite{bjo:80};Table XXXL\\
&&&& &&5.55 & 1.00 & 5.05 & 0.70  &\cite{bjo:80};Figs.104/131\\
&&&& &&5.60 & 1.25 & 5.05 & 0.42  &\cite{cra:70a}\\
&&&& &&5.40 & 1.05 & 5.30 & 0.60  &\cite{bou:13}\\
&&&& &&5.60 & 0.82 & 5.65 & 0.59  &\cite{bac:74}\\
$^{244}$Pu$^*$ & 6.02&$^{242}$Pu($t,pf$)&\cite{bac:74}~\cite{cra:70a}&Reconstructed from \cite{bac:74}&Ref.~\cite{cra:70b}, Table II& 5.30 & 1.05 & 5.25 & 0.60  &Present and \cite{bou:13}\\
&&&&&& 5.40 & 1.04 & 5.0 & 0.60  & \cite{bjo:80};Table XXXL \\
&&&& &&5.55 & 1.25 & 4.90 & 0.40  &\cite{cra:70a}\\
&&&& &&5.45 & 0.80 & 5.35 & 0.57  &\cite{bac:74}\\
\hline
\end{tabular}      
%        \caption{My caption lorem ipsum}\label{tab_b}     
    \end{varwidth}}
\end{table}

\end{document}